\documentclass[11pt]{article}
\usepackage{graphicx}
\usepackage{amssymb}
\usepackage[section]{placeins}
\usepackage{color}
\usepackage{hyperref}

\def\cadremath#1{\vbox{\hrule\hbox{\vrule\kern8pt\vbox{\kern8pt
			\hbox{ {$\displaystyle #1 $ } }\kern8pt} 
			\kern8pt\vrule}\hrule}}

\def\today{\number\day\space\ifcase\month\or Janvier \or F\'evrier \or  Mars
   \or Avril \or Mai \or Juin \or Juillet \or Ao\^ut \or Septembre \or Octobre
   \or Novembre \or D\'ecembre \fi\number \year}

\date{March 12, 2012}

\begin{document}

\title{Multi-mode solitons in the classical Dicke-Jaynes-Cummings-Gaudin Model.}
\bigskip
\author{O. Babelon, B. Dou\c{c}ot \\
Laboratoire de Physique Th\'eorique et Hautes Energies\footnote{Tour 13-14, 4\` eme \'etage, Boite 126,
4 Place Jussieu,  75252 Paris Cedex 05.} (LPTHE) \\
Unit\'e Mixte de Recherche UMR 7589 \\
Universit\'e Pierre et Marie Curie-Paris 6 and CNRS; }
\bigskip

\maketitle  

\bigskip

{\bf Abstract:} We present a detailed analysis of the classical Dicke-Jaynes-Cummings-Gaudin integrable model,
which describes a system of $n$ spins coupled to a single harmonic oscillator. We focus on the singularities
of the vector-valued moment map whose components are the $n+1$ mutually commuting conserved Hamiltonians.
The level sets of the moment map corresponding to singular values may be viewed as
degenerate and often singular Arnold-Liouville torii. 
A particularly interesting example of singularity corresponds to unstable equilibrium points where the rank of the moment map is zero,
or singular lines where the rank is one. The corresponding level sets can be described as a reunion of smooth strata of various dimensions. 
Using the Lax representation, the associated spectral curve and the separated variables,
we show how to construct explicitely these level sets.
A main difficulty in this task is to select, among possible complex solutions, the physically
admissible family for which all the spin components are real. 
We obtain explicit solutions to this problem in the rank zero and one cases. 
Remarkably this corresponds exactly to solutions obtained previously by Yuzbashyan and whose geometrical meaning
is therefore revealed. These solutions can be described
as multi-mode solitons which can live on strata of arbitrary large dimension. In these solitons, the energy initially stored
in some excited spins (or atoms) is transferred at finite times to the oscillator mode (photon) and eventually
comes back into the spin subsystem. But their multi-mode character is reflected by a large diversity in their
shape, which is controlled by the choice of the initial condition on the stratum. 

\section{Introduction.}

The Dicke model
has been used for more than fifty years in atomic physics to describe the interaction
of an ensemble of two-level atoms with the quantized electromagnetic field~\cite{Dicke}.
Recently, many experimental and theoretical works have considered systems in an optical cavity 
where matter is coupled to a single eigenmode of the field \cite{RaBrHa,BlHuWaGi}. 
These systems offer some
unprecedented opportunities to realize quantum bits and to process quantum information. The integrable version of the monomode 
Dicke model considered in this paper~\cite{Gau83,YKA} 
is obtained by making the so called rotating wave approximation in which  the non resonant terms 
(e.g. for which a photon  is created  while an atom is excited) are discarded. In the case of a single atom,
this model has been solved by Jaynes-Cummings~\cite{JC}, and has been used to study the effect of field quantization 
on Rabi oscillations \cite{BrSchMaRaHa}.
In the many atom case, besides the development of cavity QED \cite{RaBrHa} and circuit QED \cite{BlHuWaGi} already mentioned, a surge of interest 
in the Dicke-Jaynes-Cummings-Gaudin model has been motivated  in the context of cold atoms systems, 
where a sudden change in the interactions
between atoms is achieved by sweeping the external magnetic field through a Feshbach resonance \cite{Barankov04a,Barankov04b}. 
The subsequent dynamics of the atomic system is well described by the DJCG model prepared in the quantum counterpart of a classically unstable 
equilibrium state.

\begin{figure}[h]
\begin{center}
\includegraphics[height= 9cm]{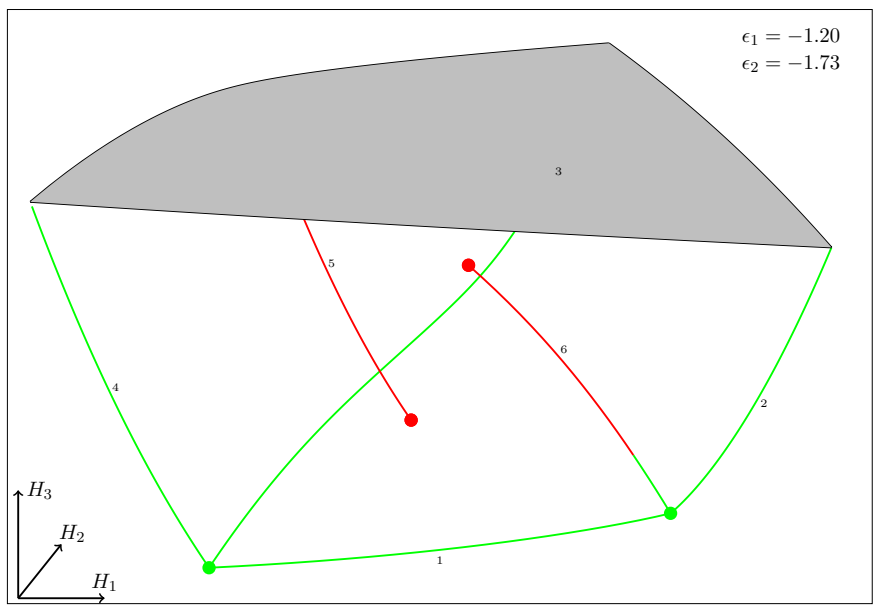}
\label{range5}
\caption{The image of the moment map for the two spins model. The green dots refer to stable (elliptic) equilibrium points. The red dots are
unstable (focus-focus) equilibrium points.  The preimage of a point on a green line is a  Liouville torus degenerated to a circle on which the rank of the moment map is one.  The preimage of a point on a red line is a pinched torus times a circle.
The parameters $(\epsilon_1,\epsilon_2)$ are chosen so that we have two elliptic  and  two focus-focus critical points.}
\end{center}
\end{figure}

This motivates the study of unstable equilibrium points in classical and quantum integrable models. 
From the classical viewpoint, an integrable system with $n$ degrees of freedom is characterized by
a collection of $n$ mutually commuting functions $H_1$,$H_2$,...,$H_n$ over the $2n$-dimensional phase-space.
Usually, the physical Hamiltonian $H$ is expressed as a function of these conserved quantities. Besides the 
classical trajectories generated by $H$, it is of great interest to consider  the
so-called moment map, which associates to any point $p$ in phase space the $n$-vector  $(H_1(p),H_2(p),...,H_n(p))$.
The image of the moment map is of great importance.  Fig.[\ref{range5}] provides an example of such an image of the moment map for the two spins 
Dicke-Jaynes-Cummings-Gaudin model.
  To each point $(H_1,H_2,\cdots, H_n)$ in this image corresponds a level set in phase space, that is to say the set of points $p$ such that
  $(H_1(p)=H_1,H_2(p)=H_2,...,H_n(p)=H_n)$.
For many systems of physical relevance these level sets are compact (as it is the case here). In the case of a regular value of the moment
map, the Arnold-Liouville theorem states that the corresponding level set is an $n$-dimensional torus.
Non regular values  of the moment map, (for which the rank of the moment map may be $< n$),   are in some sense more interesting, 
because they correspond to special torii whose dimension may be lower than $n$ or which may contain singular submanifolds. 
The image of the moment map and its critical values encode all the information on the fibration of phase space in Liouville torii. 
It is important to realize that the level set of a non-regular value of the moment map is in general a stratified manifold on which 
the rank of the moment map may jump. The dimension of each stratum is equal to the rank of the differential of the moment map on that stratum.

An extreme case is given by the level set containing an equilibrium point. By definition, the gradient
of the moment map vanishes at such point (the rank is zero), which is then an equilibrium point for any Hamiltonian
expressed in terms of  $H_1$, $H_2$,...,$H_n$.  For a purely elliptic equilibrium point, the level set is
reduced to this point.  But for an unstable point, the level set contains the equilibrium point, of course, but also a whole 
manifold whose description is far from trivial. In the vicinity of the unstable critical point, the stratification can be qualitatively understood
by the quadratic normal form~\cite{Eliasson90}, deduced from the Taylor expansion of the moment map to second order
in the deviations from the critical point. A detailed derivation of these normal forms 
in the Dicke-Jaynes-Cummings-Gaudin model has been presented in our previous contribution~\cite{BD11}. 
But the information contained in the normal form is purely local, and it is not sufficient   to describe globally the level set 
containing an unstable critical point. 
Global characteristics can be access by viewing  these level sets  as collections of trajectories, solutions of the equations of motion. 
In this paper, we shall present explicit  solutions of the physical Hamiltonian equations of motion for an arbitrary initial condition chosen on such a critical
level set. They can be described as solitonic pulses in which the energy initially stored
in some excited spins (or atoms) is transferred at finite times to the oscillator mode (photon) and eventually
comes back into the spin subsystem. 

We achieve this by using the well known algebro-geometric solution~\cite{DuKrNo90,Sk79,BaBeTa03}
of the classical Dicke-Jaynes-Cummings-Gaudin model. 
A key feature of this approach is that the
knowledge of the conserved quantities $H_1$,$H_2$,...,$H_{n+1}$, (equivalently a point in  the image
of the moment map) is encoded in a Riemann surface, called the spectral
curve. It is  of genus $g=n$, where $n$ is the number of spins. The state of the
system can be represented by a collection of $g$ points on the spectral curve.
Remarkably, the image of this collection of $g$ points by the Abel map
follows a straight line with a constant velocity on the  Jacobian torus associated to the spectral curve.
This construction linearizes the Hamiltonian flows of the model. 

It is well known that critical values of the moment map
correspond to degenerate spectral curves~\cite{Audin96,DuKrNo90}, and this seems to be the
most efficient way to locate the critical values of the moment map. This strategy has
been used extensively in our previous discussion of the moment map in the
classical Dicke-Jaynes-Cummings-Gaudin model~\cite{BD11}.  
We believe that the separated variables, which we will use in their Sklyanin presentation~\cite{Sk79},  are a powerful tool
 to uncover the stratification of a critical level set. 
In this paper, we show that each stratum of dimension $n+1-n_0$ on such level set is obtained by freezing $n_0$
separated variables on some zeros of the polynomial which defines the spectral curve.

We will aslo pay a special attention to the problem of finding the physical 
slice on each stratum. This is not a straightforward question, because the algebro-geometric
method is naturally suited to solve equations of motion in the complexified phase-space.
But only solutions for which all spin components and the two oscillator coordinates 
are real are admissible physically. The determination of the physical slice for a generic
regular value of the moment map remains an open problem. But on the critical level sets of the 
moment map, corresponding to a spectral curve of genus zero,  the problem greatly simplifies.
This corresponds to  level sets of the moment map containing critical points of rank zero or one
for which we are able to derive an explicit solution for an arbitrary number of spins.
Remarkably these solutions were already constructed by Yuzbashyan \cite{Yuzbashyan08}, and they find here their 
place  in a broader picture aiming at organizing all particular solutions of the Dicke-Jaynes-Cummings-Gaudin
model.

This paper is organized as follows. Section~\ref{sec_classical_DJCG_model}
presents the classical Dicke-Jaynes-Cummings-Gaudin model of $n$ spins
coupled to a single harmonic oscilator, together with the main features of
the algebro-geometric solution. In particular, we emphasize that the separated
variables capture in a very natural way 
the stratification of a critical level set, each stratum being characterized by a subset of frozen
variables on some double roots of the spectral polynomial.
These notions are then illustrated in section~\ref{sect_n=1} on the example of the
system with only one spin. In this case, the reality conditions can be worked
out explicitely, both on the critical level sets associated to unstable equilibrium
points and for generic values of the moment map. Section~\ref{sec_n+1_degenerate}
gives the complete determination of the real slice for the level set of an
equilibrium point, and for arbitrary values of $n$. This corresponds to the
case where all the roots of the spectral polynomial are doubly degenerate.
We recover here the {\it normal solitons} first constructed by Yuzbashyan~\cite{Yuzbashyan08}.
Section~\ref{sec_n_degenerate} addresses the same question, in the slightly more common
situation where the spectral polynomial has $n$ doubly degenerate roots and
two simple roots. In this case, we are describing a level set whose smallest
stratum has dimension 1. The corresponding values of the moment map lie on
curves in its $n$-dimensional target space. We benefit in this case from the
fact that the spectral curve remains rational. We get therefore a general
formula for the {\it anomalous solitons} in Yuzbashyan's terminology.
Section~\ref{sec_n=2} illustrates this general theory for the system with two spins,
and section~\ref{sec_n=3} for the system with three spins. We present explicit examples
of solitonic pulses corresponding to solutions living in strata of dimensions
2, 3, and 4. The presence of several directions of unstability away from the
corresponding equilibrium points is reflected by the non-monotonous time dependence
seen in these pulses, which contrasts with the simple shape observed for the
model with a single spin. Our conclusions are stated in section~\ref{sec_Conclusion}.
Finally, an Appendix provides a brief summary of the construction~\cite{BD11}
of quadratic normal forms of the moment map in the vicinity of equilibrium points.

\section{The classical Dicke-Jaynes-Cummings-Gaudin model.}
\label{sec_classical_DJCG_model}

\subsection{Integrability}

This model, describes a collection of $n$ spins  coupled to a single harmonic oscillator.
It derives from the  Hamiltonian:
\begin{equation}
H= \sum_{j=1}^{n} (2 \epsilon_j +\omega)  s_j^z  + \omega \bar{b} b + 
\sum_{j=1}^{n} \left( \bar{b} s_j^-+ b s_j^+ \right)
\label{bfconspinclas}
\end{equation}
The $\vec{s}_j$ are spin variables, and $b,\bar{b}$ is a  harmonic oscillator.
The Poisson brackets read:
\begin{equation}
\{ s_j^a , s_j^b \} = - \epsilon_{abc} s_j^c, \quad \{ b , \bar{b} \} = i
\label{poisson}
\end{equation}
The $\vec{s}_j$ brackets are degenerate. We fix the value of the Casimir functions
$$
\vec{s}_j \cdot \vec{s}_j = s^2
$$
Phase space has dimension $2(n+1)$. In the Hamiltonian we have used
$s_j^\pm = s_j^1 \pm i s_j^2$
which have Poisson brackets $ \{ s_j^z , s_j^\pm \} = \pm i s_j^\pm,\quad \{s_j^+,s_j^-\} = 2 i s_j^z$. 
The equations of motion read:
\begin{eqnarray}
\dot{b} &=&  -i \omega b - i  \sum_{j=1}^{n} s_j^-  \label{motionb} \\
\dot{s_j^z} &=&  i  ( \bar{b} s_j^- - b s_j^+ ) \label{motionsz}\\
\dot{s_j^+} &=& i(2 \epsilon_j+\omega) s_j^+ -2i \bar{b} s_j^z \label{motions+} \\
\dot{s}_j^- &=&  -i(2 \epsilon_j +\omega) s_j^- +2i b s_j^z \label{motions-}
\end{eqnarray}

Integrability is revealed after introducing the Lax matrices:
\begin{eqnarray}
L(\lambda) &=& 2 \lambda \sigma^z + 2 (b \sigma^+ + \bar{b} \sigma^-) + \sum_{j=1}^{n} {\vec{s}_j \cdot \vec{\sigma} \over \lambda-\epsilon_j} 
\label{defLaxL} \\
M(\lambda) &=& -i\lambda \sigma^z  -i{ \omega\over 2}\sigma^z 
-i (b\sigma^+ + \bar{b} \sigma^-)
\label{defLaxM}
\end{eqnarray}
where $\sigma^a$ are the Pauli matrices, $\sigma^\pm = {1\over 2} (\sigma^x \pm i \sigma^y)$.
$$
\sigma^x = \pmatrix{ 0 & 1 \cr 1 & 0}, \quad \sigma^y = \pmatrix{ 0 & -i \cr i & 0},\quad  \sigma^z = \pmatrix{1 & 0 \cr 0 & -1},\quad 
$$
It is not difficult to check that the equations of motion are equivalent to the Lax equation:
\begin{equation}
\dot{L}(\lambda) = [M(\lambda), L(\lambda) ]
\label{eqLax}
\end{equation}
Letting
$$
L(\lambda) = \pmatrix{ A(\lambda) & B(\lambda) \cr C(\lambda) & -A(\lambda) }
$$
we have:
\begin{eqnarray}
A(\lambda) &=& 2\lambda+ \sum_{j=1}^{n} {s_j^z \over \lambda - \epsilon_j } \label{defA}\\
B(\lambda) &=& 2b + \sum_{j=1}^{n} {s_j^- \over \lambda - \epsilon_j } \label{defB} \\
C(\lambda) &=& 2\bar{b} + \sum_{j=1}^{n} {s_j^+ \over \lambda - \epsilon_j }  \label{defC}
\end{eqnarray}
These generating functions have the simple  Poisson brackets:
$$
\{ L_1(\lambda) , L_2(\mu) \} =- i \left[ {P_{12}\over \lambda - \mu } , L_1(\lambda) + L_2(\mu) \right]
$$
where
$$
P_{12} = \pmatrix{ 1 & 0 & 0 & 0 \cr 0 & 0 & 1 & 0 \cr 0 & 1 & 0 & 0 \cr 0 & 0 & 0 & 1}
$$
It follows  immediately that $ \rm{Tr}\,(L^2(\lambda) ) = 2 A^2(\lambda) + 2 B(\lambda) C(\lambda)  $ Poisson commute for different values of the spectral parameter:
$$
\{ \rm{Tr}\,(L^2(\lambda_1) ), \rm{Tr}\,(L^2(\lambda_2) ) \} = 0
$$
Hence 
$$
\Lambda(\lambda)\equiv \frac{1}{2}\rm{Tr}\,(L^2(\lambda) ) = A^2(\lambda) +B(\lambda) C(\lambda)
$$ 
generates Poisson commuting quantities.   One has
\begin{eqnarray}
\Lambda(\lambda)  =  {Q_{2n+2}(\lambda)\over \prod_j (\lambda-\epsilon_j)^2}=  4\lambda^2  + 4  H_{n+1}  + 
 2 \sum_{j=1}^{n}  
{H_j  \over \lambda - \epsilon_j}  +  \sum_{j=1}^{n} {s^2 \over ( \lambda - \epsilon_j)^2 }
\label{detL}
\end{eqnarray}
where $Q_{2n+1}(\lambda)$ is a polynomial of degree $2n+2$.  The $(n+1)$ commuting Hamiltonians $H_j$, $j=1,\cdots , n+1$  read:
\begin{equation}
H_{n+1} =  b\bar{b} +  \sum_{j=1}^n s_j^z 
\label{Hn}
\end{equation}
and
\begin{equation}
H_j =  2\epsilon_j  s_j^z +  ( b s_j^+ + \bar{b} s_j^-)
+ \sum_{k\neq j} {s_j \cdot s_k \over \epsilon_j - \epsilon_k }, \quad j=1,\cdots , n
\label{Hj}
\end{equation}
The physically interesting Hamiltonian eq.(\ref{bfconspinclas}) is:
\begin{equation}
H = \omega H_{n+1} + \sum_{j=1}^{n} H_j
\label{hamphys}
\end{equation}

On the physical phase space  the complex conjugation acts as  $\overline{s_j^z}=s_j^z$ and $\overline{s_j^+} = s_j^-$, and of course $\bar{b}$ is the complex conjugate of $b$. Hence for $\lambda$  real, $A(\lambda)$ is real and $B(\lambda) = \overline{C(\lambda)}$. It follows that on the physical phase space, $Q_{2n+2}(\lambda)$ is {\em positive} when $\lambda$ is {\em real}.

\subsection{Separated variables.}
\label{subsection_separated_variables}

The Lax form eq.~(\ref{eqLax}) of the equation of motion implies that the so-called
spectral curve $\Gamma$, defined by $\det ( L(\lambda) - \mu) = 0$ is a constant of motion.  
Specifically:
\begin{equation}
\Gamma: \;   \mu^2 - A^2(\lambda) - B(\lambda) C(\lambda) = 0, \textrm{ i.e. } \mu^2 =  {Q_{2n+2}(\lambda) \over \prod_j (\lambda - \epsilon_j)^2}
\label{QABC}
\end{equation}
Defining $y = \mu  \prod_j (\lambda - \epsilon_j)$,  the equation of the curve becomes $y^2 = Q_{2n+2}(\lambda)$
which is an hyperelliptic curve. Since the polynomial  $Q_{2n+2}(\lambda)$ has degree $2n+2$,  the genus of the curve in $n$. 
Because the model is integrable, it is possible to construct action-angle coordinates (at least locally), but their connection
to the initial physical dynamical variables is rather complicated. In this work, we prefered to work with
the so-called separated variables  which have the double advantage that their
equations of motion are much simpler than the original ones and that they are not too far from   
the physical spin and oscillator coordinates. The separated variables are $g=n$ points  on the curve whose coordinates $(\lambda_k,\mu_k)$ can be taken as coordinates on phase space. 
They are defined as follows. Let us write
\begin{equation}
C(\lambda) = 2\bar{b}+ \sum_{j=1}^{n} {s_j^+ \over \lambda - \epsilon_j } \equiv 
 2\bar{b} {\prod_{k=1}^n(\lambda - \lambda_k) \over \prod_{j=1}^{n} (\lambda - \epsilon_j) }
 \label{Csepare}
\end{equation}
the separated coordinates are  the collection $(\lambda_k, \mu_k=A(\lambda_k))$. They have canonical Poisson brackets
$$
\{ \lambda_k, \mu_{k'} \} = -i \delta_{k,k'}
$$
Notice however that if $\bar{b}=0$ theses variables are not well defined.

There are only $2n$ such coordinates which turn out to be invariant under the global $U(1)$ rotation
generated by $H_{n+1}$:
$$
b\to e^{i\theta} \; b,\quad \bar{b}\to e^{-i\theta} \; \bar{b}, \quad s_j^-\to e^{i\theta} s_j^-, \quad  s_j^+\to e^{-i\theta} s_j^+
$$

So  they describe  the {\em reduced} model obtained by fixing the value of $H_{n+1}$ and taking 
into consideration only the dynamical variables invariant under this $U(1)$ action. The initial dynamical model can be recovered by adding the phase of the oscillator coordinates $\bar{b},b$ to the separated variables.
The equations of motion with respect to $H_i$ in these new variables are (no summation over $k$):
\begin{equation}
\partial_{t_i} \lambda_k = i \mu_k ({\cal B}^{-1})_{ik}, \quad {\cal B}_{kj} ={1\over \lambda_k-\epsilon_j},\quad 
({\cal B}^{-1})_{jp} = {\prod_{l\neq p} (\epsilon_j -\lambda_l) \prod_i ( \lambda_p-\epsilon_i) \over \prod_{i\neq j} (\epsilon_j-\epsilon_i) \prod_{l\neq p} (\lambda_p-\lambda_l) }
\label{motioni}
\end{equation}
Using the identities: 
$$
\sum_{i=1}^n {\epsilon_i^p \over \prod_{j\neq i} (\epsilon_i-\epsilon_j) } =\delta_{p,n-1},\quad 0\leq p \leq n-1
$$
the flow associated to the physical Hamiltonian eq.(\ref{hamphys}) reads:
\begin{equation}
\partial_{t} \lambda_k = i\mu_k {\prod_j (\lambda_k-\epsilon_j) \over \prod_{l\neq k} (\lambda_k-\lambda_l)},\quad
\mu_k = A(\lambda_k)
\label{flot}
\end{equation}

\bigskip

The equations of motion eq.(\ref{flot}) must be complemented by the equation of motion for $\bar{b},b$ 
which allows to recover the motion of the full model from the motion of the separated variables in the reduced model.
We have:
\begin{equation}
\partial_t \bar{b} = i\omega \bar{b}+ i \sum_{{\rm Res~} \epsilon_j} C(\lambda) =  i\omega \bar{b}- i \sum_{{\rm Res~}  \infty}  C(\lambda)
= i \bar{b} (\omega -2 (\Sigma_1 - \sigma_1(\epsilon)))
\label{globalu1}
\end{equation}
where $\Sigma_1=\sum_i \lambda_{i}$ and $\sigma_1(\epsilon)=\sum_i \epsilon_{i}$. Of course, we also have the complex conjugated equation of motion for $b$.

\bigskip

We show now how to reconstruct  spin coordinates from the separated variables $(\lambda_k, \mu_k)$ and the
oscillator coordinates $(\bar{b},b)$. 
For $C(\lambda)$ we have eq.(\ref{Csepare}). It is a rational fraction of $\lambda$ which has simple poles at $\lambda=\epsilon_j$
whose residue is $s_{j}^{+}$.  
For $A(\lambda)$, we write:
$$
A(\lambda) = {P_{n+1}(\lambda) \over  \prod_{j=1}^n (\lambda - \epsilon_j) }
$$
The polynomial $P_{n+1}(\lambda)- 2\lambda \prod_{j=1}^n (\lambda - \epsilon_j)$ is of degree $n-1$, 
and we know its value at the $n$ points $\lambda_j$ because:
\begin{equation}
P_{n+1}(\lambda_j) = \mu_j \prod_{k=1}^n (\lambda_j - \epsilon_k) 
\label{eqpn}
\end{equation}
Therefore, we can write:
\begin{equation}
P_{n+1}(\lambda) = 2\lambda \prod_{j=1}^n (\lambda - \epsilon_j)
+ \sum_i (\mu_i-2\lambda_i) \prod_{k=1}^n (\lambda_i - \epsilon_k){ \prod_{ l \neq i} (\lambda - \lambda_l) \over 
\prod_{l \neq i} (\lambda_i - \lambda_l)}
\label{pn+1}
\end{equation}
Once $A(\lambda)$ and $C(\lambda)$ are known, we can find the spin themselves by 
taking the residues at $\lambda = \epsilon_j$. We get:
\begin{equation}
s_j^+ = 2\bar{b} {\prod_{i=1}^n(\epsilon_j - \lambda_i) \over \prod_{k\neq j}^n (\epsilon_j - \epsilon_k) }, \quad
s_j^z = {\prod_{l=1}^n (\epsilon_j  - \lambda_l)\over \prod_{k\neq j}^n (\epsilon_j - \epsilon_k)  } 
\sum_i \left[(2\lambda_{i}-\mu_i) {\prod_{k\neq j} (\lambda_i - \epsilon_k) \over 
\prod_{l \neq i} (\lambda_i - \lambda_l)} \right]
\label{s+sz}
\end{equation}
The modulus of $b$ is invariant under the global $U(1)$ action generated by $H_{n+1}$ and is obtained from (recall that in the reduced model $H_{n+1}$ is a parameter)
$$
\bar{b} b = H_{n+1} - \sum_j s_j^z
$$
The phase of $b$ and $ \bar{b}$ is not determined in the reduced model. The same phase appears in the formula for $s_j^+$. 
Finally, the $s_{j}^{-}$ components are obtained from the constraints:
$$
(s^{z}_{i})^{2}+s^{+}_{i}s^{-}_{i}=s^{2}
$$
They are determined up to the phase of $1/\bar{b}$.
\bigskip

Note that if the $\lambda_i$'s are arbitrary complex numbers, 
$s_{j}^{-}$ is not the complex conjugate of $s_{j}^{+}$, as it is the case
for the physical phase-space. This shows that separated variables are natural
coordinates on the {\em complexified} phase-space. In these variables, the 
problem of identifying the sets $\{\lambda_i, \mu_i\}$ corresponding to physical configurations
(i.e. those for which all spin components are real) is rather non-trivial, and most of
the present work is dedicated to it.

\bigskip

In many situations, and in particular in this article, we are interested in the system with prescribed  real values of the
conserved quantities, $H_1$,...,$H_{n+1}$,  i.e. we take as coordinates the $(\lambda_j, H_j)$ instead of the $(\lambda_j, \mu_j)$.  This amounts to fixing the Liouville torus we work with, or equivalently  the spectral
polynomial $Q_{2n+2}(\lambda)$. In this setting, the $\mu_j$ are determined  by the equation
$$
\prod_{k=1}^n (\lambda_j-\epsilon_k) \mu_j = \pm \sqrt{ Q_{2n+2}(\lambda_j) }
$$
The choice of the signs  in this formula is very important and will be a recurrent theme in our subsequent discussions.
They affect the formulae for $s_j^z$ or equivalently the polynomial $P_{n+1}(\lambda)$. The values of $s_{j}^{-}$'s can then be obtained from
 $B(\lambda)$ which is determined by:
\begin{equation}
B(\lambda)= {Q_{2n+2}(\lambda) - P_{n+1}^2(\lambda) \over C(\lambda) \prod_k (\lambda - \epsilon_k)^2 }
={1\over 2\bar{b}} { Q_{2n+2}(\lambda) - P^2_{n+1}(\lambda) \over 
\prod_i (\lambda -\lambda_i) \prod_k (\lambda - \epsilon_k) }
\label{conj}
\end{equation}
The polynomial in the numerator is of degree $2n$, and moreover it is divisible by 
$\prod_i (\lambda -\lambda_i)$ because $P_{n+1}^2(\lambda_i) = Q_{2n+2}(\lambda_i)$, so we can write
\begin{equation}
B(\lambda) = 2 b {\prod_i (\lambda -\bar{\lambda}_i)  \over 
\prod_k (\lambda - \epsilon_k) }
\end{equation}
The spin components $s_j^-$ are obtained by taking the residue at $\lambda=\epsilon_j$ :
$$
s_j^- =  2 b {\prod_i (\epsilon_j -\bar{\lambda}_i)  \over 
\prod_{k\neq j} (\epsilon_j - \epsilon_k) }
$$

In this consruction, the variables $\bar{\lambda}_i$ and the corresponding $\bar{\mu}_i$ are complicated functions of the $( \lambda_i, H_i)$ and of the choice of signs used in the determination of the variables $\mu_i$.
The set of physical configurations (often refered to here as the {\em real slice}) 
is obtained by writing that the set $(\bar{\lambda_i}, \bar{\mu_i})$ is the {\em complex conjugate} of the set $(\lambda_i, \mu_i)$. This leads to a set of complicated relations whose solution is not known in general.   It is the purpose of this work to show how to implement them in particular cases.

\subsection{Moment map and degenerate spectral curves}
\label{subsec_deg_spectral_curves}

In this section, we discuss the case of a spectral polynomial having double roots, 
so that $Q_{2n+2}(\lambda)$ can be written as:
\begin{equation}
Q_{2n+2}(\lambda) =  4\; p_{2m+2}(\lambda) \prod_{i=1}^{n-m} (\lambda - E_i)^2, \quad 
p_{2m+2}(\lambda) = \lambda^{2m+2} + b_{2m+1} \lambda^{2m+1} +\cdots
\label{Qpm}
\end{equation}
The $E_i$ are either real or come in complex conjugated pairs. Moreover  $p_{2m+2}(\lambda)$ is positive for real $\lambda$.
As discussed in some previous works~\cite{YKA,BD11,Audin96}, this corresponds to critical values
of the moment map. The preimage of the moment map for such value is a stratified manifold,
whose various strata have dimensions ranging from $m+1$ to $n+1$. Separated variables provide 
a very natural access to this stratification, because each stratum of dimension $n-n_0+1$
can be obtained by freezing $n_0 \leq n-m$ separated variables $\lambda_j^{0}$
on $n_0$ double zeroes of  $Q_{2n+2}(\lambda)$. On such stratum, the rank of the moment map also
drops to $n-n_0+1$. 
As explained before~\cite{YKA}, the dynamics of the system for initial conditions lying in
this stratum is very similar to the one of an effective model with $n-n_0$ spins, but we shall
not use this physically appealing result here. 

To produce a polynomial $Q_{2n+2}(\lambda)$ of the form eq.(\ref{Qpm}) is not completely straightforward because it must 
also be compatible with eq.(\ref{detL}). Let us explain how it works.

The fact that the rational fraction $Q_{2n+2}(\lambda)/\prod_j(\lambda-\epsilon_j)^{2}$
has double poles at $\lambda-\epsilon_j$ whose weight is $s^{2}$ imposes 
$n$ conditions:
$$
\prod_{i} (\lambda - E_i) \vert_{\epsilon_j} = { s \alpha_j \over 2 \sqrt{p_{2m+2}(\epsilon_j)}} \prod_{k\neq j}(\epsilon_j - \epsilon_k)
$$
where $\alpha_j =  \pm 1$.  The  values of the polynomial $\prod_{i} (\lambda - E_i)$
at the $n$ points $\epsilon_j$ are therefore  known, and we can write by Lagrange interpolation formula:
\begin{equation}
\prod_{i} (\lambda - E_i) =   \sum_j  { s \alpha_j \over 2 \sqrt{p_{2m+2}(\epsilon_j)}}  \prod_{i\neq j} (\lambda - \epsilon_i)
\label{cond0gene}
\end{equation}
But the degree of the left hand side is  $n-m$ while  the degree on the right hand side is superficially $n-1$, 
so  we have the consistency conditions:
\begin{equation}
\sum_j  {\alpha_j  \epsilon_j^k\over 2 \sqrt{p_{2m+2}(\epsilon_j)}} = 0, \quad k= 0, \cdots ,m-2
\label{cond1gene}
\end{equation}
and:
\begin{equation}
\sum_j  {s\alpha_j  \epsilon_j^{m-1}\over 2 \sqrt{p_{2m+2}(\epsilon_j)}} = 1
\label{cond2gene}
\end{equation}
These equations are obtained by writing that:
$$
\oint {dz \over 2\pi i}{z^{k}\prod_{i=1}^{n-m} (z-E_i) \over  \prod_{i=1}^{n} (z-\epsilon_i)}= \delta_{k,m-1}, \quad k= 0, \cdots ,m-1 
$$
where the integrals are taken along a contour in the complex plane which encircles all the $\epsilon_j$'s.
These are exactly $m$ conditions on the $2m+3$ coefficients of $p_{2m+2}(\lambda)$, of which the leading
coefficient is known. 
One more condition is obtained by writing that in the right-hand side of eq.~(\ref{detL})
the coefficient of $\lambda$ vanishes. This gives:
$$
b_{2m+1}+2(\sigma_{1}(\epsilon)-\sigma_{1}(E)) = 0
$$
But we also have:
$$
\oint {dz \over 2\pi i}{z^{m}\prod_{i=1}^{n-m} (z-E_i) \over  \prod_{i=1}^{n} (z-\epsilon_i)}= \sigma_{1}(\epsilon)-\sigma_{1}(E)= 
\sum_j  {s\alpha_j  \epsilon_j^{m}\over 2 \sqrt{p_{2m+2}(\epsilon_j)}} 
$$
where the integral is again on a contour encircling all the $\epsilon_j$'s.
Assembling the previous two equations gives the consistency condition: 
\begin{equation}
b_{2m+1} + \sum_j  {s\alpha_j  \epsilon_j^{m}\over  \sqrt{p_{2m+2}(\epsilon_j)}}=0
\label{cond3gene}
\end{equation}
Altogether, we are left with $m+1$ free coefficients in $p_{2m+2}(\lambda)$
which play the role of conserved Hamiltonians of an effective system with $m$ spins \cite{YKA}. 
All the other symmetric functions  $\sigma_k(E)$, $k > 1$ are then determined by eq.(\ref{cond0gene}).

\bigskip

If $m=-1$, that is to say $p_{2m+2}(\lambda)=1$, eq.(\ref{cond0gene}) must be slightly modified as
\begin{equation}
\prod_{i} (\lambda - E_i) =  (\lambda + \beta) \prod_j (\lambda -\epsilon_j) + \sum_j  {s\alpha_j \over 2}  \prod_{i\neq j} (\lambda - \epsilon_i)
\label{cond0parti}
\end{equation}
This means
\begin{equation}
{Q_{2n+2}(\lambda) \over \prod_j (\lambda-\epsilon_j)^2} = \left( 2 \lambda +2 \beta + \sum_j {\alpha_j s \over \lambda-\epsilon_j} \right)^2
\label{cond0parti1}
\end{equation}
Comparing with eq.(\ref{detL}) at $\lambda = \infty$ we see that $\beta=0$.

\bigskip

Let us discuss now the freezing of the $\lambda_k$ at the double roots of $Q_{2n+2}(\lambda)$. The equations of motion eq.(\ref{flot})
become
\begin{equation}
\partial_{t} \lambda_k = \pm i{\prod_j (\lambda_k-E_j) \sqrt{p_{2m+2}(\lambda_k) }\over \prod_{l\neq k} (\lambda_k-\lambda_l)},
\label{flot2}
\end{equation}
so if $\lambda_k = E_i$ at $t=0$, it stays there forever, and  it is consistent with the equations of motion to freeze some 
$\lambda_k$ at the double roots of $Q_{2n+2}(\lambda)$. This however cannot be done in an arbitrary way as we  now explain.

\bigskip

Let us assume that $Q_{2n+2}(\lambda)$ has a {\em real } root at $\lambda=E$.
This means that $A^2(\lambda) + B(\lambda)C(\lambda)$ vanishes when $\lambda=E$.
But for real $\lambda$, $A(\lambda)$ is real and one has $C(\lambda)= \overline{B(\lambda)}$ 
so that $A(\lambda)$, $B(\lambda)$, $C(\lambda)$ must all vanish at $\lambda=E$.
In particular, recalling eq.(\ref{Csepare}), this means that one of the separated variables, 
say $\lambda_1$, is frozen at the value $E$, {\em provided} $\bar{b} b \neq 0$.
This implies also that $\lambda-E$ divides simultaneously $A(\lambda)$, $B(\lambda)$ and
$C(\lambda)$, and therefore $(\lambda-E)^{2}$ divides $Q_{2n+2}(\lambda)$. A single real
root of $Q_{2n+2}(\lambda)$ is necessarily a double root. Writing:
$A(\lambda)=(\lambda-E)\widetilde{A}(\lambda)$,
$B(\lambda)=(\lambda-E)\widetilde{B}(\lambda)$,
$C(\lambda)=(\lambda-E)\widetilde{C}(\lambda)$,
$Q_{2n+2}(\lambda)=(\lambda-E)^{2}\widetilde{Q}_{2n}(\lambda)$,
we see that:
$$
\widetilde{Q}_{2n}(\lambda)=\prod_{j}(\lambda-\epsilon_j)^{2}
\left(\widetilde{A}(\lambda)^{2}+\widetilde{B}(\lambda)\widetilde{C}(\lambda)\right)
$$
so we can apply the same discussion to the real roots of $\widetilde{Q}_{2n}(\lambda)$.
We deduce from this that the multiplicity $p$ of the real root $E$ of $Q_{2n+2}(\lambda)$
is even and that $(\lambda-E)^{p/2}$ divides simultaneously $A(\lambda)$, $B(\lambda)$ and
$C(\lambda)$. In this case, $p/2$ separated variables $\lambda_1$,...,$\lambda_{p/2}$
are frozen at the real value $E$.

Nothing as simple holds in the case of a simple {\em complex} root. 
So, let us assume that there is  a {\it double} complex root $E$. Of course the complex conjugate $\bar{E}$ is also a
double root. What can be said in general  
is that the values of $\lambda_1$,...,$\lambda_n$ freeze by complex conjugated pairs. 

In fact we have $\Lambda (\lambda) = A^2(\lambda) + B(\lambda) C(\lambda) $ hence
$$
\Lambda' (\lambda) = 2 A(\lambda) A'(\lambda) + B'(\lambda)C(\lambda) + B(\lambda) C'(\lambda)
$$
On a degenerate spectral curve, $\Lambda (\lambda) $ has at least one double zero $E$. Therefore
\begin{eqnarray*}
0&=&A^2(E) + B(E) C(E) \\
0&=& 2 A(E) A'(E) + B'(E)C(E) + B(E) C'(E)
\end{eqnarray*}
By contrast to the real case, we cannot infer from these equations that $C(E)=0$.
This is related to the fact that for critical spectral curves associated to
unstable critical points, the corresponding level set of the conserved Hamiltonians
has dimension equal to the number of unstable modes. So we expect that some 
separated variables $\lambda_j$ can remain unfrozen. But if one of them freezes
at $E$, that is if $C(E)=0$, the first equation implies $A(E)=0$ and the second equation implies $B(E) C'(E)=0$. Therefore if the zero
 of $C(\lambda)$ at $\lambda=E$ is 
simple, then necessarily $B(E)=0$. 
But $B(E)$ is the complex conjugate of  $C(\bar{E})$ which must therefore vanish {\em provided} $\bar{b} b \neq 0$.
From this we conclude that another separated variable $\lambda_k$ freezes at $\bar{E}$. 

In contrast to the real case, freezing is not compulsory, but it is the possibility to freeze the $\lambda_i$ by complex conjugated pairs 
that leads to the description of the real slice and the stratification of the level set.

\section{The one-spin model.}
\label{sect_n=1}

The model with a single spin coupled to the oscillator is interesting, because it illustrates
most of the points discussed so far. In this case, there are only two conserved quantities,
$H_{1}$ and $H_{2}$, which read:
\begin{eqnarray}
H_1&=& 2\epsilon_1  s_1^z + b s_1^+ + \bar{b}s_1^-   \label{H11spin} \\
H_2 &=& \bar{b} b + s_1^z \label{H21spin}
\end{eqnarray} 
The rank of the momentum map can be either 0, 1, or 2. The later case corresponds to generic
Arnold-Liouville tori of dimension 2. Let us now recall briefly the discussion of critical  
values of the moment map given in a previous work~\cite{BD11}.

\subsection{Rank zero}

\begin{figure}[h!]
\begin{center}
\includegraphics[height= 8cm]{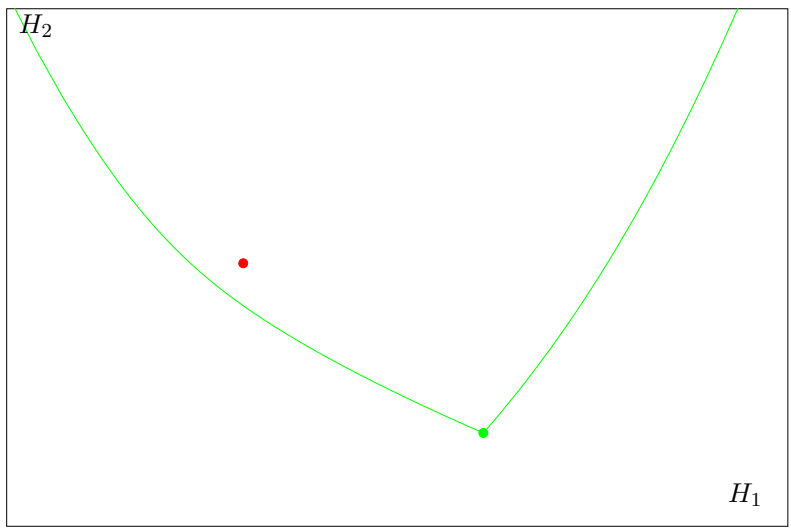}
\caption{The image of the moment map in the case of one spin, $(\epsilon_1=-0.707)$, with one unstable critical point. 
The green point is the stable (elliptic) point $P_2(\downarrow)$. The red point is the unstable (focus-focus) point $P_1(\uparrow)$.  
It is in the interior of the image of the moment map. The two boundaries of this domain correspond to spectral polynomials
whith one double real zero, so that the moment map has rank 1 on the preimage of these curves. 
}
\label{instablepolytope}
\end{center}
\nonumber
\end{figure}

The rank of the moment map vanishes on the critical points 
given by  $b=\bar{b} = 0, s^{\pm}_1 = 0$. Hence we have two points: 
$$
s_1^z = e_1 s, \quad e_1=\pm 1
$$
The corresponding values $P=(H_1,H_2)$ are:
$$
P_1(\uparrow)=\left[ 2\epsilon_1 s, s \right] ,\quad 
P_2(\downarrow)=\left[ -2\epsilon_1 s, -s \right] 
$$
To determine the type of the singularities, we look at the classical Bethe equations (see section~\ref{sec_Normal_Forms})  
which read: 
$$
2E + {se_1\over E-\epsilon_1}=0 \Leftrightarrow 2E^2 -2\epsilon_1 E + se_1 =0
$$
The discriminant of this equation is  $\epsilon_1^2 - 2 s e_1 $. When the spin is down (point $P_2(\downarrow)$), the discriminant is positive,
the two classical Bethe roots are real and this is an elliptic singularity, in agreement with the general analysis of section
\ref{sec_Normal_Forms}. When the spin is up (point $P_1(\uparrow)$) we have real roots when $\epsilon_1^2 \geq 2s$ 
(i.e. the singularity is elliptic in this case), and  a pair of complex conjugate roots $E,\bar{E}$ 
when $\epsilon_1^2 \leq 2s$ (i.e. the singularity is focus-focus in that case). The image of the moment map
is shown on Fig.[\ref{instablepolytope}] in the unstable case. We see that the stable critical point $P_1(\uparrow)$
is located at the tip of the  image of the moment map, whereas the unstable one $P_2(\downarrow)$ lies in the interior of this domain.

In the one spin case, eq.(\ref{detL}) reads :
\begin{equation}
{Q_4(\lambda)\over (\lambda-\epsilon_1)^2 } = 4\lambda^2+ 4 H_2 + 2{H_1\over \lambda-\epsilon_1}
+ {s^2 \over (\lambda-\epsilon_1)^2}
\label{detL1}
\end{equation}
For the values $H_1=2\epsilon_{1}e_{1}s , H_2=e_{1}s $ it is completely degenerate and
the spectral polynomial has two double roots $E_1$, $E_2$ 
identical to the classical Bethe roots, and reads:
$$
Q_4(\lambda)= 4(\lambda^2 -\epsilon_{1}\lambda + \frac{s}{2}e_{1})^2=4(\lambda-E_1)^2(\lambda-E_2)^{2}
$$
where
$$
E_1+E_2=\epsilon_1, \quad E_1 E_2 = {s e_1\over 2}
$$

Let us now describe in more detail the real slices corresponding to the two critical
values of $(H_1,H_2)$. 
Recall that
\begin{equation}
s_1^z = (\epsilon_1-\lambda_1)( 2\lambda_1-\mu_1)
\label{sz1spin}
\end{equation}
We express first the separated variable $\mu_1$ as a function of $\lambda_1$:
\begin{equation}
\mu_1 = \pm 2 {(\lambda_1-E_1)(\lambda_1-E_2) \over (\lambda_1-\epsilon_1)}
\label{mu1_vs_lambda1}
\end{equation}

We have  to distinguish the stable and unstable cases. In the stable case $E_1$ and $E_2$ are real and we know that 
$\lambda_1$ has to be frozen at one of them, say $E_1$. Then $\mu_1=0$, there is no problem of sign in eq.(\ref{mu1_vs_lambda1}) and
$$
s_1^z = 2 E_1 (\epsilon_1-E_1) = 2 E_1 E_2 = s e_1
$$
as it should be.  The level set in this case is composed of only one stratum consisting of the critical point itself where the rank of the moment map is zero.

Let us now assume that we are in the unstable case, that is at the point $P_2(\uparrow)$, $e_1=+1$ and $2s-\epsilon_1^2 \geq 0$. In this case $E_1= E, E_2=\bar{E}$ are complex and $\lambda_1$ remains unfrozen.
Let us choose the $+$ sign in eq.(\ref{mu1_vs_lambda1}). Inserting into eq.(\ref{sz1spin}) we obtain
$$
s_1^z = s e_1
$$
so that, in the spin variables, we are precisely at the critical point. On this stratum of the level set the rank of the moment map is zero. 
It is very interesting to remark that the variable $\lambda_1$ completely disappears in this case (hence the rank zero). 
In terms of the separated variable,
this stratum of the level set 
seems therefore to consists of the whole $\lambda_1$-plane. This is due to the fact that at the critical point $\bar{b} b =0$ and  the separated variables are not well defined and appear as singular coordinates. However, as we will show below, the $\lambda_1$ coordinate has the magic property of realizing a blowup of the singularity.

Choosing now the $-$ sign in eq.~(\ref{mu1_vs_lambda1}) yields: 
$$
s_1^z = -se_1 -4 \lambda_1(\lambda_1-\epsilon_1)
$$
Writing that $s_1^z$ is real, $s_1^z = \bar{s_1^z}$,  gives
\begin{equation}
\lambda_1+ \bar{\lambda}_1=\epsilon_1
\label{realline}
\end{equation}
Setting $\lambda_1 = x + i y, \bar{\lambda}_1 =  x - i y$, we see that the real slice is given by the vertical line located at
$x=\epsilon_1/2$. It is not the whole  line however. We find from $\bar{b} b = H_2-s_1^z $ 
$$
\bar{b}b = 2s -\epsilon_1^2 -4 y^2 \geq 0
$$
The real slice 
of the reduced model corresponds in this case 
to the segment parallel to the imaginary axis delimited by the two double roots $E$
and $\bar{E} $ of the spectral polynomial. This manifold has real dimension one.

As we have discussed before~\cite{BD11,BaDoCa09}, this corresponds to a real slice of the  model reduced by the global $U(1)$ action eq.(\ref{globalu1}).
From the viewpoint of the original model, we have to reintroduce this global angle and the real slice becomes a 
two dimensional torus pinched at the unstable critical point. 

Hence in this simple case we see that the level set of the unstable critical point is 
composed of two strata : the critical point itself where the rank of the moment map is zero and the pinched torus where the rank is two.

The dynamics of the model on this torus is the composition of a large motion in which the oscillator amplitude $\bar{b}b$ goes to zero at time
$\pm \infty$ but reaches a finite maximum at a finite time, and a global $U(1)$ rotation. Only the former
movement is captured by the reduced model, and it maps into the finite segment of the $\lambda_1$ variable
which we have just described. To see this, let us consider the 
equation of motion for the flow generated by $H_1$:
\begin{equation}
\partial_{t_1} \lambda_1 = 2i (\lambda_1 -E)(\lambda_1-\bar{E})
\label{flot1spin}
\end{equation}
whose solution is: 
\begin{equation}
\lambda_1 = {E-\bar{E}X \over 1 - X}, \quad X= X_0 e^{2i(E-\bar{E})t_1} 
\label{solution_flot1spin}
\end{equation}
The reality condition eq.(\ref{realline}) becomes:
$$
(E-\bar{E})(X-\bar{X})=0
$$
which imposes $\bar{X_0} = X_0$ so that $X_0$ is real. Its absolute value can be absorbed  
in the origin of time $t_1$. Only its sign matters. The constraint $s_1^z \leq s$ is equivalent to $X_0<0$, and we recover the fact 
that $\lambda_1$ runs along the line interval joining $E$ and $\bar{E}$. 

\subsection{Rank one}

Physical configurations at which the rank of the moment map is equal to one
correspond to a spectral polynomial with one double zero, which is necessarily
real. We therefore write:
\begin{equation}
Q_4(\lambda) = (\lambda+a_0)^2 (b_2 \lambda^2 + b_1 \lambda + b_0)
\label{Q4rang1}
\end{equation}
where $a_0$ and $b_i$ are real. We denote:
$$
a_0={1\over 2} \;x - \epsilon_1 
$$
Taking into account the usual constraints on $Q_4(\lambda)$ saying that it depends only on two free parameters $H_1$
and $H_2$, see eq.(~\ref{detL1}),  
we can determine $b_0$, $b_1$, and $b_2$ in terms of $x$ by solving linear equations. We find:
$$
b_2= 4, \quad
b_1= -4 x, \quad
b_0= 4\;{  \epsilon_1 x^3 - \epsilon_1^2 x^2 +  s^2 \over  x^2 }
$$
\begin{eqnarray*}
H_1 &=& -{x^4 -2\epsilon_{1} x^3 -4 s^2 \over 2 x} \\
H_2 &=& -{3 x^4 -8\epsilon_{1} x^3 + 4 \epsilon_{1}^2 x^2 -4 s^2 \over 4 x^2}
\end{eqnarray*}
These are parametric equations for a curve which coincides with the boundaries of the image of the moment
map in the $(H_1,H_2)$ plane, for which an illustration can be found on Fig.[\ref{instablepolytope}].

The determination of the real slice is straightforward in this case. Indeed,
because we have a real double root, the separated variable is frozen at the
value $\lambda_1=-a_0, \mu_1=0$. This single point of the reduced model corresponds to the
one-dimensional orbits under the global rotations generated by $H_2$ characterized by  a common phase on $\bar{b}$ and $s_1^+$ :
\begin{eqnarray*}
s_1^z & = & -{1\over 2} x(x-2\epsilon_1) \\
\bar{b}b & = & {(2s -x^2 +2\epsilon_1 x )(2s +x^2 -2\epsilon_1 x )\over 4 x^2} \\
s_1^{+} & = & x \bar{b} 
\end{eqnarray*}
Hence, we have shown that the preimage of a point on the boundary consists of only one stratum which is a circle $S^1$.
Notice that when 
\begin{equation}
x^2 -2\epsilon_1 x + 2 e s = 0, \quad e=\pm 1
\label{xbound}
\end{equation}
we have $s_1^z = es$ and $\bar{b} b=0$ and this corresponds to the points $P_1(\downarrow)$ and $P_2(\uparrow)$.  

In fact there is a simple relation between rank 0 and rank 1.  The spectral curves eq.(\ref{Q4rang1}) degenerate when the polynomial $b_2 \lambda^2+ b_1 \lambda + b_0$ has a double real root. Its discriminant is:
$$
b_1^2-4 b_0 b_2 = 16\; { (x^2 -2\epsilon_1 x -2s)(x^2 -2\epsilon_1 x +2s)\over x^2}=-64\; \bar{b} b \leq 0
$$
It vanishes precisely when eqs.(\ref{xbound}) are satisfied. 
These equations are nothing but the classical Bethe equations eq.(\ref{classicalBethe}) as can be seen by setting
$x=2 E$. Hence in the stable case, the discriminant vanishes for real values of $x$, and the critical points are on the boundary lines of rank one. In the unstable case 
however the discriminant vanishes for complex values of $x$ and the critical point cannot lie on the boundary.

\subsection{Reality conditions on a generic torus.}

In this simple case of a single spin it turns out that one can work out explicitly the exact reality conditions,
for an arbitrary choice of the conserved quantities $H_1$ and $H_2$. For this, express everything in terms of $\lambda_1$ and $\bar{\lambda}_1$.
First, we have
$$
s_{1}^+=2\bar{b}(\epsilon_{1}-\lambda_1),\quad s_{1}^-=2b(\epsilon_{1}-\bar{\lambda}_1),
$$
In the reduced model, $H_2$ is a constant and we have
$$
(H_2-\bar{b} b)^2 = (s_1^z)^2 = s^2-s_1^+s_1^-
$$
which yields
$$
(\bar{b} b)^2 + (4 X-2 H_2) \bar{b} b + H_2 -s^2 =0
$$
where $X = (\lambda_1-\epsilon_1)(\bar{\lambda}_1-\epsilon_1)$.  Finally eqs.(\ref{H11spin}) and (\ref{H21spin}) give:
\begin{equation}
H_1 = 2 \epsilon_1 H_2 - 2 \bar{b} b ( \lambda_1+\bar{\lambda}_1-\epsilon),\quad 
\label{H1bbarlambda1}
\end{equation}
Eliminating  $\bar{b} b$ between these two equations, we get :
\begin{eqnarray}
R&:& (H_1-2\epsilon_1 H_2)^2 + 4(H_1-2\epsilon_1 H_2)(\lambda_1+\bar{\lambda}_1-\epsilon_1) (H_2-2\lambda_1 \bar{\lambda}_1 +2\epsilon_1 (\lambda_1 +\bar{\lambda}_1 -\epsilon_1)) \nonumber\\
&&+ 4 (H_2^2-s^2)(\lambda_1+ \bar{\lambda}_1 -\epsilon_1)^2 =0
\label{R2}
\end{eqnarray}
To obtain the equation of the real slice we ask that $\lambda_1$ and $\bar{\lambda}_1$  are mutually complex conjugates. Setting $\lambda_1=x+i y,\bar{\lambda}_1 = x-i y$, the expression $R$  is quadratic in $y$ and cubic in $x$.  It can be uniformized by 
Weierstrass functions.  
Setting: 
$$
x={-6 \epsilon_1 u + 10 \epsilon_1 H_2 - 3 H_1 -2\epsilon_1^2 \over 4(-3 u + 2 H_2 -\epsilon_1^2)} , \quad
y={-3 v \over 4(-3 u + 2 H_2 -\epsilon_1^2)} 
$$
the reality condition for $\lambda_{1}$ takes the standard form for a planar cubic:
$$
v^2=4 u^3 - g_2 u-g_3
$$
with
$$
g_2={4\over 3}( H_2^2 + 2\epsilon_1^2 H_2 -3 \epsilon_1 H_1 + \epsilon_1^4 + 3 s^2)
$$
$$
g_3={1\over 27}(8 H_2^3 + 24 \epsilon_1^2 H_2^2 - 36 \epsilon_1 H_1 H_2
-72 s^2 H_2 + 24 \epsilon_1^4 H_2 + 27 H_1^2 -36 \epsilon_1^3 H_1
+ 36 \epsilon_1^2 s^2 + 8 \epsilon_1^6 )
$$

The range of physically acceptable values of $x$ is determined by imposing:
$$
\bar{b} b = -{H_1-2\epsilon_1 H_2 \over 2(\lambda_1+\bar{\lambda}_1- \epsilon_1)} \geq 0, \textrm{ or }
s_1^z = H_2-\bar{b}b = H_2 +{H_1-2\epsilon_1 H_2 \over 2(\lambda_1+\bar{\lambda}_1- \epsilon_1)} \in \{-s,+s\}
$$
where we have used eq.~(\ref{H1bbarlambda1}).

\begin{figure}[ht]
\begin{center}
\includegraphics[height= 8cm]{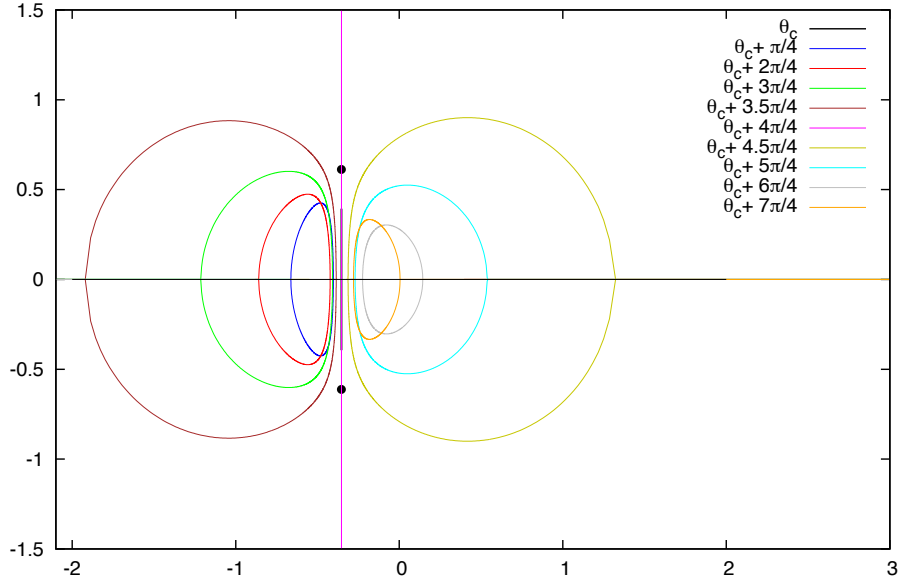}
\caption{The real slices when we run around the small circle around the singularity.  $\epsilon_{1}=-0.707$, $r=0.3$.}
\label{realslices_instable3}
\end{center}
\nonumber
\end{figure}

To help visualize these real slices in the $\lambda_{1}$ complex plane, we consider
a small circle around the image of the unstable point in the $(H_1,H_2)$ plane: 
\begin{equation}
H_1 = 2 s \epsilon_1 + r\sin \theta,\quad H_2=s + r \cos \theta
\label{parametrized_moment}
\end{equation}
and we obtain a family of real slices shown in Fig.[\ref{realslices_instable3}]. 
When $H_1=2\epsilon_1 H_2$, $H_2 \neq \pm s$, the reality condition eq.(\ref{R2})  simplifies to
\begin{equation}
\lambda_1+ \bar{\lambda}_1 - \epsilon_1 = 0
\label{straightline}
\end{equation}
In fact, on the line $H_1=2\epsilon_1 H_2$, the polynomial $Q_4(\lambda)$ depends on the variable $\lambda(\lambda-\epsilon_{1})$ only
which is invariant under the involution $\lambda\to \epsilon_1-\lambda$. 
The slices thus become the vertical real line $x=\epsilon_1/2$ when  $\theta=\theta_c$ where $\tan \theta_c=2\epsilon_1$.

The two critical points lie on the line $H_1=2\epsilon_1 H_2$. When we are close to them,
$$
H_1=2\epsilon_1 s e_1 + \delta H_1, \quad H_2= s e_1 + \alpha \; \delta H_1
$$
Keeping the first order term in eq.(\ref{R2}) we see that the real slice is the product of the straight line: 
$$
\lambda_1+ \bar{\lambda}_1 -\epsilon_1 =0
$$
by the circle:
\begin{equation}
(1-2\epsilon_1 \alpha)[2 (\lambda_1-\epsilon_1)(\bar{\lambda}_1 - \epsilon_1) -s e_1] - 2s e_1\; \alpha \; (\lambda_1+ \bar{\lambda}_1-\epsilon )  =0
\label{pencil1}
\end{equation}
We can rewrite the above equation as:
$$
(\lambda_{1} - \lambda_\alpha)(\bar{\lambda}_{1}-\lambda_\alpha) = (E_1-\lambda_\alpha)(E_2-\lambda_\alpha), \quad
 \lambda_\alpha = \epsilon_1 + { \alpha \; s e_1 \over 1-2\epsilon_1  \alpha } 
$$ 
where $E_i$ are the solutions of the classical Bethe equation:
$$
2 E(E-\epsilon_1) + s e_1 =0
$$

\begin{figure}[ht]
\begin{center}
\includegraphics[height=8cm,width=16cm]{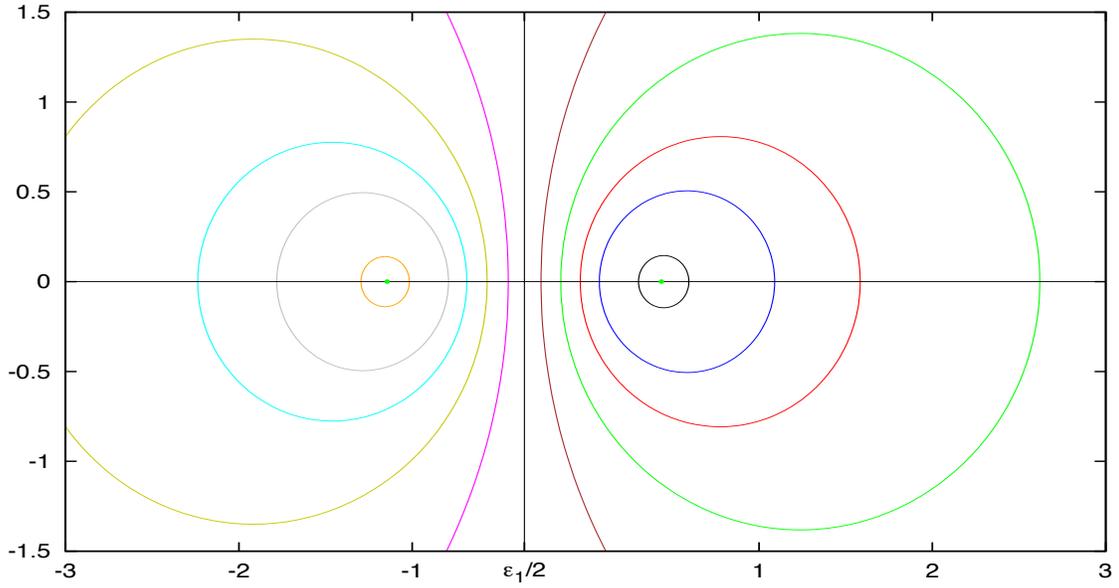}
\caption{The pencil of circles in the stable case  $\epsilon_1= -0.707$}
\label{realslices_stable}
\end{center}
\nonumber
\end{figure}

In the stable case $e_1=-1$, the $E_i$ are real and when $\alpha$ varies, 
we have a pencil of circles with limit points located at $\lambda_\alpha=E_1$ 
and $\lambda_\alpha=E_2$, see Fig.[\ref{realslices_stable}]. 

\begin{figure}[ht]
\begin{center}
\includegraphics[height= 8cm,width=16cm]{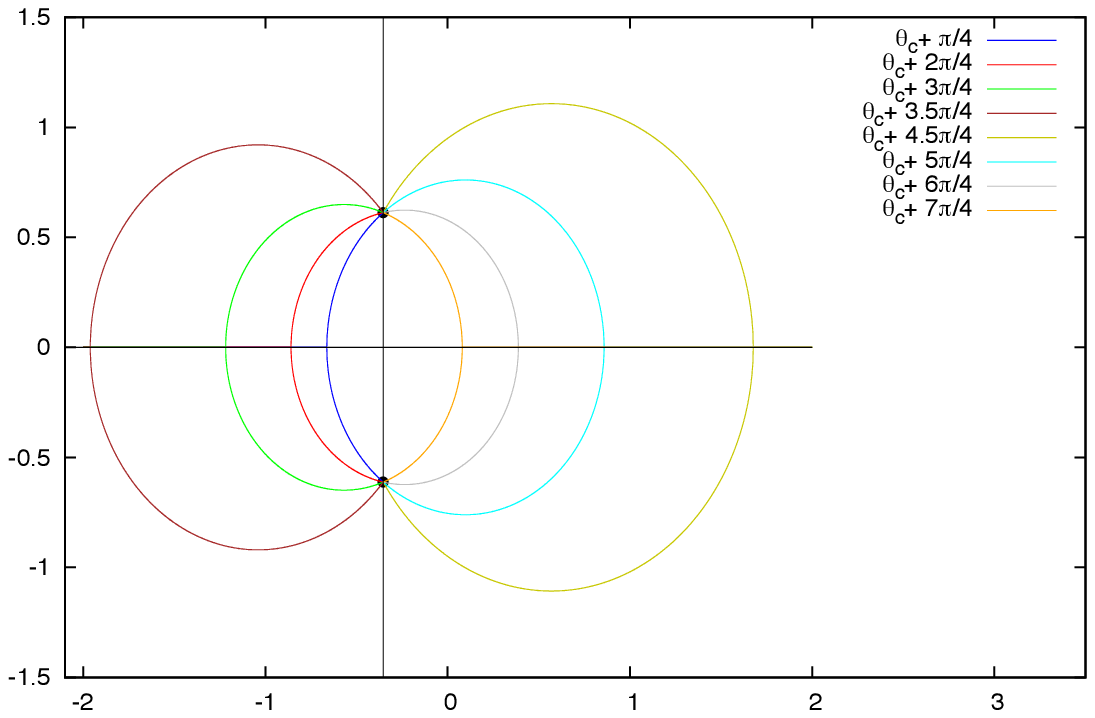}
\caption{The pencil of circles in the stable case ($\epsilon_1=-0.707.)$}
\label{realslices_instable2}
\end{center}
\nonumber
\end{figure}

In the unstable case obtained when $e_1=+1$ and $\epsilon_{1}^{2}<2s$, 
the $E_i$ are complex conjugate to each other and we have a pencil of circles with base points 
located at  $\lambda_\alpha=E$ and $\lambda_\alpha=\bar{E}$, see Fig.[\ref{realslices_instable2}].

\begin{figure}[ht]
\begin{center}
\includegraphics[height= 8cm,width=10cm]{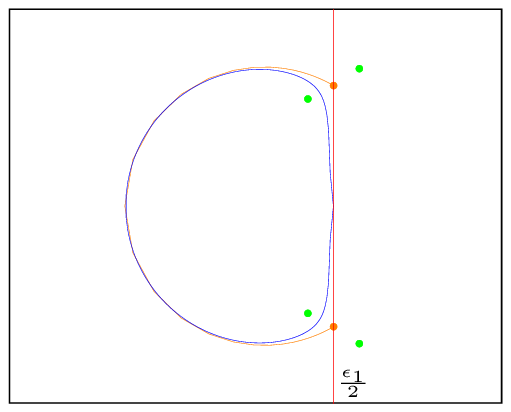}
\caption{ The exact real slice (in blue) compared to the corresponding circle (in orange) of the pencil in the unstable case. The green dots are the branch points of the spectral curve. The vertical line is the line $\lambda_1 + \bar{\lambda}_1=\epsilon_1$. The orange points are the classical Bethe roots.
($\epsilon_1=-0.707$, $H_1= 2 \epsilon_1 H_2 + 0.1$, $H_2=1$.)}
\label{realslices_instable}
\end{center}
\nonumber
\end{figure}

This expansion of eq.~(\ref{R2}) gives a very reasonable
approximation of the real slice for $(H_1,H_2)$ close to the image of 
the unstable critical point $P_1(\uparrow)$ under the moment map. In Fig.[\ref{realslices_instable}],
we see that for such values, the exact real slice is composed of two parts. One part
is close to the vertical segment joining the two double roots $E$ and $\bar{E}$ of
the spectral polynomial associated to $P_1(\uparrow)$. The second part is close to 
a circle belonging to the pencil just described. Here, we see a rather striking consequence
of the fact that separated variables are singular in the vicinity of the critical points.
Whereas the pinched torus containing the unstable critical point $P_1(\uparrow)$ appears as a {\em segment} of the vertical line eq.(\ref{straightline})
in the $\lambda_{1}$ plane, any generic invariant torus arbitrary close to it is projected into a {\em closed 
curve}, whose shape depends crucially on the {\em direction} in the $(H_1,H_2)$ plane along which we are
approaching the critical value. The additional arc, well approximated by an element of the pencil of circles,
corresponds to physical configurations which remain close to the unstable point $P_1(\uparrow)$. To establish this,
it is useful to use the quadratic normal form of the moment map in the vicinity of $P_1(\uparrow)$~\cite{BD11}, whose construction
is recalled in section~\ref{sec_Normal_Forms}.    
 
\subsection{Normal coordinates.}
\label{subsec_normal_coordinates_n=1}

To study the vicinity of a critical point, we can use the  normal coordinates defined in  eq.(\ref{CNormal}). We have
$$
C(\lambda) ={2\over \lambda-\epsilon_1} \left[ \left( {C_{1}\over a'_{1}} + {C_{2}\over a'_{2}}\right) \lambda
- \left({C_{1}\over a'_{1}}E_{2}+{C_{2}\over a'_{2}}E_{1} \right) \right]
$$
The zero of $C(\lambda)$ is the separated variable $\lambda_1$. We have
\begin{equation}
\lambda_1 = {  E_2 a'_2 C_1 + E_1 a'_1 C_2 \over a'_2 C_1 +a'_1 C_2 }
 \label{normaltoseparate}
\end{equation}
It is important to remark that since $\lambda_1$ depends only on the ratio $C_1/C_2$ a small neighborhood of the critical point is mapped 
to the whole $\lambda_1$ plane. 

\bigskip

We can parametrize the normal coordinates in terms of action-angle coordinates making explicit the reality conditions.

In the stable case, we set:
$$
C_j = A_j e^{i \theta_j}, \quad B_j =  A_j e^{-i \theta_j}
$$
with real parameters $A_j$, $\theta_j$. The real slice in the $\lambda_1$ variables is therefore given  by:
$$
\lambda_1(\theta) =  {   E_2 a'_2 A_1 + E_1 a'_1 A_2 e^{i\theta} \over
 a'_2 A_1 +a'_1 A_2 e^{i\theta }}, \quad \theta = \theta_2-\theta_1
 $$
Taking the complex conjugate, we can eliminate $\theta$ between $\lambda_1$ and $\bar{\lambda}_1$. We find:
$$
(a'_{1})^{2} A_{2}^{2} (\lambda_1-E_1)(\bar{\lambda}_1-E_1) - (a'_{2})^{2} A_{1}^{2} (\lambda_1-E_2)(\bar{\lambda}_1-E_2) =0
$$
and we recover our circle pencil with limit points $E_1$ and $E_2$ corresponding to $A_1=0$ and $A_2=0$ respectively.

In the unstable case, we set:
$$
B_1=A_1 \rho \; e^{i\theta}, \quad C_1= A_1 \rho^{-1} e^{-i\theta},\quad 
B_2= \bar{A}_1 \rho^{-1} e^{i\theta},\quad C_2 =  \bar{A}_1 \rho \; e^{-i\theta}
$$
where $\rho$ is a real positive number. Then:
$$
\lambda_1(\rho) = { E_2 a'_2 A_1 + E_1 a'_1 \bar{A}_1 \rho^2 \over 
  a'_2 A_1 +  a'_1 \bar{A}_1 \rho^2}
$$
Taking  the complex conjugate, remembering that $\bar{E}_1 = E_2$, we can eliminate $\rho^2$ and get:
$$
(a'_1)^2 \bar{A}_1^2 (\lambda_1-E_1)(\bar{\lambda}_1-E_1) - (a'_2)^2 A_1^2 (\lambda_1-E_2)(\bar{\lambda}_1-E_2) =0
$$
This is a circle pencil with base points $E_1$ and $E_2$.  

The conserved quantities are given by:
$$
H_2=s + {A_1^2\over 2 a'_1} + {\bar{A}_1^2\over 2 a'_2}, \quad
H_1= 2s\epsilon_1 +{s A_1^2\over 2 (\epsilon_1-E_1) a'_1} + {s\bar{A}_1^2\over 2 (\epsilon_1-E_2)a'_2}
$$
so that using the parametrization eq.(\ref{parametrized_moment}), the equation of the pencil  becomes:
$$
\sin \theta \; [ 2(\lambda_1-\epsilon_1)(\bar{\lambda}_1-\epsilon_1) -s] -2\cos \theta  \;
[2\epsilon_1(\lambda_1-\epsilon_1)(\bar{\lambda}_1-\epsilon_1) + s(\lambda_1+\bar{\lambda}_1 - 2\epsilon_1)] =0
$$
It coincides with the pencil eq.(\ref{pencil1}) as can be seen by setting $\alpha=\cot  \theta$. It is shown in Fig.[\ref{realslices_instable2}].
Again, we stress that a specific circle of the pencil is specified by the direction in the $(H_1,H_2)$ plane through which we reach the 
critical point. This is a typical blowup of a singularity.

As we  see on  Fig.[\ref{realslices_instable}], if we consider a generic torus close 
to the pinched torus containing the unstable point $P_1(\uparrow)$, the corresponding   
exact real slice in the $\lambda_1$ plane
comprises a large circle and a segment very close to the line $\lambda_1+\bar{\lambda}_1-\epsilon_1=0$. 
This segment is absent from what we obtained from the quadratic normal form, which give us only the arc
of circle belonging to the pencil.

To understand this, we first emphasize that replacing the moment map by its quadratic normal form 
has a dramatic effect on the real slice corresponding to the pinched torus: instead of the segment
joining $E$ and $\bar{E}$ in the $\lambda_1$ plane, only the two end-points $E$ and $\bar{E}$ 
survive after making this quadratic approximation. Indeed,  
in the immediate vicinity of the critical point, the pinched torus has the shape of two cones that meet precisely at the critical point. 
One of these cones is associated to the unstable small perturbations, namely those which are exponentially amplified, 
whereas the other cone corresponds to perturbations which are exponentially attenuated. 
Taking into account the reality conditions, we can parametrize as follows:
$$
C_1= A_1,\quad B_1= \bar{A}_2, \quad C_2= A_2,\quad B_2= \bar{A}_1
$$
The action variables are:
$$
B_1 C_1 = A_1 \bar{A}_2, \quad B_2 C_2 = \bar{A}_1 A_2
$$
When we are sitting right on the pinched torus, these two action variables should vanish. We have two ways to do it corresponding to the 
stable and unstable cones:
$$
A_1=0,\quad A_2\neq 0,  \textrm{  or  } A_1\neq 0,\quad A_2=0
$$
Inserting into eq.(\ref{normaltoseparate}) we see that they correspond to $\lambda_1=E_1$ and $\lambda_1=E_2$ respectively. 
Hence the variable $\lambda_{1}$ remains trapped at these points in this approximation.

Now, consider a solution of the equations of motion, and
suppose we pick an initial condition close to the critical point $P_{1}(\uparrow)$, but not exactly on the
pinched torus. Let us further assume that this initial condition is much closer to the stable cone than
to the unstable one. The point representing the system in configuration space will first move towards $P_1(\uparrow)$
along the stable cone, but after a finite time, while remaining close to $P_{1}(\uparrow)$, it will be ejected away
from it along the unstable cone. This part of the movement takes place  in the vicinity of $P_{1}(\uparrow)$ and is well captured
by the quadratic normal form. It corresponds to the arc of circle that connects the complex root $E_{s}$ associated
to the stable cone to its complex conjugate $E_{u}$ associated to the unstable cone. This part of the trajectory has 
no counterpart when we are exactly on the pinched torus. 

After this, the point representing the system goes far away from
$P_{1}(\uparrow)$ along the unstable cone, and the quadratic normal form is no longer valid. But of course, the two cones are connected
and any initial condition located on the unstable cone gives rise to a trajectory which reaches eventually 
the stable cone in a finite time. This is manifested by the part of the trajectory in the  $\lambda_{1}$ plane which
remains close to the finite segment of the line eq.(\ref{straightline}) that connects  $E_{u}$ to $E_{s}$. This part is fully non-perturbative, in the sense
that it cannot be captured by the quadratic normal form, but it is very well approximated by the large motion along the
pinched torus. 

In a sense the separated variable $\lambda_1$ and the normal coordinate $C_i$ play complementary role and are each 
well adapted to describe a different part of the trajectory.
The need to glue these two qualitatively different parts on a trajectory close to a prinched torus
is an important feature which played a crucial role in our semi-classical treatment of the model~\cite{BaDoCa09}.

\section{Critical torii associated to equilibrium points for general $n$.}
\label{sec_n+1_degenerate}

In this section, we study the level set of a critical point where the rank of the moment map is zero. In Fig.[\ref{range5}] these are the pre-images of the 
green (stable) and red (unstable) points. Hence, we  consider  the critical points of the moment map (rank zero). 
$$
b=\bar{b}=0,\quad s_j^z = s e_j,\quad s_j^\pm = 0, \quad e_j=\pm 1
$$
We know that they correspond to  maximally degenerate spectral curves  where:
\begin{equation}
Q_{2n+2}(\lambda) =4 \prod_{l=1}^{n+1} (\lambda - E_l)^2
\label{Qdegenerate}
\end{equation}
From eq.(\ref{cond0parti1}) we also know that the zeroes of $Q_{2n+2}(\lambda)$ are  the roots of the classical Bethe equation:
\begin{equation}
2E + \sum_{j=1}^{n} {s e_j\over E -\epsilon_j} =0
\label{Ei0}
\end{equation}
Notice that we have  the identities:
$$
2 { \prod_k (\epsilon_j- E_k) \over \prod_{k\neq j} (\epsilon_j-\epsilon_k)} = s e_j, \quad 
\sum_l E_l  = \sigma_1(\epsilon)
$$
The variables $\mu_i$ are given by
$$
\mu_i = \pm 2 {\prod_l (\lambda_i- E_l) \over \prod_{j} (\lambda_i-\epsilon_j)}
$$
As we have seen in the one spin case, the choice of sign here plays a  crucial role in the description of the various strata of the level set.

We will study the strata of the level set by constructing  the solutions of the equations of motion with generic initial conditions on the level set.
The equations of motion eq.(\ref{flot}) become
$$
\dot{\lambda}_i = \pm 2i{ \prod_l (\lambda-E_l) \over \prod_{j\neq i} (\lambda_i-\lambda_j)}
$$
We wish to solve these equations  with the proper choice of signs and initial conditions so that the spins variables are real.

\bigskip

We first  reconstruct the spin variables. The $s_j^+$ are given by:
\begin{equation}
s_j^+(t) =   {2\bar{b}(t)  \over \prod_{k\neq j} (\epsilon_j - \epsilon_k) }  {\cal P}(\epsilon_j,t) 
\label{s-exact}
\end{equation}
where we have introduced the polynomial:
$$
{\cal P}(\lambda,t) = \prod_j (\lambda - \lambda_j(t))
$$
To reconstruct $s_j^z$,  we have to build the polynomial $P_{n+1}(\lambda)$. From eqs.(\ref{eqpn},\ref{Qdegenerate}), we have:
$$
P_{n+1}(\lambda_i) = \pm \sqrt{ Q_{2n+2}(\lambda_i)} =  \pm 2 \prod_l (\lambda_i - E_l), \quad P_{n+1}(\lambda) = 2(\lambda^{n+1} - \sigma_1(\epsilon) \lambda^n + \cdots )
$$
The choice of signs here is crucial. It is the same signs which appear in the equations of motion.

\bigskip

If we take the $+$ sign for all $i$, then obviously $P_{n+1}(\lambda) = 2 \prod_l (\lambda - E_l)$, 
and taking the residue at $\lambda=\epsilon_j$ in $P_{n+1}(\lambda)/\prod_i (\lambda-\epsilon_i)$, we find $s_j^z=se_j$. This is the static solution corresponding to the critical point.

\bigskip

To go beyond this trivial solution, we divide the $\lambda_i$ into three sets $\lambda_i^{+}\in {\cal E}^{+}$, $\lambda_i^{0}\in {\cal E}^{0}$ and $\lambda_i^{-}\in {\cal E}^{-}$ depending on the sign in this formula. In particular ${\cal E}^{0}$ is the set of $\lambda_i$ frozen at some $E_l$.  The equations of motion become:
\begin{equation}
\partial_t \lambda_i^{-} = - 2 i {\prod_l' (\lambda_i^{-}- E_l) \over \prod_{j}' (\lambda_i^{-}-\lambda_j)} , \quad
\partial_t \lambda_i^{+} = 2 i {\prod_l' (\lambda_i^{+}- E_l) \over \prod_{j}' (\lambda_i^{+}-\lambda_j)}
\end{equation}
where the prime means that elements in the set ${\cal E}^{0}$ are excluded because they cancel between numerator and denominator.
Therefore, for each $E_l \notin {\cal E}^{0}$ we can write:
$$
\sum_{\lambda_i^{-} } {\dot{\lambda}_i^{-}  \over \lambda_i^{-}  - E_l } -\sum_{\lambda_i^{+} } {\dot{\lambda}_i^{+}  \over \lambda_i^{+}  - E_l } 
=- 2i  \int_{C_\infty} {dz\over 2i\pi} {\prod_{k\neq l}' (z - E_k) \over \prod_{j }' (z-\lambda_j)} = - 2i (\Sigma_1' - \sigma_1'(E) + E_l)
$$
where $C_\infty$ is a big circle at infinity surrounding all the $\lambda_j$ and we have defined $\Sigma_1' = \sum_j' \lambda_j$, $\sigma_1'(E)=\sum_k' E_k$. Hence:
$$
\log {{\cal P}_{-}(E_l) \over {\cal P}_{+}(E_l) } = - 2i E_l t  - 2i \int^t dt (\Sigma_1' - \sigma_1'(E) ) 
$$
where we have defined the polynomials: 
$$
{\cal P}_{\pm}(\lambda) =   \prod_{\lambda_i^{(\pm)}\in {\cal E}^{(\pm)}}  (\lambda -\lambda_i^{(\pm)}),\quad {\cal P}_{0}(\lambda) =   \prod_{\lambda_i^{0}\in {\cal E}^{0}} (\lambda -\lambda_i^{0})
$$

Remembering that
$$
\dot{\bar{b}} = -2i \bar{b} (\Sigma_1' - \sigma_1'(E)-\omega/2)
$$
we arrive at:
\begin{equation}
{\cal P}_{-}(E_l) =  \bar{b}(t) X_l {\cal P}_{+}(E_l) , \quad X_l= X_l(0) e^{-i(2 E_l + \omega) t }, \quad E_l \notin {\cal E}^{0}
\label{cond1}
\end{equation}

Denoting $n_\pm,n_0$ the number of elements in ${\cal E}^{(\pm)}, {\cal E}^{0}$ respectively,
we get  a set of $n-n_0+1$ {\em linear} equations for the $n_+ + n_- + 1=n-n_0+1$ unknown coefficients of  
the polynomials ${\cal P}_{-}(\lambda)$ and $\bar{b}(t){\cal P}_{+}(\lambda)$. 
As discussed in subsection~\ref{subsec_deg_spectral_curves}, we expect that freezing $n_0$ separated variables $\lambda_j$ 
yields a stratum of dimension $n-n_0+1$ on the pre-image of the moment map corresponding to the degenerate spectral curve.

Simple examination of the quadratic normal form of section~\ref{sec_Normal_Forms}
shows that the complex numbers $2 E_l + \omega$ are the eigenfrequencies for the linearized equations of motion
in the vicinity of the critical point. Eq.~(\ref{cond1}) can therefore be  interpreted as a non linear superposition of normal modes 
yielding a  global motion along the critical torus.
 
The next step is to build  the polynomial $P_{n+1}(\lambda)$, using the fact that $P_{n+1}(\lambda_i) = \pm 2 \prod_l (\lambda_i- E_l)$: 
$$
P_{n+1}(\lambda) = 2 \prod_l  (\lambda -E_l) - 4 \sum_{\lambda_i^{-}} \prod_{E_l\notin {\cal E}^{0}} (\lambda_i^{-} -E_l) {\prod_{\lambda_l^{-} \neq \lambda_i^{-} } (\lambda- \lambda_l^{-}) \over \prod_{\lambda_l^{-} \neq \lambda_i^{-} }  ({\lambda_i^{-} - \lambda_l^{-} } )}
{{\cal P}_+(\lambda){\cal P}_0(\lambda) \over {\cal P}_+(\lambda_i^{-}) }
$$
To derive this formula, we used the fact that the polynomials $P_{n+1}(\lambda)$
and $2 \prod_l  (\lambda -E_l)$ have the same terms of degrees $n+1$ and $n$; the last
statement comes from the relation $\sum_{j}\epsilon_{j}=\sum_{l}E_{l}$ which is a direct consequence
of the classical Bethe equation. 
Similarly, we can also write:
\begin{eqnarray*}
P_{n+1}(\lambda) &=& -2 \prod_l (\lambda -E_l) + 4 \sum_{\lambda_i^{+}} \prod_{E_l\notin {\cal E}^{0}} (\lambda_i^{+} -E_l) {\prod_{\lambda_l^{+} \neq \lambda_i^{+} } (\lambda- \lambda_l^{+}) \over \prod_{\lambda_l^{+} \neq \lambda_i^{+} }  ({\lambda_i^{+} - \lambda_l^{+} } )}
{{\cal P}_-(\lambda){\cal P}_0(\lambda) \over {\cal P}_-(\lambda_i^{+}) } \\
&&+4(\lambda + \Sigma_1' -\sigma_1'(E) ) {\cal P}_+(\lambda){\cal P}_-(\lambda){\cal P}_0(\lambda)
\end{eqnarray*}
Note that the last term in the right hand side is necessary to adjust the terms of degree $n+1$
and $n$ in $\lambda$ between the two sides of the equation. Again, we used the relation
$\sum_{j}\epsilon_{j}=\sum_{l}E_{l}$.
These two expressions for  $P_{n+1}(\lambda)$ motivate the following definitions of polynomials ${\cal S}_\pm(\lambda)$:
\begin{eqnarray*}
{\cal S}_+(\lambda) &=& \sum_{\lambda_i^{-}} \prod_{E_l\notin {\cal E}^{0}} (\lambda_i^{-} -E_l) {\prod_{\lambda_l^{-} \neq \lambda_i^{-} } (\lambda- \lambda_l^{-}) \over {\cal P}'_-(\lambda_i^{-}){\cal P}_+(\lambda_i^{-})  } \\
{\cal S}_-(\lambda) &=& \sum_{\lambda_i^{+}} \prod_{E_l\notin {\cal E}^{0}} (\lambda_i^{+} -E_l) {\prod_{\lambda_l^{+} \neq \lambda_i^{+} } (\lambda- \lambda_l^{+}) \over {\cal P}'_+(\lambda_i^{+})  {\cal P}_-(\lambda_i^{+})}
+(\lambda + \Sigma_1' -\sigma_1'(E) ) {\cal P}_+(\lambda)
\end{eqnarray*}
so that we can write:
\begin{eqnarray}
P_{n+1}(\lambda) &=& 2 \prod_l (\lambda -E_l) - 4 {\cal S}_+(\lambda) {\cal P}_+(\lambda) {\cal P}_0(\lambda) \label{Pn+1+} \\
P_{n+1}(\lambda) &=&- 2 \prod_l (\lambda -E_l) + 4 {\cal S}_-(\lambda) {\cal P}_-(\lambda)  {\cal P}_0(\lambda)\label{Pn+1-} 
\end{eqnarray}
Note that ${\cal S}_+(\lambda)$ has degree $n_{-}-1$ and  ${\cal S}_{-}(\lambda)$ has degree $n_{+}+1$.
Now, we have:
$$
4 \prod_l (\lambda -E_l)^2 - P_{n+1}^2(\lambda) =
16\; {\cal S}_-(\lambda) {\cal S}_+(\lambda)  {\cal P}_0(\lambda) {\cal P}(\lambda) = 4 \bar{b} b  {\cal P}(\lambda)  \bar{\cal P}(\lambda)
$$
where ${\cal P}(\lambda)=\prod_{j}(\lambda-\lambda_{j}) = {\cal P}_-(\lambda){\cal P}_0(\lambda){\cal P}_+(\lambda)$ and 
$\bar{\cal P}(\lambda)=\prod_{j}(\lambda-\bar{\lambda}_{j})$ is the complex conjugate of ${\cal P}(\lambda)$, that is
$\bar{\cal P}(\lambda) = \overline{ {\cal P}(\bar{\lambda})} $.  Therefore:
\begin{equation}
\bar{b} b \;  \bar{\cal P}(\lambda) =4\; {\cal S}_-(\lambda) {\cal S}_+(\lambda)  {\cal P}_0(\lambda)
\label{PbarSS1}
\end{equation}
So the zeroes $\bar{\lambda}_i$ of $\bar{\cal P}(\lambda)$ split into the zeroes 
$\bar{\lambda}_i^{+}$, $\bar{\lambda}_i^{-}$ and $\bar{\lambda}_i^{0}$ of 
$ {\cal S}_+(\lambda)$, $ {\cal S}_-(\lambda)$ and $ {\cal P}_0(\lambda)$ respectively. 
A direct consequence of these definitions and of eqs.~(\ref{Pn+1+}) and (\ref{Pn+1-}) is that:
\begin{eqnarray}
P_{n+1}(\bar{\lambda}_i^{+}) & = & + 2 \prod_l (\bar{\lambda}_i^{+} -E_l) \label{signPn+1bar+} \\
P_{n+1}(\bar{\lambda}_i^{-}) & = & - 2 \prod_l (\bar{\lambda}_i^{-} -E_l) \label{signPn+1bar-}
\end{eqnarray}

By the definition of the $\bar{\lambda}_i^{0}$'s, we see that 
the set ${\cal E}^{0}$ is self conjugate (see section \ref{subsec_deg_spectral_curves}) and that ${\cal P}_0(\lambda)=\bar {\cal P}_0(\lambda)$.
The above definition of ${\cal S}_{-}(\lambda)$ shows that the coefficient of its term of highest
degree is equal to one (this is {\em not} the case for ${\cal S}_{+}(\lambda)$). Because of this:
\begin{equation}
{\cal S}_-(\lambda) = \bar{\cal P}_-(\lambda)
\label{Sm}
\end{equation}
Combining this with eq.~(\ref{PbarSS1}) we get also:
\begin{equation}
{\cal S}_+(\lambda) = {1 \over 4} \bar{b} b\; \bar{\cal P}_+(\lambda)
\label{Sp}
\end{equation}
Comparing the terms of highest degrees in ${\cal S}_+(\lambda)$ and $\bar{\cal P}_+(\lambda)$ gives: 
\begin{equation}
\bar{b} b = 4  \sum_{\lambda_i^{-}} { \prod_{E_l\notin {\cal E}^{0}} (\lambda_i^{-} -E_l) \over { {\cal P}'_-(\lambda_i^{-}){\cal P}_+(\lambda_i^{-})  }}
\label{bbar}
\end{equation}
It is expressed only in terms of $\lambda_i^{-}$'s. So, if $n_-=0$, we recover the fact already mentioned that $\bar{b} b =0$
and the system remains at the critical point.

At this stage, we are ready to enforce the reality condition. As discussed in subsection~\ref{subsection_separated_variables},
the real slice is obtained by imposing that the set $\{\bar{\lambda}_{i}\}$
be the same as the set $\{\lambda_{i}^{*}\}$, where in the rest of this section we denote by $z^{*}$ the complex conjugate
of $z$. Equivalently: 
$$
\bar{\cal P}(\lambda^{*})={\cal P}(\lambda)^{*}
$$
for any $\lambda$. 
From the discussion in subsection~ \ref{subsec_deg_spectral_curves}, we know that the frozen variables
$\lambda_{i}^{0}$ appear in complex conjugate pairs so that 
${\cal P}_{0}(\lambda^{*})={\cal P}_{0}(\lambda)^{*}$.
We also know that  ${\cal P}_0(\lambda)=\bar {\cal P}_0(\lambda)$ so that 
$\bar{\cal P}_{0}(\lambda^{*})={\cal P}_{0}(\lambda)^{*}$.
The above reality condition becomes then:
$$
\bar{\cal P}_{-}(\lambda^{*})\bar{\cal P}_{+}(\lambda^{*})
={\cal P}_{-}(\lambda)^{*}{\cal P}_{+}(\lambda)^{*}
$$
It is clearly sufficient to impose simultaneously:
\begin{equation}
\label{necessary_sufficient}
\bar{\cal P}_{-}(\lambda^{*}) = {\cal P}_{-}(\lambda)^{*} ,\quad \bar{\cal P}_{+}(\lambda^{*}) = {\cal P}_{+}(\lambda)^{*}
\end{equation}
But we now claim that this condition is also necessary.
This comes from the fact already noted that the sign of $P_{n+1}(\lambda)/\prod_{l}(\lambda-E_{l})$ is positive
for $\lambda=\lambda_i^{+}$ or $\lambda=\bar{\lambda}_{i}^{+}$ and negative for $\lambda=\lambda_i^{-}$ or $\lambda=\bar{\lambda}_{i}^{-}$.
So the roots $\bar{\lambda}_i^{+}$ have to be complex conjugates of $\lambda_i^{+}$ and likewise,
the roots $\bar{\lambda}_i^{-}$ have to be complex conjugates of $\lambda_i^{-}$. An interesting and useful
consequence of this is that
we must have ${\rm deg } \;\;{\cal P}_\pm = {\rm deg }\;\; {\cal S}_\pm = {\rm deg }\;\; \bar{\cal P}_\pm$, which
requires $n_+=n_- -1$. Since $n_++n_0+n_-=n$ we find: 
$$
n_+={1\over 2}(n-1-n_0),  \quad n_-= {1\over 2} (n+1-n_0)
$$

Because the coefficients of highest degrees
of ${\cal P}_{-}(\lambda)$  and  ${\cal P}_{+}(\lambda)$ are set equal to one,
the above constraints~(\ref{necessary_sufficient}) are equivalent  to (assuming of course that $\bar{\cal P}_{+}(\lambda)$ and $\bar{\cal P}_{-}(\lambda)$ are mutually prime
and similarly for ${\cal P}_{+}(\lambda)$ and ${\cal P}_{-}(\lambda)$). 
$$
{\bar{\cal P}_{-}(\lambda^{*}) \over \bar{\cal P}_{+}(\lambda^{*})} =
\left({{\cal P}_{-}(\lambda) \over {\cal P}_{+}(\lambda)}\right)^{*} 
$$
We also note that we should add the constraint $\bar{b}=b^{*}$.
This plus the fact that these two polynomials involve a total of $n_{+}+n_{-}$ unknown coefficients,
shows that it is necessary and sufficient to enforce
the following conditions for the $n+1-n_{0}=n_{+}+n_{-}+1$ roots $E_{l}$ which don't belong to ${\cal E}^{0}$: 
\begin{equation}
{\bar{\cal P}_{-}(E_{l}^{*}) \over b\bar{\cal P}_{+}(E_{l}^{*})} =
\left({{\cal P}_{-}(E_{l}) \over \bar{b}{\cal P}_{+}(E_{l})}\right)^{*} 
\label{criterion_cal_P}
\end{equation}

As we have seen, the general solution of the Hamiltonian evolution on the critical torus, eq.~(\ref{cond1}) implies:
\begin{equation}
{{\cal P}_{-}(E_{l}) \over \bar{b}{\cal P}_{+}(E_{l})} = X_l 
\label{RHS_criterion}
\end{equation}
To evaluate the left-hand side of the conditions (\ref{criterion_cal_P}), we set $\lambda=E_{l}$
in eqs.(\ref{Pn+1+}, \ref{Pn+1-}), which gives:
$$
P_{n+1}(E_l) = -4 {\cal S}_+(E_l) {\cal P}_+(E_l) {\cal P}_0(E_l),\quad P_{n+1}(E_l) = 4 {\cal S}_-(E_l) {\cal P}_-(E_l) {\cal P}_0(E_l),
\quad E_l \notin {\cal E}^{0}
$$
and therefore:
\begin{equation}
 {\cal S}_+(E_l) {\cal P}_+(E_l) = - {\cal S}_-(E_l) {\cal P}_-(E_l)
 \label{cond2}
 \end{equation}
Since ${\cal P}_+(E_l) \neq 0$, and 
remembering eq.(\ref{cond1}) this implies: 
$$
{{\cal S}_{-}(E_l) \over {\cal S}_{+}(E_l)} = - {1 \over \bar{b} X_l}  
$$
But using eqs.~(\ref{Sm}) and (\ref{Sp}), we get:
\begin{equation}
{\bar{\cal P}_{-}(E_{l}^{*}) \over b\bar{\cal P}_{+}(E_{l}^{*})}=-{1\over 4 X_{\bar{l}}}
\label{LHS_criterion}
\end{equation}
where we define the index $\bar{l}$ to be such that $E_{\bar{l}}=E_{l}^{*}$.
Given eqs.~(\ref{RHS_criterion}) and (\ref{LHS_criterion}), the reality conditions (\ref{criterion_cal_P})
and the condition $\bar{b}=b^{*}$ are satisfied if and only if:
\begin{equation}
X_l^{*} \; X_{\bar{l}} = -{1\over 4}
\label{explicit_reality_conditions}
\end{equation}

Let us check that the above conditions also imply the fact that $s_j^{z}$ is real for any $j$.
For this, we have to satisfy the relation $P_{n+1}(\lambda^{*})=P_{n+1}(\lambda)^{*}$.
But, as we have just seen,  conditions~(\ref{explicit_reality_conditions}) imply that 
$\bar{\lambda}_{i}^{(\pm)}=(\lambda_{i}^{(\pm)})^{*}$. From this, combined with  
eqs.~(\ref{signPn+1bar+}), (\ref{signPn+1bar-}) and the fact that the polynomial 
$\prod_i(\lambda-E_i)$ has real coefficients we see that: 
$$
P_{n+1}((\lambda_{i}^{(\pm)})^{*})=P_{n+1}(\lambda_{i}^{(\pm)})^{*}
$$
But since:
$$
P_{n+1}((\lambda_{i}^{0})^{*})=P_{n+1}(\lambda_{i}^{0})^{*}=0
$$
we get:
$$
P_{n+1}(\lambda_{i}^{*})=P_{n+1}(\lambda_{i})^{*}
$$
for all the $n$ separated variables. And because $P_{n+1}(\lambda)=2\lambda^{n+1}-2\sigma_{1}(\epsilon)\lambda^{n}+...$,
these $n$ relations are sufficient to impose the fact that $P_{n+1}$ has real coefficients.

Together with the solution~(\ref{cond1}) of the Hamiltonian evolution, 
the explicit form eq.~(\ref{explicit_reality_conditions}) of the reality conditions characterize completely
a class of solitons, which have been first constructed by Yuzbashyan~\cite{Yuzbashyan08}, and called
by him {\it normal solitons}.
Applications will be given in sections~\ref{sec_n=2} and \ref{sec_n=3}
for systems with two and three spins.
As expected, time disappears from these conditions so that they reduce to constraints on the integration constants:
$$
X_l(0)^{*} \; X_{\bar{l}}(0) = -{1\over 4}   
$$
This characterizes a stratum of dimension $n-n_0+1$ on the real slice. 

\section{Case of $n$ double zeroes.}
\label{sec_n_degenerate}

Let us consider now critical values of the moment map whose pre-image contains configurations
where the moment map has rank one.  These are the points on  the green and red lines in Fig.[\ref{range5}].
In this case, the spectral polynomial $Q_{2n+2}(\lambda)$ can be written as:
$$
Q_{2n+2}(\lambda) = 4 \;p_2(\lambda) \prod_{i=1}^n (\lambda - E_i)^2, \quad 
p_2(\lambda) = \lambda^2 + b_1 \lambda +b_0 
$$
We have seen that $p_2(\lambda)$ is positive for real $\lambda$. Hence 
$$
\Delta = b_1^2-4 b_0 < 0
$$
If $n_r$ is the number of {\em real} zeroes, we must have $n=n_r+ 2m$ because the complex $E_i$ must come in complex conjugated pairs. 
As was explained in subsection~\ref{subsec_deg_spectral_curves}, 
we must satisfy the relations:
\begin{equation}
b_1=2 (\sigma_1(E)-\sigma_1(\epsilon))
\label{condition_on_b1}
\end{equation}
$$
2\sigma_1(E)=2\sigma_1(\epsilon) -\sum_j {\alpha_j s \over \sqrt{p_2(\epsilon_j)} }
$$
and
\begin{equation}
b_1 + \sum_j {\alpha_j s \over \sqrt{p_2(\epsilon_j)} } =0
\label{fameuse}
\end{equation}
This has the same form as eq.~(\ref{cond3gene}) and plays the role of
a compatibility relation between the {\em real} parameters $b_1,b_0$. We get therefore a one
parameter family of such spectral curves. As we have seen before, in the case of one spin,
this family corresponds to the boundary of the image of the moment map in the $(H_1,H_2)$ plane.

We set now:
$$
\lambda = -{b_1\over 2} + {\sqrt{ | \Delta | } \over 4} ( \Lambda - \Lambda^{-1} )
$$
so that $p_2(\lambda)$ becomes a perfect square:
$$
p_2(\lambda) = {|\Delta |\over 16} (  \Lambda + \Lambda^{-1} )^2
$$
and: 
$$
Q_{2n+2}(\lambda)= 4 \left( {\sqrt{ | \Delta | } \over 4} \right)^{2n+2}  (  \Lambda + \Lambda^{-1} )^2
\prod_{i=1}^n ( \Lambda - \Lambda^{-1} - 2 A_i)^2
$$
where:
$$
A_i = {b_1 + 2 E_i \over \sqrt{ | \Delta | } } 
$$
so that:
$$
\sqrt{Q_{2n+2}(\lambda)} =  2 \left( {\sqrt{ | \Delta | } \over 4} \right)^{n+1}  (  \Lambda + \Lambda^{-1} )
\prod_{i=1}^n ( \Lambda - \Lambda^{-1} - 2 A_i) \equiv Q(\Lambda)
$$
The sign of the square root is changed by the transformation $\Lambda \to - \Lambda^{-1}$. 
This transformation leaves $\lambda$ invariant and corresponds therefore  
to the hyperelliptic involution on the spectral curve, which has been uniformized in the complex plane $\Lambda$.

The points $(\lambda_k,\pm \mu_k)$ correspond to the points $(\Lambda_k,  Q(\Lambda_k))$ and $(- \Lambda_k^{-1},  Q(- \Lambda_k^{-1})= - Q(\Lambda_k)) $ in the complex $\Lambda$ plane. Among these $2n$ points only $n$ of them correspond to poles of the eigenvector.
We denote them by $\Lambda_k$ and the other ones are their image by the hyperelliptic involution $\Lambda_k^\eta = - \Lambda_k^{-1}$.
As we know, when we have a double {\em real} zero $E_i$, one of the $\lambda_i$'s is frozen at $E_i$. 
Some other variables  $\lambda_i$'s can be frozen by complex conjugate pairs on complex double roots.
Let us denote by ${\cal E}^{0}$ the set of $n_0 \geq n_r$ values $\Lambda_i$ corresponding to the frozen roots. 
We have $n_0-n_r$ even and as a result, $n-n_0$ is also even. 

The equations of motion eq.(\ref{flot}) become:
$$
{\dot{\Lambda}_k \over \Lambda_k} ={\sqrt{ | \Delta | } \over 2} { \prod_i' (\Lambda_k - \Lambda_k^{-1}- 2 A_i )
\over \prod_{l\neq k}' (\Lambda_k - \Lambda_k^{-1}-\Lambda_l + \Lambda_l^{-1} )}
$$
where the $'$ means that the terms corresponding to frozen roots are excluded because they cancel between numerator and denominator.
From this, we deduce:
$$
{\sum_k}' {d\Lambda_k  \over \Lambda_k^{2}  - 2 A_j \Lambda_k -1 } 
=   i {\sqrt{ | \Delta | } \over 2} dt, \quad \forall j  \notin {\cal E}^{0}
$$
Introducing the roots of the denominators:
$$
B_j = A_j + \sqrt{ A_j^2 +1},\quad B_{j}^\eta= - B_j^{-1}= A_j - \sqrt{ A_j^2 +1}
$$
the solution of the equations of motion is:
$$
{\prod_k' (\Lambda_k - B_j) 
\over \prod_k' (\Lambda_k - B_j^\eta) } = X_j , \quad X_j = X_j(0) e^{ i \sqrt{|\Delta]} \sqrt{ A_j^2+1} \; t }, \quad 
j \notin {\cal E}^{0}
$$
Note that the eigenfrequencies $\sqrt{|\Delta]} \sqrt{ A_j^2+1}=2\sqrt{p_{2}(E_j)}$ have a slightly more complicated
relation to the double roots $E_j$ than in the case of a spectral polynomial with $n+1$ double roots studied before. 

We introduce now the polynomials:
$$
{\cal P}_{+}(\Lambda) = {\prod_k}' (\Lambda - \Lambda_k),\quad {\cal P}_{+}^\eta(\Lambda) = {\prod_k}' (\Lambda - \Lambda_k^{\eta})
$$
satisfying the useful formulae:
\begin{equation}
{\cal P}_{+}(\Lambda^\eta)= (-\Lambda)^{-n+n_0} {\prod_k}' \Lambda_k \;\; {\cal P}_{+}^\eta(\Lambda), \quad
{\cal P}_{+}^\eta(\Lambda^\eta)= ( \Lambda)^{-n+n_0} {\prod_k}' (\Lambda_k)^{-1} \;\; {\cal P}_{+}(\Lambda)
\label{usefulformulae}
\end{equation}
With these notations, the solution of the equation of motion becomes:
\begin{equation}
{\cal P}_{+}(B_j)= X_j \;  {\cal P}_{+}(B_j^\eta) = X_j B_j^{-n+n_0}\; {\prod_k}' \Lambda_k  \; {\cal P}_{+}^\eta(B_j)
\label{cond3}
\end{equation}
These are $n-n_0$ equations for the normalised degree $n-n_0$ polynomial ${\cal P}_{+}(\Lambda)$ which is therefore completely determined.
Note that unlike the case of a spectral polynomial with $n+1$ double roots, we have access only to the evolution of separated variables (the variables $b, \bar{b}$ do not appear in these formulae). In other
words, this gives the dynamics of the reduced system, and the collective motion associated to the global $U(1)$ 
action has to be determined separately.

The next step is to find the polynomial $P_{n+1}(\lambda)$, which we express in the variable $\Lambda$. 
To simplify the notation, $P_{n+1}(\lambda(\Lambda))$ is a rational fraction in $\Lambda$
which will be abusively denoted by $P_{n+1}(\Lambda)$. 
Note that with this notation  $\Lambda^{n+1}P_{n+1}(\Lambda)$ is a polynomial
of degree $2(n+1)$ in $\Lambda$.  
We have the constraints:
$$
P_{n+1}(\Lambda_k) = P_{n+1}(\Lambda_k^{\eta})  = Q(\Lambda_k) = -Q(\Lambda_k^{\eta})
$$
By analogy to eqs.~(\ref{Pn+1+}), (\ref{Pn+1-}), we can write:
\begin{eqnarray}
\Lambda^{n+1}P_{n+1}(\Lambda) &=& \Lambda^{n+1}Q(\Lambda) - 2 {\cal S}_+(\Lambda) {\cal P}_{+}(\Lambda) 
{\cal P}_{0}(\Lambda){\cal P}_{0}^\eta(\Lambda)
\label{Pn+1+bis} \\
\Lambda^{n+1}P_{n+1}(\Lambda) &=&-\Lambda^{n+1}Q(\Lambda) + 2 {\cal S}_-(\Lambda)  {\cal P}_{+}^\eta(\Lambda) 
{\cal P}_{0}(\Lambda){\cal P}_{0}^\eta(\Lambda)
\label{Pn+1-bis} 
\end{eqnarray}
The polynomials ${\cal P}_{0}(\Lambda){\cal P}_{0}^\eta(\Lambda)$ are introduced because both 
$P_{n+1}(\Lambda)$ and $Q(\Lambda)$ vanish at those points.
The coefficients of degree $2n+2$ of $\Lambda^{n+1}P_{n+1}(\Lambda)$ and $\Lambda^{n+1}Q(\Lambda)$
are identical, and the same is true for the coefficients of degree $2n+1$, because of the relation
(\ref{condition_on_b1}). This leads to:
$$ 
\textrm{deg  }{\cal S}_+= n-n_0,\quad   \textrm{deg  } {\cal S}_- = n-n_0+2
$$ 
These two polynomials are determined by the  $n-n_0$ conditions:
$$
{\cal S}_+(\Lambda_k^\eta) = - { (\Lambda_k^{\eta})^{n+1} Q(\Lambda_k )\over 
{\cal P}_{+}(\Lambda_k^{\eta})  {\cal P}_{0}(\Lambda_k^\eta){\cal P}_{0}^\eta(\Lambda_k^\eta) } , \quad
{\cal S}_-(\Lambda_k) =  { (\Lambda_k)^{n+1} Q(\Lambda_k )\over 
{\cal P}^{\eta}_{+}(\Lambda_k){\cal P}_{0}(\Lambda_k){\cal P}_{0}^\eta(\Lambda_k)  }
$$
and additional conditions at infinity. For example:
$$
{\cal S}_-(\Lambda ) = 2  \left( {\sqrt{ | \Delta | } \over 4} \right)^{n+1}  \Lambda^{n-n_0+2} + \cdots
$$
From $P_{n+1}(\Lambda)=P_{n+1}(\Lambda^\eta)$, and $Q(\Lambda) = -Q(\Lambda^{\eta})$,
we get:
\begin{equation}
 {\cal S}_-(\Lambda)= {\prod_k}' \Lambda_k\;\;
 \Lambda^{n-n_0+2} {\cal S}_+(\Lambda^\eta)
 \label{S+S-}
 \end{equation}
Note that the quantity ${\prod_k}' \Lambda_k$ is given by the constant term in the polynomial
${\cal P}_{+}(\Lambda)$, which is reconstructed from the relations~(\ref{cond3}).
Eq.~(\ref{S+S-}) implies that  ${\cal S}_-(\Lambda) \simeq \Lambda^2$ when $\Lambda \to 0$.
Furthermore, the known coefficient of highest degree in ${\cal S}_-(\Lambda)$ determines the coefficient
of lowest degree in ${\cal S}_{+}(\Lambda)$. The $n-n_0$ values of ${\cal S}_+(\Lambda_k^\eta)$ are then sufficient
to reconstruct the polynomial ${\cal S}_{+}(\Lambda)$ because its degree is $n-n_0$.

Proceeding as before, we have:
$$
\Lambda^{2n+2}(Q^2(\Lambda) - P_{n+1}^2(\Lambda) )=
4\; {\cal S}_-(\Lambda) {\cal S}_+(\Lambda)   {\cal P}(\Lambda)  {\cal P}_{0}(\Lambda){\cal P}_{0}^\eta(\Lambda) = 
4  \left( {\sqrt{ | \Delta | } \over 4} \right)^{2n} \bar{b} b \; \Lambda^2 \; {\cal P}(\Lambda)  \bar{\cal P}(\Lambda)
$$
where: 
$$
 { \cal P}(\Lambda) =  {\cal P}_{+}(\Lambda)    {\cal P}_{+}^\eta(\Lambda)  {\cal P}_{0}(\Lambda){\cal P}_{0}^\eta(\Lambda)   ,\quad
  \bar{\cal P}(\Lambda) =  \bar{\cal P}_{+}(\Lambda)    \bar{\cal P}_{+}^\eta(\Lambda)\bar{\cal P}_{0}(\Lambda)\bar{\cal P}_{0}^\eta(\Lambda) 
$$
Therefore:
$$
 \left( {\sqrt{ | \Delta | } \over 4} \right)^{2n}  \bar{b} b \; \Lambda^2\;\bar{\cal P}(\Lambda)=
{\cal S}_-(\Lambda) {\cal S}_+(\Lambda){\cal P}_{0}(\Lambda){\cal P}_{0}^\eta(\Lambda)
$$
So we conclude that the frozen zeroes of $\bar{\cal P}(\Lambda)$ are the same as those of ${\cal P}(\Lambda)$, that
is:
$$
\bar{\cal P}_{0}(\Lambda)\bar{\cal P}_{0}^\eta(\Lambda)={\cal P}_{0}(\Lambda){\cal P}_{0}^\eta(\Lambda)
$$
We have then:
\begin{equation}
 \left( {\sqrt{ | \Delta | } \over 4} \right)^{2n}  \bar{b} b \; \Lambda^2\;  \bar{\cal P}_{+}(\Lambda)    
\bar{\cal P}_{+}^\eta(\Lambda)  = {\cal S}_-(\Lambda) {\cal S}_+(\Lambda) 
\label{PbarSSbis}
\end{equation}
Hence, the non frozen zeroes  of $\bar{\cal P}(\Lambda)$ split into the zeroes $\bar{\Lambda}_k$ of 
$ {\cal S}_+(\Lambda)$ and   $\bar{\Lambda}_k^\eta$  of  $ {\cal S}_-(\Lambda)$ and 
these two sets of zeroes are related by the hyperelliptic involution due to eq.(\ref{S+S-}).
Notice that by eqs.~(\ref{Pn+1+bis}), (\ref{Pn+1-bis}), we have:
$$
P_{n+1}(\bar{\Lambda}_k) = + Q(\bar{\Lambda}_k), \quad
P_{n+1}(\bar{\Lambda}_k^\eta) = -Q (\bar{\Lambda}_k^\eta)
$$
which are indeed compatible. On the real slice, this will imply 
that $\bar{\Lambda}_k$ is the complex conjugate of $\Lambda_k$. 
Because ${\cal S}_-(\Lambda)$ is equal to $\Lambda^{2}$ times a polynomial
of degree $n-n_0$ whose coerfficient of highest degree is known, we
get:
\begin{equation}
{\cal S}_-(\Lambda) =  2 \left( {\sqrt{ | \Delta | } \over 4} \right)^{n+1}  \Lambda^2\; \bar{\cal P}_{+}^\eta(\Lambda)
\label{Smbis}
\end{equation}
Substituting this equality in eq.(\ref{PbarSSbis}), we find:
\begin{equation}
{\cal S}_+(\Lambda) = {1\over 2} \left( {\sqrt{ | \Delta | } \over 4} \right)^{n-1}  \bar{b} b\; \bar{\cal P}_{+}(\Lambda),\quad 
\label{Spbis}
\end{equation}
Inserting into eq.(\ref{S+S-}) we find the compatibility relation:
$$
\bar{b} b =  { |\Delta|\over 4} \; {1 \over \prod_k' \Lambda_k  \prod_k' \bar{\Lambda}_k}
$$
where have used the fact that $n-n_0$ is even so that $(-1)^{n-n_0} = 1$.

Let us now turn to the real slice. Reasoning in the same way as in section~\ref{sec_n+1_degenerate},
it is necessary and sufficient to impose the $n-n_0$ conditions:
$$
{{\cal P}_{+}(B_{l}^{*})\over {\cal P}_{+}(B_{l}^{\eta *})}=
\left({\bar{\cal P}_{+}(B_{l})\over \bar{\cal P}_{+}(B_{l}^{\eta})}\right)^{*}
$$
which are equivalent to (recall that ${\cal P}_+(0) = \prod' \Lambda_k$):
\begin{equation}
{{\cal P}_{+}(B_{l}^{*})\over {\cal P}_{+}(0){\cal P}_{\eta}^{+}(B_{l}^{*})}=
\left({\bar{\cal P}_{+}(B_{l})\over \bar{\cal P}_{+}(0)\bar{\cal P}_{\eta}^{+}(B_{l})}\right)^{*}
\label{criterion_calP_bis}
\end{equation}
From eq.~(\ref{cond3}), we have:
\begin{equation}
{{\cal P}_{+}(B_{l}^{*})\over {\cal P}_{+}(0){\cal P}_{\eta}^{+}(B_{l}^{*})}=X_{\bar{l}}(B_{l}^{*})^{-n+n_0}
\label{LHS_criterion_bis}
\end{equation}
To express the right-hand side in eq.~(\ref{criterion_calP_bis}), 
we write  eqs.(\ref{Pn+1+bis}), (\ref{Pn+1-bis}) for $\Lambda=B_l$,
where $l\notin {\cal E}^{0}$:
\begin{eqnarray*}
B_l^{n+1}P_{n+1}(B_l) &=& -2 {\cal S}_+(B_l)  {\cal P}_{+}(B_l){\cal P}_{0}(B_l){\cal P}_{0}^\eta(B_l)  \\
B_l^{n+1} P_{n+1}(B_l) &=& 2 {\cal S}_-(B_l) 
 {\cal P}_{+}^\eta(B_l){\cal P}_{0}(B_l){\cal P}_{0}^\eta(B_l) 
\end{eqnarray*}
and therefore:
\begin{equation}
 {\cal S}_+(B_l){\cal P}_{+}(B_l)  = - {\cal S}_-(B_l)   {\cal P}_{+}^\eta(B_l), \quad l\notin {\cal E}^{0}
 \label{cond2bis}
 \end{equation}
Remembering eq.(\ref{cond3}) this implies:
$$
{\cal S}_-(B_l) = (-1)^{n-n_0+1} B_l^{-n+n_0} X_l\; {\prod_k}' \Lambda_k\;  {\cal S}_+(B_l)
$$
Using eqs.(\ref{Smbis}), (\ref{Spbis}), we arrive at:
\begin{equation}
{\bar{\cal P}_{+}(B_{l})\over \bar{\cal P}_{+}(0)\bar{\cal P}_{\eta}^{+}(B_{l})}=-
{B_{l}^{n-n_0+2} \over X_{l}}
\label{RHS_criterion_bis}
\end{equation}
Taking into account eqs.~(\ref{LHS_criterion_bis}) and (\ref{RHS_criterion_bis}), 
the reality conditions~(\ref{criterion_calP_bis}) become:
\begin{equation}
X_l^{*} \; X_{\bar{l}} = -  (B_l^{*})^{2n-2 n_0+2}
\label{explicit_reality_conditions_bis}
\end{equation}
Time disappears from these conditions which reduce to constraints on the integration constants:
$$
X_l(0)^{*} \; X_{\bar{l}}(0) =  -  (B_l^{*})^{2n-2 n_0+2}, \quad l\notin {\cal E}^{0}
$$
Together with the solution~(\ref{cond3}) of the Hamiltonian flow, these equations characterize
completely any stratum of dimension $n-n_0+1$ on the real slice where $n_0$ separated variables
are frozen on the double roots of the spectral polynomial. They correspond to the {\it anomalous
solitons} in Yuzbashyan's terminology~\cite{Yuzbashyan08}.

\section{The two-spins model.}
\label{sec_n=2}

We now give some details on the  two spins model. Let us first write explicitly the Hamiltonians:
\begin{eqnarray*}
H_1&=& 2\epsilon_1 s_1^z + b s_1^+ + \bar{b}s_1^-  + {\vec{s_1}\cdot \vec{s_2} \over \epsilon_1-\epsilon_2} \\
H_2&=& 2\epsilon_2  s_2^z + b s_2^+ + \bar{b}s_2^-   -  {\vec{s_1}\cdot \vec{s_2} \over \epsilon_1-\epsilon_2} \\
H_3 &=& \bar{b} b + s_1^z + s_2^z
\end{eqnarray*}
where $\vec{s_1}\cdot \vec{s_2}={s_1^z}\,{s_2^z}+{1\over{2}} ({{s_1^-}\,{s_2^+}+{s_1^+}\,{s_2^-}})$.
The singular points are given by  $b=\bar{b} = 0, s^{\pm}_1 = s^{\pm}_2 =0$ so that we have four of them: 
$s_1^z = \pm s,\quad s_2^z = \pm s$. The corresponding values $P=(H_1,H_2,H_3)$ are:
$$ 
 \begin{array}{ccllr}
P_1(\uparrow,\uparrow)&=&\left[\right. {{s^2}\over{\epsilon_1-\epsilon_2}}+2\,\epsilon_1\,s, & 2\,\epsilon_2\,s-{{s^2
 }\over{\epsilon_1-\epsilon_2}}, & 2\,s \left.\right]  \label{defP1} \\
P_2(\uparrow,\downarrow)&=&\left[\right. 2\,\epsilon_1\,s-{{s^2}\over{
 \epsilon_1-\epsilon_2}}, & {{s^2}\over{\epsilon_1-\epsilon_2}}
 -2\,\epsilon_2\,s, & 0  \left.\right]  \label{defP2} \\
P_3(\downarrow,\uparrow)&=&\left[ \right.-2\,\epsilon_1\,s-{{s^2}\over{
 \epsilon_1-\epsilon_2}}, &  {{s^2}\over{\epsilon_1-\epsilon_2}}+
 2\,\epsilon_2\,s, & 0  \left.\right]  \label{defP3} \\
 P_4(\downarrow,\downarrow)&=&\left[\right. {{s^2}\over{\epsilon_1-\epsilon_2}}
 -2\, \epsilon_1\,s, & -2\,\epsilon_2\,s-{{s
 ^2}\over{\epsilon_1-\epsilon_2}}, & -2\,s  \left.\right] \label{defP4}
 \end{array}
$$

In order to determine the type of the singularities we  write the classical Bethe equations:
 \begin{equation}
a(E)= 2 E +   { s e_1\over E-\epsilon_1} + {s e_2\over E-\epsilon_2} =0
 \label{eqE1}
 \end{equation}
These are polynomial equations of degree 3. The type of the singulariy is determined by the 
number of real roots of these equations. If we have three real roots we have a stable elliptic singularity and 
if we have one real root we have an unstable focus-focus singularity.
We easily see that $P_4(\downarrow,\downarrow)$ is always stable. See \cite{BD11} for a detailed discussion of the parameter space $(\epsilon_1,\epsilon_2)$.

\subsection{Image of the moment map.} 

The spectral curve reads in this case:
\begin{equation}
\mu^2= {Q_6(\lambda)\over (\lambda-\epsilon_1)^2 (\lambda-\epsilon_2)^2 } = 4\lambda^2+ 4 H_3 + {2H_1\over \lambda-\epsilon_1}
+ {2H_2\over \lambda-\epsilon_2}
+ {s^2 \over (\lambda-\epsilon_1)^2}+ {s^2 \over (\lambda-\epsilon_2)^2}
\label{detL2_n=1}
\end{equation}
The image of the moment map can be drawn from the degenerations of this curve. An example is shown in Fig.[\ref{range5}].
In this example the parameters $(\epsilon_1,\epsilon_2)$ have been chosen such that the points $P_2(\uparrow,\downarrow)$, $P_4(\downarrow,\downarrow)$ are stable while the points $P_1(\uparrow,\uparrow)$ and $P_3(\downarrow,\uparrow)$ are unstable. 

The spin variables are reconstructed from the separated variables $(\lambda_i,\mu_i), i=1,2$.
In particular, we have:
\begin{eqnarray}
s_1^z &=& 2{(\epsilon_1-\lambda_1)(\epsilon_1-\lambda_2)\over \epsilon_1-\epsilon_2} (\lambda_1+\lambda_2 -\epsilon_2) -\nu_1 {\lambda_2-\epsilon_1 \over
(\lambda_1-\lambda_2)(\epsilon_1-\epsilon_2)} -\nu_2 {\lambda_1-\epsilon_1 \over
(\lambda_2-\lambda_1)(\epsilon_1-\epsilon_2)}
\label{s1z2spins} \\
s_2^z &=& 2{(\epsilon_2-\lambda_1)(\epsilon_2-\lambda_2)\over \epsilon_2-\epsilon_1} (\lambda_1+\lambda_2 -\epsilon_1) -\nu_1 {\lambda_2-\epsilon_2 \over
(\lambda_1-\lambda_2)(\epsilon_2-\epsilon_1)} -\nu_2 {\lambda_1-\epsilon_2 \over
(\lambda_2-\lambda_1)(\epsilon_2-\epsilon_1)} 
\label{s2z2spins}
\end{eqnarray}
where $\nu_i=(\lambda_i-\epsilon_1)(\lambda_i-\epsilon_2)\mu_i$.

\subsection{Rank zero.}

The critical points are the points where the rank of the moment map is zero. They correspond to the complete degeneracy of the spectral curve:
$$
\mu^2 = a^2(\lambda)
$$
where $a(\lambda)$ is given  eq.(\ref{classicalBethe}). The pre-images of these points are reduced to a point, in the case of a stable critical point, 
or to a pinched torus in the case of an unstable critical point. In that case non trivial solutions of the equations of motion exist which we now study. 

So, let
\begin{equation}
Q_6(\lambda) = 4 (\lambda -E_0)^2 (\lambda-E)^2(\lambda-\bar{E})^2
\label{deuxspinsinstable}
 \end{equation}
 where  $E_0, E,\bar{E}$ are the three roots of eq.(\ref{eqE1}). We assume that  $E_0$ is real $E$ is  complex and $\bar{E}$ is the complex conjugate of $E$. 
We have the following relations coming from eq.(\ref{eqE1}):
\begin{eqnarray*}
E_0+E+\bar{E} &=& \epsilon_1+\epsilon_2 \\ 
E\bar{E}+ E_0 E +E_0\bar{E} &=& \epsilon_1\epsilon_2 + {s\over 2} (e_1+e_2) \\
E_0E\bar{E} &=& {s\over 2} (e_1\epsilon_2+ e_2\epsilon_1)
\end{eqnarray*}
Since there is a double real zero, we must freeze one point of the divisor : 
$$
\lambda_2=E_0, \quad \mu_2=0
$$
then there remains only $(\lambda_1,\mu_1)$ with
$$
\mu_1 = \pm 2{ (\lambda_1 -E_0) (\lambda_1-E)(\lambda_1-\bar{E})\over (\lambda_1-\epsilon_1)(\lambda_1-\epsilon_2)   }
$$
then from eqs.(\ref{s1z2spins}), (\ref{s2z2spins}) we get immediately:
\begin{eqnarray*}
s_1^z &=& -2{\epsilon_1-E_0 \over \epsilon_1-\epsilon_2}\Big(( \lambda_1+ E_0-\epsilon_2)(\lambda_1-\epsilon_1)\mp (\lambda_1-E)(\lambda_1-\bar{E}) \Big)\\
s_2^z &=& -2 {\epsilon_2-E_0 \over \epsilon_2-\epsilon_1}\Big(( \lambda_1+ E_0-\epsilon_1)(\lambda_1-\epsilon_2)\mp (\lambda_1-E)(\lambda_1-\bar{E}) \Big)\\
\end{eqnarray*}
The upper sign leads to the constant solution $s_i^z=se_i$ as we already know.  So we choose the lower sign. Reality of $s_i^z$ requires:
$$
\lambda_1+\bar{\lambda}_1=E+\bar{E}
$$
and we can write:
\begin{eqnarray*}
s_1^z &=& -2{\epsilon_1-E_0 \over \epsilon_1-\epsilon_2}\Big(-2\lambda_1\bar{\lambda}_1 + \epsilon_1(\epsilon_2-E_0) +E\bar{E} \Big)\\
s_2^z &=& -2{\epsilon_2-E_0 \over \epsilon_2-\epsilon_1}\Big(-2\lambda_1\bar{\lambda}_1 + \epsilon_2(\epsilon_1-E_0) +E\bar{E} \Big)\\
\end{eqnarray*}

%

The equation of motion becomes:
$$
{d\lambda_1 \over  dt} =  2i(\lambda_1-E)(\lambda_1-\bar{E})
$$
This is exactly the same equation as eq.(\ref{flot1spin}), whose solution is 
given by eq.(\ref{solution_flot1spin}). It reads: 
$$
\lambda_1 = {E-\bar{E}X \over 1 - X}, \quad X= X(0) e^{2i(E-\bar{E})t} ,\quad X(0) {\rm ~~real} <0.
$$
These solutions exist on the pre-images of the red and blue dots in Fig.[\ref{range5}].

\subsection{Rank one.}

 The lines of rank one  correspond to the following degeneracy of the spectral curve:
\begin{equation}
Q_6(\lambda) = 4 (\lambda^2 + a_1\lambda + a_0)^2 (\lambda^2+b_1\lambda+b_0)
\label{Q6R1}
\end{equation}
We have four coefficients and three conditions on $Q_6(\lambda)$. Hence we have a dimension one manifold of solutions. The coefficients $b_j$ are completely determined and there is one constraint between $(a_0,a_1)$.
Note that the case of rank 0 is obtained as a special case of rank 1, when the polynomial
$\lambda^2+b_1\lambda+b_0$ has a doubly degenerate root, which is necessarily real.
Let us parametrize $(a_0,a_1)$ in terms of $(x,y)$ as follows:
$$
a_0=-{{\epsilon_1\,y+\epsilon_2\,x-2\,\epsilon_1\,\epsilon_2 }\over{2}}, \quad 
a_1={{y+ x-2\,\epsilon_2-2\,\epsilon_1}\over{2}}
$$
Imposing the vanishing of the coefficient of $\lambda$ in eq.(\ref{detL2_n=1}),  we find:
$$
b_1=-\,\left(y+x\right)
$$
Imposing that the coefficient of the double pole at $\lambda=\epsilon_2$ is $s^2$ we find:
$$
b_0={{\left(\epsilon_2\,y^3+\epsilon_2 \,x\,y^2-\epsilon_2^2\,y
 ^2+s^2\right)}\over{y^2}}
 $$
Imposing that the coefficient of the double pole at $\lambda=\epsilon_1$ is $s^2$ we find that the parameters $x$ and $y$ are tied together by the relation:
 \begin{equation}
 S_1:\quad   s^2( y^2 - x^2 )+ (\epsilon_1-\epsilon_2) (x+y-\epsilon_1-\epsilon_2)x^2 y^2=0
 \label{S1}
 \end{equation}
This is nothing but eq.(\ref{fameuse}) when we set:
$$
x={e_1s\over \sqrt{p_2(\epsilon_1)}},\quad y={e_2s\over \sqrt{p_2(\epsilon_2)}}
$$

%
Once the spectral curve is known, we can read the values of the Hamiltonians $H_i$ by writing it in the form eq.(\ref{detL2_n=1}). 
They read:
 \begin{eqnarray}
 H_1&=&-{1\over 2} x^3 -{1\over 2} y x^2 + \epsilon_1 x^2 +{2 s^2 \over (\epsilon_1-\epsilon_2) x} (y +2 \epsilon_1-2\epsilon_2)   \label{H1XY} \\
 H_2&=&-{1\over 2} y^3 -{1\over 2} x y^2 + \epsilon_2 y^2 -{2 s^2 \over (\epsilon_1-\epsilon_2) y} (x -2 \epsilon_1+2\epsilon_2) \label{H2XY} \\
 H_3 &=&-{3\over 4} (x+y-\epsilon_1-\epsilon_2)^2 -\epsilon_2 x - \epsilon_1 y + \epsilon_1\epsilon_2
 +{1\over 4} (\epsilon_1+ \epsilon_2)^2 + {s^2\over  2 x^2}  + {s^2\over 2 y^2} \label{H3XY}
 \end{eqnarray}
Together with eq.(\ref{S1}) these are the parametric equations of the lines of rank one of the moment map.

%

As we have seen, the discriminant $b_1^2-4 b_0 b_2 \leq 0$. It is zero when we are at a critical point, 
in which case $\lambda^2 + b_1\lambda + b_0= (\lambda-E_0)^2$ with $E_0$ {\em real}. 
But as soon as we leave the critical point it becomes strictly negative so that: 
$$
\lambda^2+b_1\lambda+b_0 = (\lambda-E_0) (\lambda-\bar{E}_0)
$$
with $E_0$ and $\bar{E}_0$ being complex conjugate. The difference between the green lines and the red lines in Fig.[\ref{range5}] 
comes from the sign of the other discriminant $a_1^2-4 a_0$. This leads to two very different situations which we  describe below.

First when $a_1^2-4 a_0$ is {\em positive}  the polynomial $\lambda^2 + a_1\lambda + a_0=(\lambda -E_1)(\lambda-E_2)$ has two {\em real} roots, so that $Q_6(\lambda)$ has two double real roots. According to the general discussion we must freeze two separate variables on these roots, 
and hence $\lambda_1$ and $\lambda_2$ are completely frozen. 
This corresponds to a static solution of the reduced model and therefore to a  Liouville torus reduced to a circle $S^1$ in the original model. 
This circle  can be easily described. We can compute the components $s_j^z$ of the spins by setting  
$(\lambda_1= E_1, \mu_1=0)$, $(\lambda_2= E_2, \mu_2=0)$ in  eqs.(\ref{s1z2spins}), (\ref{s2z2spins}). We find:
\begin{eqnarray}
s_1^z&=& 2{(\epsilon_1-E_1)(\epsilon_1-E_2) \over (\epsilon_1-\epsilon_2)} (E_1+E_2-\epsilon_2) = {1\over 2} x (2\epsilon_1 -x-y)
\label{s1z2spinsbis}\\
s_2^z&=&  2{(\epsilon_2-E_1)(\epsilon_2-E_2) \over (\epsilon_2-\epsilon_1)} (E_1+E_2-\epsilon_1)=  {1\over 2} y (2\epsilon_2 -x-y)
\label{s2z2spinsbis}
\end{eqnarray}
The spins simply have a uniform rotation around the $z$-axis. The quantity $\bar{b} b$ is  constant and can be calculated from
$\bar{b} b = H_2-s_1^z-s_2^z$. This is the situation that prevails when we move along a green line in Fig.[\ref{range5}].

The other situation occurs  when $a_1^2-4 a_0$ is {\em negative}.  
The polynomial $\lambda^2 + a_1\lambda + a_0=(\lambda -E)(\lambda-\bar{E})$ has now two  
{\em complex} conjugated roots, so that $Q_6(\lambda)$ has pair of complex conjugated double roots. 
This is the situation when we move on a red line in Fig.[\ref{range5}] until the discriminant eventually vanishes. 
At this  point the two roots $(E,\bar{E})$ become a double real root (so that $Q_6(\lambda)$ has a quadruple real root there). 
Beyond that point we have two real roots and the color of the line turns green.

The pre-image of a point on a rank one line can be conveniently understood in the vicinity of a critical point by using normal forms.
Around a stable critical point, we have three elliptic normal modes. At the critical point the three action variables corresponding 
to these modes are equal to zero. When we leave the critical point on a rank one line, 
two action variables are kept equal to zero, and a third one becomes non zero. We get the three 
lines leaving the critical point from the three possible choices of the non zero action variable. 
The pre-image of a point on these lines is a circle.

Around an unstable point, we have one focus-focus mode and one elliptic mode. The neighborhood of the critical point 
on the two dimensional face on which it lies is obtained 
by the activation  of the action variables of the focus-focus mode, 
keeping the action variables of the elliptic mode equal to zero.  
On the other hand, the rank one line leaving the critical point corresponds 
to the action variable associated to the elliptic mode becoming non zero while 
the two action variables associated to the focus-focus mode remain zero. 
In phase space, the pre-image associated to the elliptic mode is a circle, 
while the pre-image associated to the focus-focus modes is a pinched torus. 
Hence the pre-image of a point on the rank one line is the product of a circle and a pinched torus. 

As long as $a_1^2-4 a_0$ is {\em negative}, there exist non trivial solutions of the equations of motion. So, let us concentrate on that case:
$$
Q_6(\lambda) = 4 (\lambda -E)^2 (\lambda-\bar{E})^2 (\lambda-E_0) (\lambda-\bar{E}_0) 
$$

One simple solution in this case is to freeze the separated variables on the pair of complex conjugated 
double roots $(\lambda_1= E, \mu_1=0)$, $(\lambda_2= \bar{E}, \mu_2=0)$. This static solution corresponds precisely 
to a circle in the original model, and eqs.(\ref{s1z2spinsbis},\ref{s2z2spinsbis}) are still valid in this case. 
This is the stratum $n=2, n_0=2$ in the general theory.

But we know that there exists another stratum, $n=2, n_0=0$, of dimension $3$ in the original model and $2$ in the
reduced model, corresponding to big motions on the pinched torus mentioned above. 
To uniformize the spectral curve, we set as before:
\begin{equation}
\lambda = {E_0+\bar{E_0} \over 2} + {E_0-\bar{E_0 } \over 4i } (\Lambda - \Lambda^{-1})= -{b_1\over 2} + {\sqrt{|\Delta|}\over 4} (\Lambda -\Lambda^{-1} ) \label{laToLa}
\end{equation}
so that:
$$
(\lambda -E)(\lambda-\bar{E}) ={|\Delta|\over 16} {1\over \Lambda^2} \left(\Lambda^2 -A \Lambda -1 \right)
\left(\Lambda^2 -\bar{A} \Lambda -1 \right)
$$
where we have set:
$$
A = {b_1 + 2 E \over \sqrt{|\Delta|}} ,\quad \bar{A}={b_1 + 2 \bar{E} \over \sqrt{|\Delta|}} 
$$
We introduce the coordinates $(\Lambda_1,\Lambda_2)$ associated to   $(\lambda_1,\lambda_2)$ and the polynomial 
${\cal P}_{+}(\Lambda) = (\Lambda -\Lambda_1)(\Lambda-\Lambda_2)$. The solution of the equations of motion is given by:
\begin{eqnarray}
(\Lambda_1 - B_{1}) (\Lambda_2 - B_{1}) &=& X_{1}(t) (\Lambda_1 - B_{1}^\eta) (\Lambda_2 - B_{1}^\eta)
\label{solu1} \\
(\Lambda_1 - B_{\bar{1}}) (\Lambda_2 - B_{\bar{1}}) &=& X_{\bar{1}}(t) (\Lambda_1 - B_{\bar{1}}^\eta) (\Lambda_2 - B_{\bar{1}}^\eta)
\label{solu2}
\end{eqnarray}
where:
$$
B_{1} = A + \sqrt{A^2+1}, \quad B_{1}^\eta = A - \sqrt{A^2+1}, \quad B_{\bar{1}} = \bar{A} + \sqrt{\bar{A}^2+1}, \quad B_{\bar{1}}^\eta = \bar{A} - \sqrt{\bar{A}^2+1}
$$
and:
$$
X_{1}(t) = X_{1}(0) e^{i \sqrt{|\Delta|} \sqrt{A^2+1} t },\quad X_{\bar{1}}(t) = X_{\bar{1}}(0) e^{i \sqrt{|\Delta|} \sqrt{\bar{A}^2+1} t }
$$
The reality condition reads:
\begin{equation}
\overline{X_{1}(0)} X_{\bar{1}}(0) = -(\overline{B_{1}})^6
\label{solureal}
\end{equation}
We can express the quantity $\bar{b}b$ :
$$
\bar{b}b ={|\Delta|\over 4} {1\over \Lambda_1\Lambda_2 \bar{\Lambda}_1 \bar{\Lambda}_2}
$$
The spins can be reconstructed through the formula $s_j^z = {P_{n+1}(\epsilon_j)\over \prod_{k\neq j} (\epsilon_j-\epsilon_k)}$. 
For this we need the polynomial $P_{n+1}(\lambda)$. Adding eqs.(\ref{Pn+1+bis},\ref{Pn+1-bis}) and using eq.(\ref{S+S-}) 
and eq.(\ref{usefulformulae}), we get:
\begin{eqnarray*}
\Lambda^{3} P_{3}(\Lambda) &=& 
2 \left( {\sqrt{|\Delta |} \over 4} \right)^3 { \Lambda^3 \over \Lambda_1 \Lambda_2  \bar{ \Lambda}_1 \bar{\Lambda}_2 }    \left[ \Lambda^3 \bar{\cal P}_{+}(\Lambda^\eta) {\cal P}_{+}(\Lambda^\eta)  + (\Lambda^\eta)^3 \bar{\cal P}_{+}(\Lambda)  {\cal P}_{+}(\Lambda)  \right] 
\end{eqnarray*}
hence:
\begin{eqnarray*}
 P_{3}(\Lambda)&=& 2 \left( {\sqrt{|\Delta |} \over 4} \right)^3 { 1 \over P \bar{P} } \Big[ P\bar{P} (\Lambda-\Lambda^{-1})^3 
 + (\bar{P} S + P \bar{S} )  (\Lambda-\Lambda^{-1})^2 \\
 &&  + (S\bar{S} + 3 P \bar{P} + P + \bar{P}-1)  (\Lambda-\Lambda^{-1})
 + 2( \bar{P} S + P \bar{S} + S + \bar{S} ) \Big]
\end{eqnarray*}
where:
 $$
S=\Lambda_1+\Lambda_2, \quad P=\Lambda_1 \Lambda_2, \quad \bar{S}=\bar{\Lambda}_1+\bar{\Lambda}_2, \quad \bar{P}=\bar{\Lambda}_1 \bar{\Lambda}_2
$$
From eq.(\ref{laToLa}), we see that  this is polynomial of degree $3$ in $\lambda$, as it should be.

\begin{figure}[ht]
\begin{center}
\includegraphics[height= 8cm]{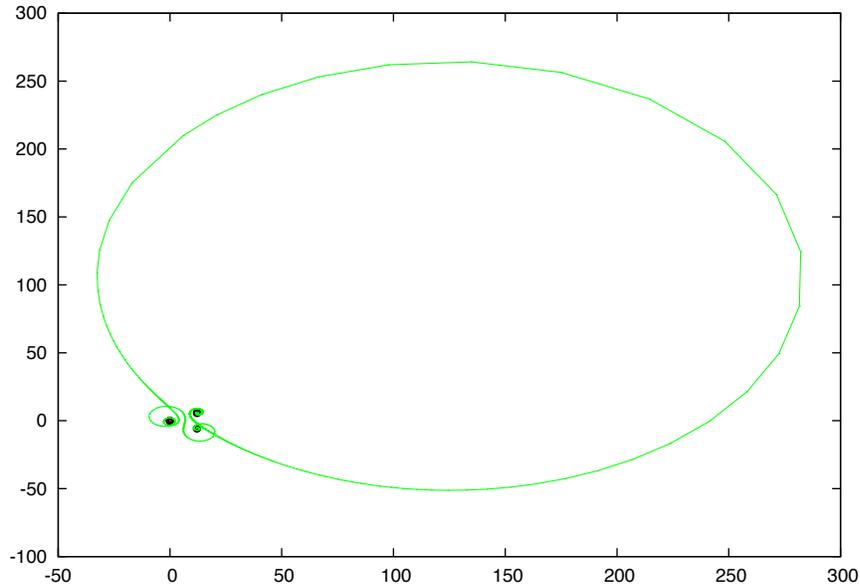}
\caption{Motion of the point say $\Lambda_1$ of the divisor. The second point $\Lambda_2$ is hidden into the black dot.  $(\epsilon_1=-1.2, \epsilon_2= -1.735)$.}
\label{La1V}
\end{center}
\nonumber
\end{figure}

\begin{figure}[ht]
\begin{center}
\includegraphics[height= 5cm]{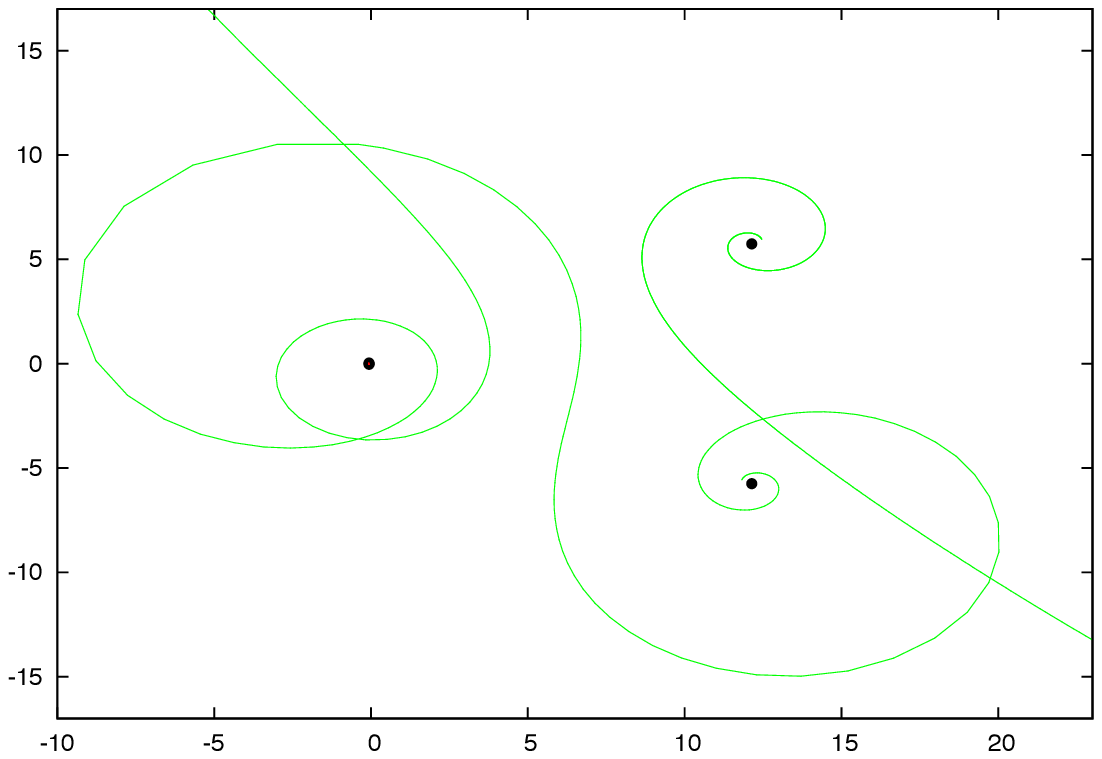}
\includegraphics[height= 5cm]{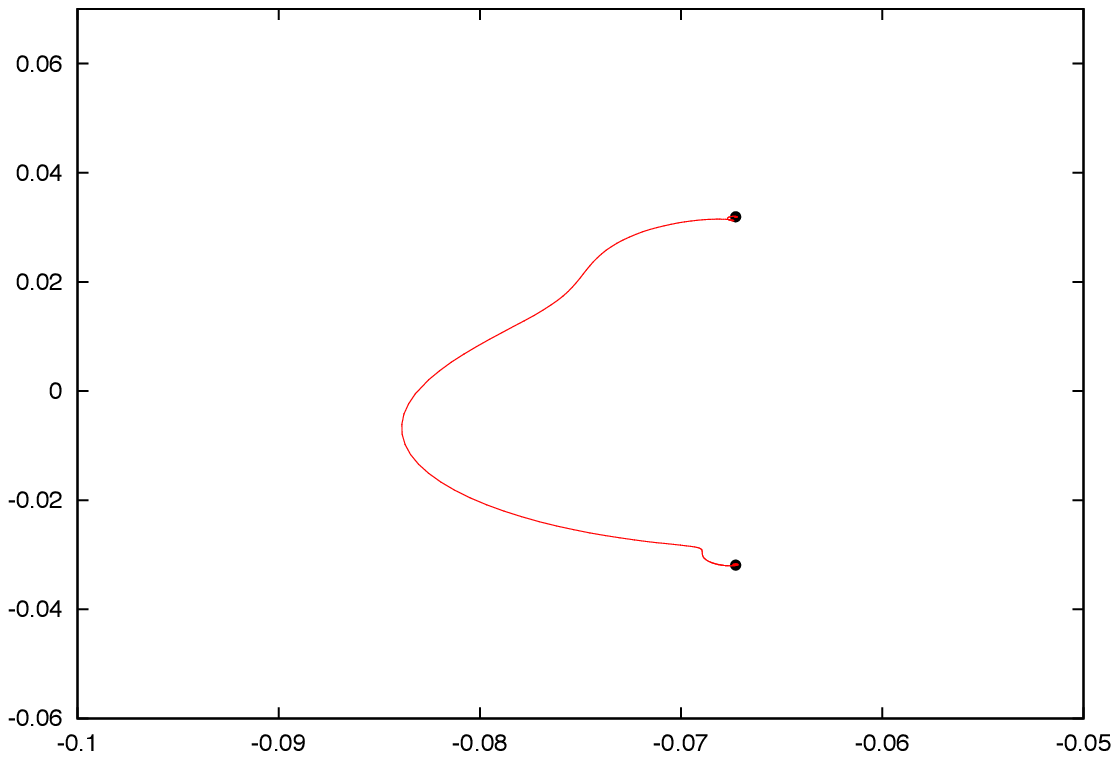}
\caption{A zoom of the motion of the divisor $\Lambda_1,\Lambda_2$.  $(\epsilon_1=-1.2, \epsilon_2= -1.735)$.}
\label{La1Vzoom}
\end{center}
\nonumber
\end{figure}

\begin{figure}[ht]
\begin{center}
\includegraphics[width=12cm,height= 8cm]{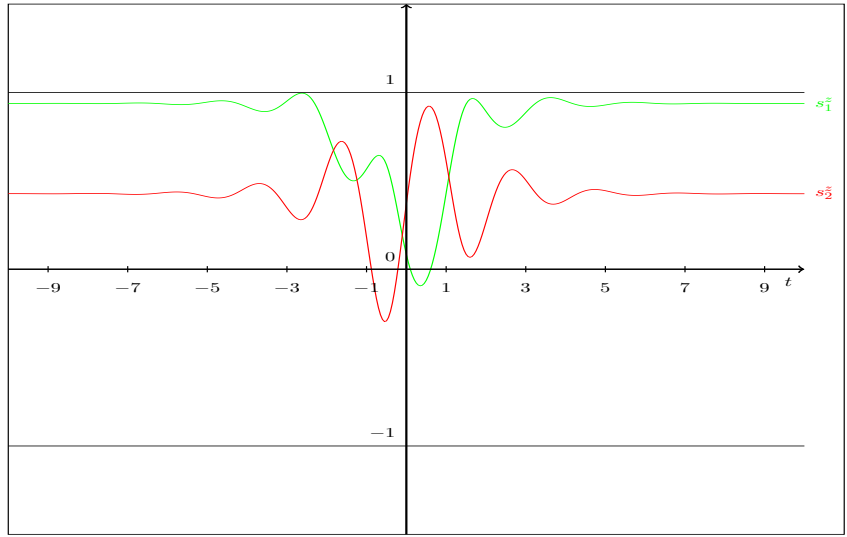}
\caption{The components $s_1^z$ and $s_2^z$ of the spins.    $(\epsilon_1=-1.2, \epsilon_2= -1.735)$.}
\label{spin12V}
\end{center}
\nonumber
\end{figure}

\begin{figure}[ht]
\begin{center}
\includegraphics[height= 8cm]{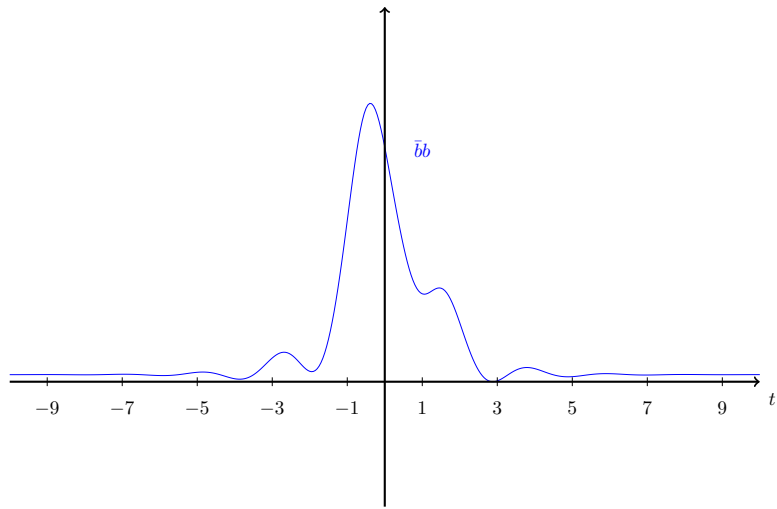}
\caption{The quantity $\bar{b}b$.    $(\epsilon_1=-1.2, \epsilon_2= -1.735)$.}
\label{bbarV}
\end{center}
\nonumber
\end{figure}

One can obtain the reality condition expressed directly on $(\Lambda_1,\Lambda_2)$. 
Eqs.(\ref{solu1},\ref{solu2}) are a linear system for the symmetric functions $S,P$.
One can eliminate $X$ and $\bar{X}$ between these equations and their complex conjugate, taking into account eq.(\ref{solureal}). 
The result can be written in the form:
\begin{equation}
{S\over P}+{\bar{S}\over \bar{P}} = -2 (A+\bar{A}),\quad \left(1 +{1\over P}\right)\left(1+ {1\over \bar{P}}\right) + {S\over P}{\bar{S}\over \bar{P}}= 4A \bar{A}
\label{globalconjugation}
\end{equation}
An example of the motion of the variables $\Lambda_1,\Lambda_2$ is shown in Fig.[\ref{La1V},\ref{La1Vzoom}], 
and the associated motions of $s_1^z,s_2^z$ and $\bar{b}b$ are shown in Figs.(\ref{spin12V},\ref{bbarV}).
We recover the soliton-like behavior of the large motion already discussed for the rank zero case. One difference
is that now, the spins are no longer along the $z$ axis as times goes to $\pm \infty$. But we also see that the
pulse in the oscillator energy $\bar{b}b(t)$ develops a non-trivial time dependence, with oscillations
superimposed to a typical solitonic enveloppe. This reflects an important qualitative difference between the
rank zero and the rank one case. In the former case, the pre-image is just a two-dimensional pinched torus,
whose small cycle corresponds to the global $U(1)$ action generated by $H_3$. This cycle is not seen in the
reduced model, for which the large motion connecting the unstable manifold to the stable one is one-dimensional.
But for a point on a line of rank one, the pre-image is the product of a circle by a two-dimensional pinched torus.
As a result, the winding motion around the small cycle of this torus becomes observable for the reduced model, as
manifested by Figs.(\ref{spin12V},\ref{bbarV}).  

\section{The three spin model.}
\label{sec_n=3}

\subsection{Critical stratum of dimension 4} 

Let us examine in some  detail the example  of three spins, for which we focus on the case of
fully degenerate spectral curves associated to the critical points: 
$$
b=\bar{b} =0, \quad s_j^{\pm} = 0, \quad s_j^z = e_j s, \quad e_j = \pm 1
$$
Therefore, the spectral polynomial can be written as:
$$
Q_8(\lambda) = 4\prod_{l=1}^4 (\lambda-E_l)^{2}
$$
According to the general analysis presented in section~\ref{sec_n+1_degenerate},
the different strata in the associated degnerate torus 
are obtained when $n_0=0$ and $n_0=2$.
Consider first the case  $n_0=0$, (or in the general case $n_0=n-3$).  We have:
\begin{eqnarray*}
{\rm deg } \;{\cal P}_+ = ~ {\rm deg } \;{\cal S}_+ =1 \\
 {\rm deg } \;{\cal P}_- = ~ {\rm deg } \;{\cal S}_- =2 
 \end{eqnarray*}
 
We set:
$$
{\cal P}_-(\lambda) = \lambda^2 -\sigma_1 \lambda + \sigma_2,\quad {\cal P}_+(\lambda) = \lambda-\lambda_0
$$
Equations (\ref{cond1}) become:
$$
{\sigma_2\over \bar{b}} -E_l \;{\sigma_1\over \bar{b}} + E_l^2 \; {1\over \bar{b}} + X_l\; \lambda_0 = E_l X_l, \quad l=1\cdots 4.
$$
or, in matrix form:
$$
\pmatrix{1&E_1&E_1^2&X_1\cr
1&E_2&E_2^2&X_2\cr
1&E_3&E_3^2&X_3\cr
1&E_4&E_4^2&X_4} 
\pmatrix{\sigma_2/ \bar{b} \cr -\sigma_1 / \bar{b} \cr  1/ \bar{b} \cr  \lambda_0 }
=\pmatrix{E_1 X_1\cr E_2 X_2 \cr E_3 X_3 \cr E_4 X_4}
$$
We introduce the matrices:
$$
M_0=\pmatrix{1&E_1&E_1^2&X_1\cr 1&E_2&E_2^2&X_2\cr 1&E_3&E_3^2&X_3\cr 1&E_4&E_4^2&X_4}  , \quad
M_1=\pmatrix{1&E_1&E_1^2&E_1X_1\cr 1&E_2&E_2^2&E_2 X_2\cr 1&E_3&E_3^2&E_3 X_3\cr 1&E_4&E_4^2&E_4 X_4} 
$$
$$
M_2=\pmatrix{E_1&E_1^2&X_1&E_1 X_1\cr E_2&E_2^2&X_2& E_2 X_2\cr E_3&E_3^2&X_3&E_3 X_3\cr E_4&E_4^2&X_4&E_4 X_4},
\quad
M_3=\pmatrix{1&E_1^2&X_1&E_1 X_1\cr 1&E_2^2&X_2& E_2 X_2\cr 1&E_3^2&X_3&E_3 X_3\cr 1&E_4^2&X_4&E_4 X_4},
\quad
M_4=\pmatrix{1&E_1&X_1&E_1 X_1\cr 1&E_2 &X_2& E_2 X_2\cr 1&E_3 &X_3&E_3 X_3\cr 1&E_4 &X_4&E_4 X_4}
$$
Denoting $D_i = \det M_i$ we have:
\begin{equation}
\lambda_0={D_1\over D_0},\quad \sigma_2={D_2\over D_4},\quad \sigma_1={D_3\over D_4},\quad \bar{b} = -{D_0\over D_4}
\label{exactsolution}
\end{equation}
Moreover, eq.(\ref{bbar}) becomes:
$$
\bar{b} b = {4\over {\cal P}_-(\lambda_0)} \left(
-{Q(\lambda_1^{-})-Q(\lambda_2^{-}) \over \lambda_1^{-}-\lambda_2^{-}}\; \lambda_0 + {\lambda_2^{-} Q(\lambda_1^{-})-\lambda_1^{-} Q(\lambda_2^{-}) \over \lambda_1^{-}-\lambda_2^{-}} \right)
$$
where we have introduced the polynomial $Q(\lambda)$ (not to be confused with $Q_8(\lambda)$) defined by:
$$
Q(\lambda) = \prod_{l=1}^4 (\lambda-E_l)
$$
The right hand side in the expression of $\bar{b}b$ 
is a function of $\sigma_1$ and $\sigma_2$ since $\lambda_1^{-}$ and  $\lambda_2^{-}$ are the roots of 
${\cal P}_-(\lambda)$.

\begin{figure}[h!]
\begin{center}
\includegraphics[height= 8cm]{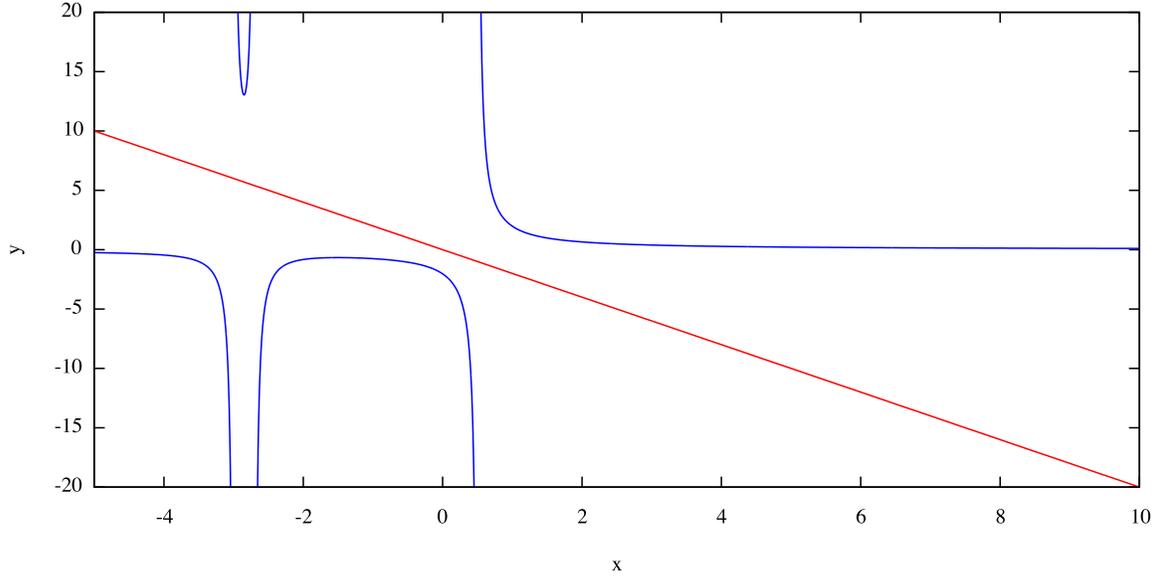}
\caption{The 3 spins model. Determination  of the parameters $\epsilon_i$ such that the classical Bethe equation has 4 complex roots.
The blue curve is $y={se_1/ (x-\epsilon_1)} +{se_2/ (x-\epsilon_2)} + {se_3/ (x-\epsilon_3)}$, the red curve is the straight line $y=-2x$.
$(e_1,e_2,e_3)=(1,-1,1)$, $(\epsilon_{1},\epsilon_{2},\epsilon_{3})=(-3,-2.7,0.5)$.}
\label{3spinscomb}
\end{center}
\nonumber
\end{figure}

In Fig.[\ref{3spinscomb}] we show an example of parameters $e_j$, $\epsilon_j$ such that the classical Bethe equation has four complex solutions. 
Note that if we choose for our critical point the ground-state of the diagonal part
of the Hamitonian $\sum_j2\epsilon_{j}s_{j}^{z}$,
we have $e_{j}\epsilon_{j}$ negative for all $j$, and a simple graphical construction shows that the Bethe equations
have at most a single pair of complex conjugate roots. In this case, according to our general discussion, 
$n_0=n-1$ and the preimage is composed of a two dimensional pinched torus, independently of the value of $n$.
In the example of Fig.[\ref{3spinscomb}], 
$(e_1,e_2,e_3)=(1,-1,1)$ and $(\epsilon_{1},\epsilon_{2},\epsilon_{3})=(-3,-2.7,0.5)$ so that only $e_{1}\epsilon_{1}$
is negative, and the critical point is a doubly excited state of $\sum_j2\epsilon_{j}s_{j}^{z}$.

\begin{figure}[h!]
\begin{center}
\includegraphics[height= 8cm]{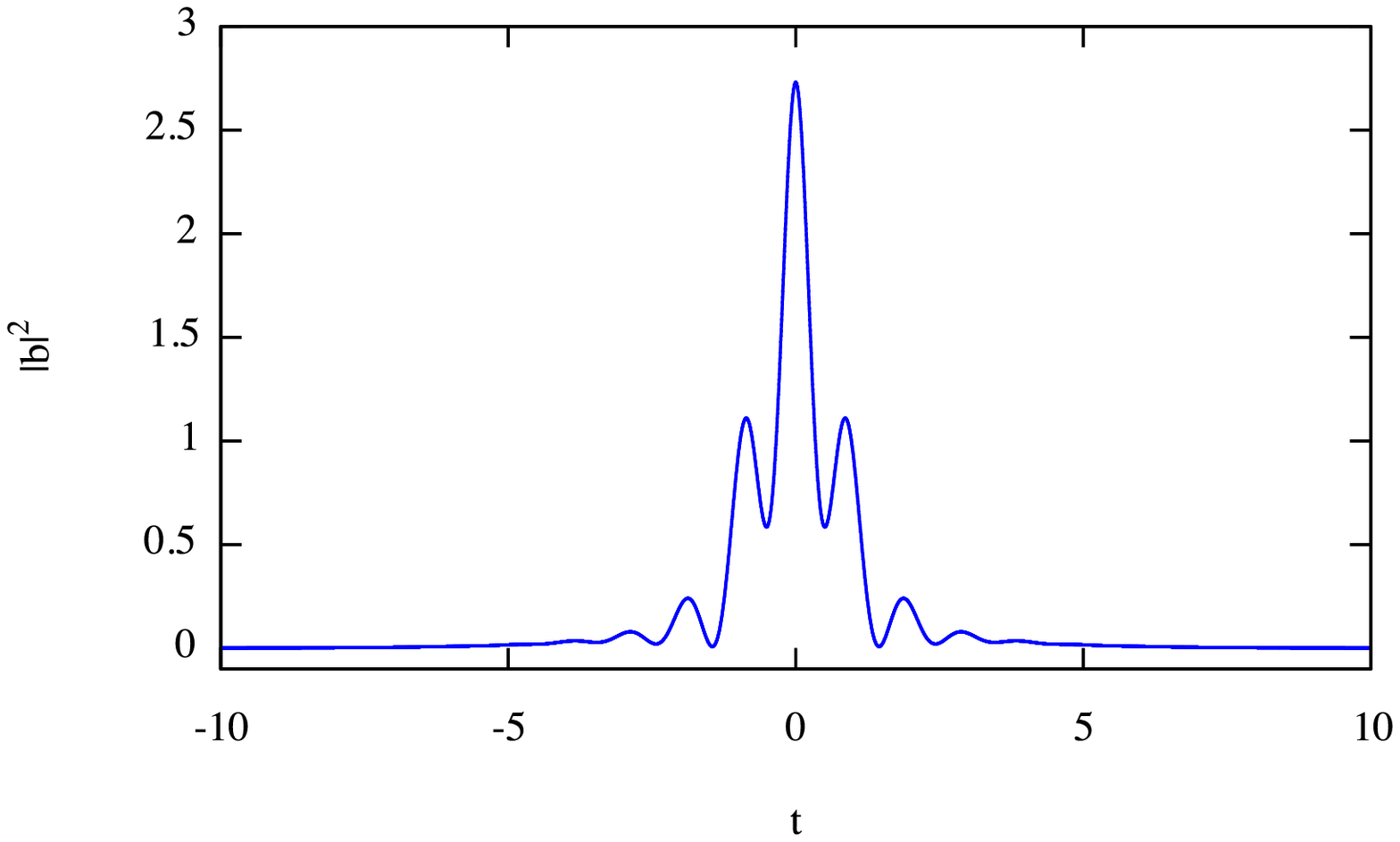}
\caption{The 3 spins model. $\bar{b}b(t)$ as function of time. $X_1(0)=X_2(0)=1/2$}
\label{3spinsphotons}
\end{center}
\nonumber
\end{figure}

\begin{figure}[h!]
\begin{center}
\includegraphics[height= 8cm]{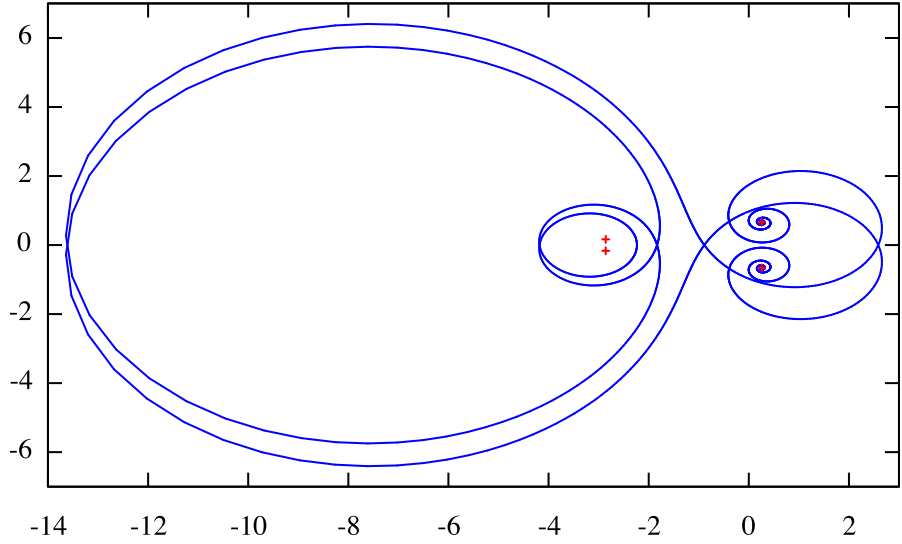}
\caption{The 3 spins model. The trajectory of $\lambda_0$.}
\label{3spinscarla}
\end{center}
\nonumber
\end{figure}

\begin{figure}[h!]
\begin{center}
\includegraphics[height= 8cm]{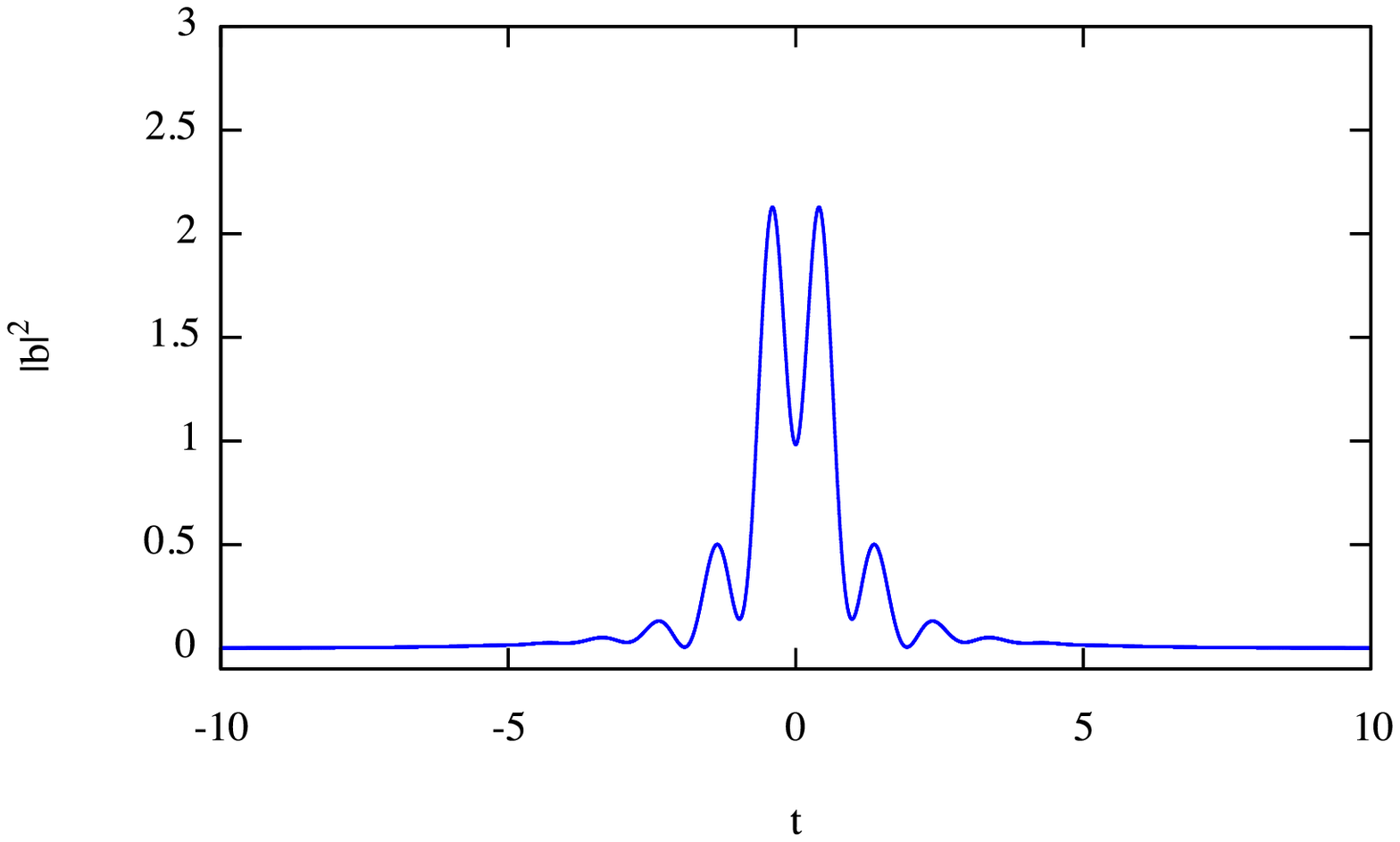}
\caption{The 3 spins model. $\bar{b}b(t)$ as function of time. $X_1(0)=1/2, X_2(0)=-1/2$}
\label{3spinsphotons2}
\end{center}
\nonumber
\end{figure}

From the general form of the time evolution~(\ref{cond1}), we see that:
$$
\bar{X}_{l}(t)={\bar{X}_{l}(0) \over X_{\bar{l}}(0)}X_{\bar{l}}(-t) 
$$
and the reality conditions eq.~(\ref{explicit_reality_conditions}) imply then that:
$$
\bar{X}_{l}(t)=-4\bar{X}_{l}(0)^{2}X_{\bar{l}}(-t) 
$$
As a result, if $X_i^2(0) = 1/4$ then we have $\bar{b} b(-t) = \bar{b} b(t)$ so that $ \bar{b} b$ has an extremum at $t=0$.  
Only the relative signs of $X_i(0)$ matters. In Fig.[\ref{3spinsphotons}] we draw the function $\bar{b} b(t)$ when $X_1(0)=X_2(0)=1/2$. 
The maximum is at $\bar{b}b(0)= | E_1+E_2-\bar{E}_1-\bar{E}_2|^2$. The time evolution of $\bar{b} b(t)$ shows a rather
rich internal structure on top of an overal solitonic shape. Note that this trajectory lies on a stratum of dimension 4,
which can be described, at least in the vicinity of the unstable critical point $(e_1,e_2,e_3)=(1,-1,1)$ 
where the system is equivalent to its quadratic normal form, as a product
of two pinched torii of dimension 2. The complexity of the motion is illustrated in Fig.[\ref{3spinscarla}]
which shows the trajectory $\lambda_0(t)$ in the case $X_1(0)=X_2(0)=1/2$. We see that it runs from one
of the roots $E_{i}$ to its complex conjugate. 
Because  the dimension of this stratum is larger than two,
it is quite easy to change the shape of these solitonic pulses. An other example is shown   
in Fig.[\ref{3spinsphotons2}] when $X_1(0)=-X_2(0)=1/2$. 
The extremum at $\bar{b}b(0)= | E_1-E_2-\bar{E}_1+\bar{E}_2|^2$ is now a local minimum.

\subsection{Critical stratum of dimension $N$}

These observations can be extended for a system with an arbitrary number of spins. Let us 
consider a stratum of dimension $N$ containing a critical point, where $N$ is
the number of roots which don't correspond to any frozen separated variable $\lambda_i$.
We have $N=n+1-n_0=n_{-}+n_{+}+1$ and $N$ is even because $n_{-}-n_{+}=1$. 
In general, $\bar{b}$ can be written as the ratio of two determinants
of size $N$, that is $\bar{b} = (-1)^{n_+} D_{0}/D_{n+1}$ where:
\begin{equation}
D_0 = 1\wedge E\wedge E^2\cdots  \wedge E^{n_- } \wedge X \wedge X E  \cdots \wedge X E^{(n_+ -1)}
\label{D0}
\end{equation}
\begin{equation}
D_{n+1} = 1\wedge E\wedge E^2\cdots  \wedge E^{(n_- -1)} \wedge X \wedge X E  \cdots \wedge X E^{n_+} 
\label{Dn+1}
\end{equation}
Notice that the degree in energy of these determinants are:
$$
d_0= { n_-(n_-+1)\over 2 } + { n_+(n_+-1)\over 2 }
$$
$$
d_{n+1}= { n_-(n_--1)\over 2 } + { n_+(n_++1)\over 2 }
$$
so that
$$
d_0-d_{n+1}= n_{-}-n_{+}=1
$$
At the symmetric point $X_i(t=0)= X_{i}(0)=1/2$, we have:
$$
D_{n+1} = \prod_{i < j} (E_i-E_j) (\bar{E}_i-\bar{E}_j)
$$
$$
D_{0} = D_{n+1} \sum_i (E_i-\bar{E}_i ) = 2i D_{n+1} \sum_i {\textrm Im} E_i
$$
All ${\textrm Im} E_i$ are positive, so assuming that they are of order one:
$$
\bar{b}b(0) \geq N^2 | \textrm{min } \textrm{ Im} E_i |^2
$$
and we recover the super radiance phenomenon of Dicke.

\subsection{Critical stratum of minimal dimension}

We consider now the case with $n_0=2$ frozen roots. This gives a smaller stratum of dimension
two. In the general case we would take $n_0=n-1$. We have:
\begin{eqnarray*}
 {\rm deg } \;{\cal P}_+ = ~ {\rm deg } \;{\cal S}_+  =0 \\
 {\rm deg } \;{\cal P}_- =~  {\rm deg } \;{\cal S}_- =1 
 \end{eqnarray*}
We choose to freeze $(\lambda_{2},\lambda_{3})$ at $(E_2,\bar{E}_2)$. So $\lambda_1$ is
the only  separated variable which is not frozen. We can write:
$$
{\cal P}_-(\lambda) = (\lambda-\lambda_1),\quad {\cal P}_+(\lambda) = 1,\quad  {\cal P}_0(\lambda)= (\lambda -E_2)(\lambda - \bar{E}_2)
$$
The conditions eq.(\ref{cond1}) give
\begin{equation}
{\cal P}_-(E_1)= \bar{b} X_1,\quad {\cal P}_-(\bar{E}_1)= \bar{b} X_{\bar{1}}
\label{cond3bis}
\end{equation}
which are solved by:
\begin{equation}
\lambda_1={\bar{E}_1 X_1 - E_1  X_{\bar{1}} \over  X_1 -X_{\bar{1}} },\quad
 \bar{b} = {   E_1  - \bar{E}_1   \over X_1 -X_{\bar{1}} }
 \label{singlexactmode}
\end{equation}
where: 
$$
\overline{X_1(0)} \; X_{\bar{1}}(0) = -{1\over 4}   
$$

\begin{figure}[h!]
\begin{center}
\includegraphics[height= 8cm]{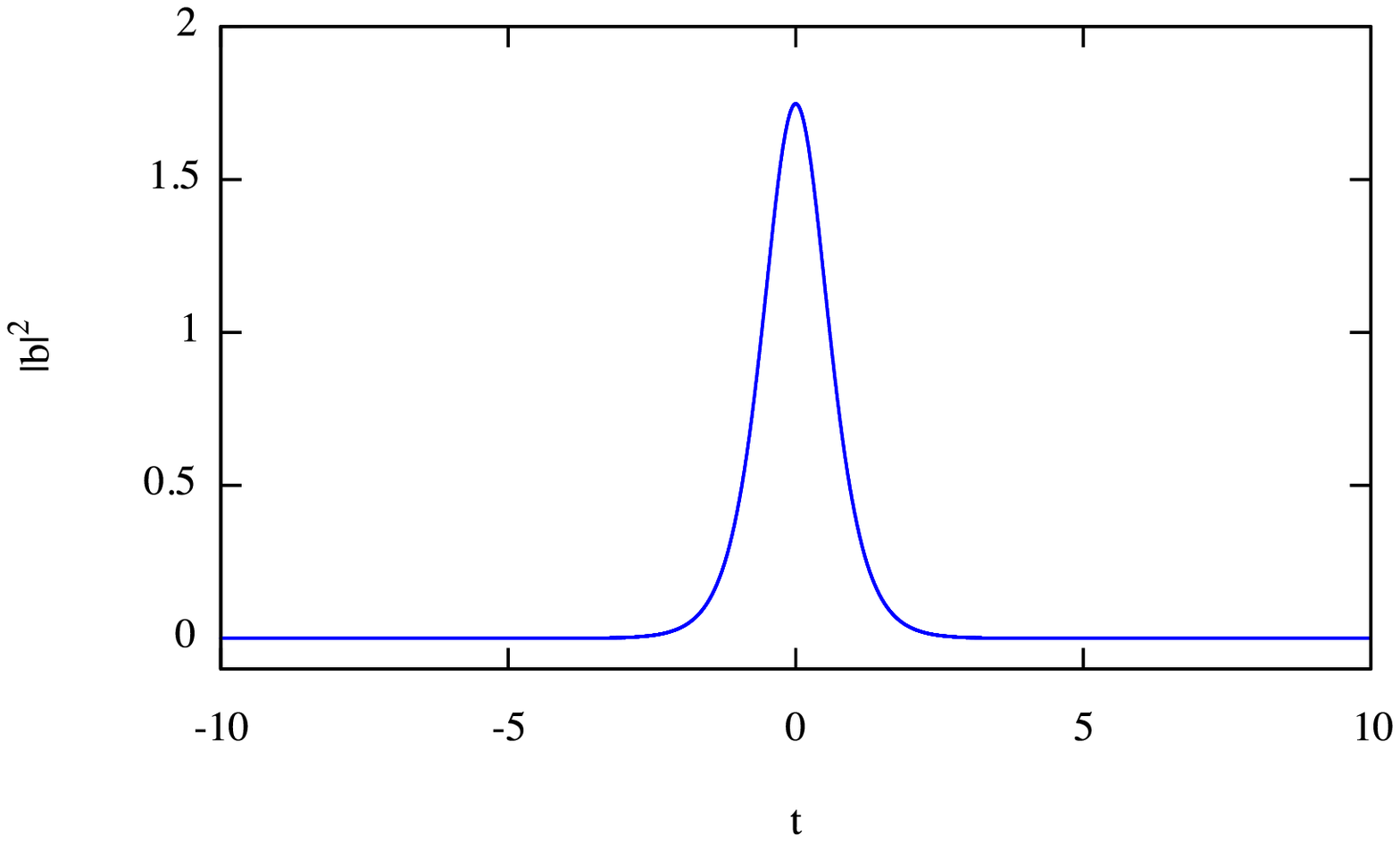}
\caption{The 3 spins model. The curve $\bar{b}b(t)$ on the small stratum.  $X_2(0)=0.5$.}
\label{3spinsmallstrata}
\end{center}
\nonumber
\end{figure}

The last equality yields the familiar compatibility relation:
\begin{equation}
\lambda_1 + \bar{\lambda}_1 = E_1 + \bar{E}_1
\label{cond4}
\end{equation}
Since the unfrozen pair can be $(E_i,\bar{E}_i)$, $i=1\cdots n$, we have $n$ such solutions. 
We show an example of the small strata in Fig.[\ref{3spinsmallstrata}]. 
We recover a simple solitonic pulse, in agreement with the general idea, discussed
in subsection~\ref{subsec_deg_spectral_curves} that in this case, the system dynamics can be mapped into the one of
an effective model with a single spin.

\subsection{Relation with Normal coordinates.}

A natural question is  to compare 
the exact solution Eq.(\ref{exactsolution}) to the normal modes expansion Eq.(\ref{normalbbar}). 
Normal modes are defined near a critical point which is precisely the pinch of the torus. So, we have to examine the exact 
solution near the pinch, that is to say when  time $t\to \pm \infty$. 

As we describe in section~\ref{sec_Normal_Forms},  normal coordinates are of the form
$B_i\equiv B(E_{i})$, $C_i\equiv C(E_{i})$, where $E_{i}$ are the $n+1$ double roots of the spectral polynomial.
There, we show that:
$$
C_i(t) = C_i(0) e^{i(\omega + 2 E_i)t}
$$
This implies that $C_i(t)$ remains proportional to $1/X_i(t)$, and we can choose
the initial condition so that 
$$
C_i(t) = {  1\over X_i(t) }
$$
hence, using Eq.(\ref{normalbbar}):
$$
\bar{b}(t) = \sum_i {  1 \over a'_i} { 1 \over X_i(t) }
$$
When $t\to \infty$, $X_i(t) \to 0$ or $\infty$. Since $\bar{b} b$ remains bounded, terms with $X_i(t) \to 0$ cannot appear when we take 
the limit of an exact solution. So, when $t\to -\infty$ only terms with $\textrm{Im }E_i <0$ contribute, and when $t\to +\infty$ only terms 
with $\textrm{Im } E_i >0$ contribute. By definition, we call $E_i$ those classical Bethe roots with $\textrm{Im } E_i >0$. Then $E_{\bar{i}}=\bar{E}_i$ has $\textrm{Im } E_{\bar{i}} <0$.

The simplest case is given by the stratum of minimal dimension Eq.(\ref{singlexactmode}):
$$
 \bar{b}_{\textrm{min}}(t) = {   E_{i}  - E_{\bar{i}}   \over X_i(t) -X_{\bar{i}} (t)}
 $$
Then
$$
\bar{b}_{\textrm{min}}(t)\vert_{t\to -\infty} = -{ E_i-E_{\bar{i}} \over X_{\bar{i}}(t) }, \quad
\bar{b}_{\textrm{min}}(t)\vert_{t\to +\infty} = { E_i  - E_{\bar{i}}  \over X_i(t) }
$$ 
So, we find the remarkable result that $ \bar{b}_{\textrm{min}}(t)$ is an exact solution which interpolates between the single normal modes 
$C_i$ and $C_{\bar{i}}$ at $t=\pm \infty$. In terms of separated variables, this corresponds to $n-1$ variables  $\lambda_i$ frozen on $n-1$ of the $E_i$, and the last $\lambda_i$ moving between $E_i$ and $E_{\bar{i}}$. As $t\to \pm \infty$, $\lambda_i \to E_i, E_{\bar{i}}$ and the normal modes correspond to all $\lambda_i$ frozen.

More generally, when we consider the limit at $t\to -\infty$ in Eqs.(\ref{D0},\ref{Dn+1}), we should keep the dominant terms, 
which amounts to replacing (assuming for simplicity that all roots are complex):
$$
X \to \pmatrix{ 0 \cr \vdots \cr 0 \cr X_{\bar{1}} \cr \vdots \cr X_{\overline{n+1}}}
$$
Then:
$$
\bar{b}(t)\vert_{t \to -\infty} = \sum_{\bar{i}} {\prod_{j} (E_{\bar{i}}-E_{j}) \over \prod_{\bar{j}\neq \bar{i}} (E_{\bar{i}}-E_{\bar{j}}) } {1\over X_{\bar{i}}(t)},\quad
\bar{b}(t)\vert_{t \to +\infty} = \sum_i {\prod_{\bar{j}} (E_i-E_{\bar{j}}) \over \prod_{j\neq i} (E_i-E_j) } {1\over X_i(t)}
$$
Hence, the solution appears as a superposition of normal modes at $t\to -\infty$  
and the above formulae determine  the asymptotic behavior at $t=+ \infty$. 
Of course, in between, the solution is a non linear superposition of  normal modes which is completely
beyond the scope of the quadratic normal form, because it describes a motion that starts 
from the unstable manifold on the critical torus and eventually returns to the stable manifold, after a big motion on the torus.
The quadratic normal form is on the other hand very useful
to describe trajectories in the vicinity of the critical point but on a different pre-image of the moment map, which
begin close to the stable manifold, and get kicked towards the unstable one. As for the single spin case, a generic trajectory 
on a torus close to the critical one will consist in two parts, one well described by the quadratic normal
form, and the other well approximated by the generalized solitonic pulses constructed in this section.

The various strata of the real slice are characterized by the number of modes appearing in the expansion of the solution at say $t=-\infty$.
In principle this could be controlled by sending some initial conditions $X_{\bar{i}}(0)$ to $\infty$. 
But because of the reality conditions, this is a rather 
singular limit, and that is why an independent description of each stratum is necessary.

\section{Conclusion}
\label{sec_Conclusion}

In this work, we have analyzed in detail singular torii in the classical
integrable Dicke-Jaynes-Cummings-Gaudin model which describes a system
of $n$ inequivalent spins coupled to a single harmonic oscillator. 
These singular torii correspond
to critical values of the moment map constructed from the conserved quantities
under the Hamiltonian dynamics.  The level sets associated to such critical values
appear to have a natural stratification, the dimension of each stratum being
equal to the rank of the differential of the moment map. Although the complexified model is 
the natural stage for the algebro-geometric method used to solve Hamilton's equations of motion,
from a physical perspective, it is crucial to look at these level sets on the {\it real slice},
whose precise identification has been one the main goals of this work.

The dimension of a stratum on a singular level set is given by $n+1-n_0$ where $n_0$
is the number of separated variables~\cite{Sk79} which are frozen on some
double roots of the associated spectral polynomial $Q_{2n+2}(\lambda)$. The values of $n_0$
vary among the integers ranging between the number of real double roots
of  $Q_{2n+2}(\lambda)$ and the total number of double roots of this
polynomial, given the fact that frozen roots occur in complex conjugate 
pairs. The fact that the possible dimensions of strata belonging to a given
level set vary by multiples of 2 has a simple interpretation. 
We know that singularities in the
Dicke-Jaynes-Cummings-Gaudin model have a quadratic normal form
which can be either of elliptic or focus-focus type~\cite{BD11}.
The quadratic normal form gives a picture of a stratum on a level set 
that is, in the vicinity of a critical point, 
the product of an $m$-dimensional torus and $p$ two-dimensional pinched torii.
This gives a stratum of dimension $m+2p$ where 
$p$ ranges from zero to the number of distinct focus-focus singularities
appearing in the quadratic normal form, which is equal to the number
of complex conjugate pairs of double roots of the spectral polynomial.
The total number of double roots of this polynomial is then equal to $n+1-m$.

We gave a complete parametrization
of the real slice for generic torii only in the simplest case of a single spin.
Fortunately, these real slices can be explicitely constructed for arbitrary
$n$ on level sets of the moment map containing very {\it small} strata, of 
minimal dimension $m=0$ (unstable critical points) or $m=1$. 
The former case corresponds to isolated critical
values and the later case to curves in the target space of the moment map.  
For these two situations, we have shown that it is possible to identify 
the physical strata of arbitrary dimensions in the corresponding level sets.
In the case $m=0$, the spectral polynomial has $n+1$ double roots, and we recover
the {\it normal solitons} constructed by Yuzbashyan~\cite{Yuzbashyan08}.
When $m=1$, $Q_{2n+2}(\lambda)$ has only $n$ double roots, but it defines a rational spectral
curve, which allowed us to obtain the general form of the  
{\it anomalous solitons} discussed by Yuzbashyan~\cite{Yuzbashyan08}.
We emphasize that these solitons can involve an arbitrary number of degrees
of freedom, equal to the number of independent focus-focus singularites
present in the quadratic normal form in the vicinity of the stratum of minimal
dimension. 

This work raises several open questions. On the theoretical side, the most
difficult seems to be to construct the real slice for level sets which
contain a stratum of minimal dimension $m \geq 2$. For $m=2$, the spectral
curve can be uniformized by a non-singular curve of genus one, so we may
hope, using Weierstrass functions, to find an explicit solution generalizing
those we found for $m=0$ and $m=1$. 

Another question is the quantum counterpart
of these multi-mode solitons. If the quantum system is prepared in the separable 
state where each spin $j$ is in the eigenstate of $s_j^{z}$ with the eigenvalue
$se_j$, $e_j=\pm 1$, and the oscillator in the ground-state of $b^{+}b$, how will
such state evolve quantum-mechanically ? The problem is to decompose this initial
state in the eigenvector basis of the quantum Hamiltonian. Some previous works
have addressed this question in the case of a classical pinched torus of dimension 
2~\cite{Bonifacio,BaDoCa09,Keeling09},
where the problem can be mapped into the evolution of a quantum wave packet 
for a single degree of freedom, prepared initially at the top of an unstable potential barrier.  
A complicated aperiodic sequence of pulses is obtained, each of them being
close to the classical monomode pulse of the $n=1$ model. We suspect that the 
qualitative difference between classical and quantum evolutions will be more
pronounced is the case of a critical level set of dimension larger than 2.
Intuitively, we expect that quantum fluctuations perform some averaging over
the configuration space of the multimode pulses whose result may be quite
different from any classical pulse chosen in this ensemble. The theoretical challenge
here is to extend the previous semi-classical analysis~\cite{Bonifacio,BaDoCa09,Keeling09} 
to a system with several degrees of freedom. 

Finally, we hope that some of the aspects of the integrable dynamics discussed here
will be evidenced experimentally.
As already mentioned in the Introduction, one possibility is to consider cold gases of
fermionic atoms in which an attractive interaction is switched on
suddenly by sweeping the external magnetic field through a Feshbach resonance~\cite{Barankov04a,Barankov04b}.
The problem with this system is that it is most conveniently prepared in the ground-state of
a weakly interacting Fermi liquid. After crossing the Feshbach resonance, this state
becomes an unstable equilibrium point with only one focus-focus singularity in its normal
form. So this gives rise only to the elementary single mode soliton. As shown in section~\ref{sec_n=3},
we need to start with an excited eigenstate of the diagonal part of the Hamiltonian to have a chance
to observe multi-mode solitons. In this respect, another class of systems looks promising. Recently, several
experimental groups have succeeded to couple a single quantum oscillator to a collection of electronic spins~\cite{Kubo10,Schuster10,Amsuss11,Bushev11}.
These experiments are in a regime where the spin ensemble behaves nearly as a macroscopic oscillator, so
most-likely in a regime close to the ground-state of the coupled system. But it seems that such systems
allow various manipulations~\cite{Kubo10,Wu10}, potentially useful for the long time storage of quantum
information. We hope that they will provide a route to access also some of the interesting physics
controlled by unstable critical points, which has been the main focus of the present work.

\section{Appendix: Normal form around critical points}
\label{sec_Normal_Forms}

Critical points are equilibrium points for all the Hamiltonians $H_j$, $j=1,\cdots, n+1$. 
At such points the derivatives with respect of all coordinates on phase space vanish. 
We have $2^n$   critical points  located at:
\begin{equation}
b=\bar{b} =0, \quad s_j^{\pm} = 0, \quad s_j^z = e_j s, \quad e_j = \pm 1
\label{static}
\end{equation}
When we expand around a configuration given by eq.(\ref{static}), all the quantities ($b$, $\bar{b}$, $s_j^+$, $s_j^-$) 
are first order, but $s_j^z$ is second order because $ s_j^z = s e_j -{e_j\over 2s} s_j^+ s_j^-  + \cdots, \quad e_j = \pm 1$. 
It is then simple to see that all first order terms in the expansions of the Hamiltonians $H_j$ vanish. 
In particular, we can replace the dynamical generating function $A(\lambda)$  defined by eq.~(\ref{defA})
by its non-dynamical approximation: 
$$
a(\lambda)=2\lambda+\sum_j {se_j\over \lambda-\epsilon_j}
$$
We can expand the Hamiltonians $H_j$ around the equilibrium points eq.(\ref{static}) and write them  
in normal form. Normal modes are obtained as follows. Consider the equation:
\begin{equation}
a(E)=2E + \sum_j {se_j\over E-\epsilon_j}=0, \quad  \mbox{ ``Classical Bethe Equation''}
\label{classicalBethe}
\end{equation}
This is an equation of degree $n+1$ for $E$. Calling $E_i$ its solutions,  we construct in this way $n+1$ variables 
$C_i \equiv C(E_i)$ and conjugated variables $B_i \equiv B(E_i)$.  We have:
\begin{equation}
\{ C_i, C_j \} = 0,\quad \{ B_i, B_j \} = 0
\label{CiCj}
\end{equation}
and 
\begin{equation}
\{ B_i, C_j\} =  2i a'_{i} \delta_{ij}
\label{BiCi}
\end{equation}
where $a'_{i}$ stands for $a'(E_{i})$.
Up to normalisation, these are canonical coordinates.
It is simple to express the quadratic Hamiltonians in theses coordinates:
\begin{equation}
{1\over 2}{\rm Tr}\; L^2(\lambda) = a^2(\lambda) + \sum_j { a(\lambda)\over a'_{j} (\lambda-E_j) } B_j C_j
\label{benoit}
\end{equation}
This has the correct analytical properties in $\lambda$ and in particular,  
there is no pole at $\lambda=E_j$ because $a(E_j)=0$.  Expanding around $\lambda=\infty$ we get:
$$
H_{n+1}= s\sum_k e_k + \sum_i {1\over 2 a'_{i}} B_i C_i
$$
and computing the residue at $\lambda=\epsilon_j$, we find:
$$
H_j= se_j\left[ 2\epsilon_j + \sum_k {se_k \over \epsilon_j-\epsilon_k}  \right] + \sum_i {1\over 2 a'_{i}} { se_j \over \epsilon_j-E_i} B_i C_i 
$$
The physical Hamiltonian is then:
\begin{equation}
H=H_{cp}+\sum_i {\omega + 2E_{i} \over 2 a'_{i}} B_i C_i
\label{HphysNormal}
\end{equation}
where $H_{cp}$ is the total energy at the critical point. This expression, together with Poisson brackets (\ref{CiCj}) and (\ref{BiCi}),
shows that that eigenfrequencies for the linearized equations of motion under the Hamiltonian $H$ are $\pm (\omega+2E_i)$. 

We can invert these formulae: devide eq.(\ref{benoit}) by $\lambda-E_j$ and take the residue at $\lambda=E_j$. Since $a(E_j)=0$ we get:
$$
{1\over 2}{\rm Tr}\; L^2(E_j) =  B_j C_j
$$
or explicitly:
$$
B_j C_j = 4 E_j^2 + 4 H_{n+1} + \sum_{k=1}^n {2H_k\over E_j-\epsilon_k} + \sum_{k=1}^n {s^2\over (E_j-\epsilon_k)^2}
$$

Let us express now the first order quantities $b$, $\bar{b}$, $s_j^+$, $s_j^-$ in terms of the normal coordinates
$B_i$, $C_i$. For this, we can reconstruct the generating functions $B(\lambda)$ and $C(\lambda)$. In fact:
$$
B(\lambda) = {N(\lambda)  \over \prod_k (\lambda-\epsilon_k) }
$$
where $N(\lambda)$ is a polynomial of degree $n$ and we know its values at the  $n+1$ classical Bethe roots:
$$
N(E_i) = B_i \prod_k (E_i-\epsilon_k)
$$
Hence we can reconstruct it by Lagrange interpolation formula:
$$
N(\lambda) = \sum_i B_i {\prod_k (E_i-\epsilon_k) \prod_{j\neq i} (\lambda-E_j) \over \prod_{j\neq i} (E_i-E_j) }
$$
Now:
$$
a'_{i} = 2 {\prod_{j\neq i}(E_i-E_j)\over \prod_k (E_i-\epsilon_k) }
$$
so we can write:
\begin{equation}
B(\lambda) = 2 \sum_i B_i {1\over a'_{i}} {\prod_{k\neq i} (\lambda-E_k) \over \prod_j (\lambda-\epsilon_j) }
\label{BNormal}
\end{equation}
and similarly:
\begin{equation}
C(\lambda) = 2 \sum_i C_i {1\over a'_{i}} {\prod_{k\neq i} (\lambda-E_k) \over \prod_j (\lambda-\epsilon_j) }
\label{CNormal}
\end{equation}
Computing the leading terms at $\infty$ we find:
\begin{equation}
b= \sum_i {1\over a'_{i}} B_i, \quad \bar{b} = \sum_i {1\over a'_{i}} C_i
\label{normalbbar}
\end{equation}
Computing the residue at $\lambda=\epsilon_j$ we get:
$$
s_j^- = 2 {\prod_k (\epsilon_j-E_k) \over \prod_{k\neq j} (\epsilon_j-\epsilon_k) } \sum_i {1\over a'_{i}(\epsilon_j-E_i)} B_i
$$
We can simplify this formula. Using the definition of classical Bethe roots
$$
a(\lambda) = 2\lambda + \sum_j {se_j\over \lambda-\epsilon_j} = 2 { \prod_i (\lambda-E_i)\over \prod_k(\lambda-\epsilon_k)}
$$
and calculating the residue at $\lambda=\epsilon_j$ we find:
$$
se_j = 2 { \prod_k (\epsilon_j-E_k)\over \prod_{k\neq j}(\epsilon_j-\epsilon_k)}
$$
hence:
$$
s_j^- = se_j\sum_i {1\over a'_{i}(\epsilon_j-E_i)} B_i, \quad s_j^+ = se_j\sum_i {1\over a'_{i}(\epsilon_j-E_i)} C_i
$$
The reality conditions are easily expressed in this approximation. From $B(\lambda)=C(\lambda^{*})^{*}$, we
get:
$$
B_{i}=C_{\bar{i}}^{*}
$$
where $C_{\bar{i}}$ stands for $C(\bar{E}_i)$. From Eq.~(\ref{CNormal}), it is possible to describe the pattern of
separated variables $\lambda_{\alpha}$ corresponding to the vicinity of the critical point on the critical torus.
Using Eq.~(\ref{HphysNormal}), we see that this critical torus is defined, within the quadratic approximation, by:
$$
B_iC_i=0
$$
for any $i$. For a real root $E_i$, the reality condition implies $B_i=C_{i}^{*}$ so $B_i=C_i=0$
and we have one separated variable $\lambda_{\alpha}$ frozen at $E_i$. Let us now consider a complex root $E_i$.
Using the decomposition Eq.(\ref{HphysNormal}) of the physical Hamiltonian in normal modes, and 
Eqs.(\ref{CiCj},\ref{BiCi}) for Poisson brackets,
we find:
$$
\partial_t C_i = \{ H, C_i\} = i(\omega + 2 E_i ) C_i
$$
so that:
$$
C_i(t) = C_i(0) e^{i(\omega + 2 E_i)t}
$$
The tangent cone to the critical torus at the critical point is composed of two hyperplanes.
The first one corresponds to unstable directions, for which $C_i(t)$ grows as time increases, that is
$\Im E_i<0$. The second one is associated to stable directions, given by $\Im E_i>0$. Generically,
on the unstable hyperplane, the expression~(\ref{CNormal}) for $C(\lambda)$ is dominated
at large times by the term such that $\Im E_i<0$ with largest absolute value. Eq.~(\ref{CNormal})
shows that in this case, all the separated variables are located on the $n$ roots $E_j$
different from $E_i$. On a codimension one subspace on the unstable hyperplane, the coefficient
$C_i(0)$ of this leading eigenvalue vanishes, and the subleading root becomes non-frozen.
More generally, $C(\lambda)$ is dominated at large times by the term for which $\textrm{Im } E_i$ is negative
and of maximal absolute value among the $i$'s such that $C_i(0)\neq 0$. 

Similarly, the stable hyperplane exhibits a nested sequence of linear subspaces, each of which
corresponds to freezing the separated variables on all the roots $E_j$ excepted one root $E_i$
with positive imaginary part.


\begin{thebibliography}{XXXXX}

\bibitem{Dicke} R. H. Dicke, {\it Coherence in spontaneous radiation processes}, Phys. Rev. {\bf 93}, 99 (1954).

\bibitem{RaBrHa}   J. M. Raimond, M. Brune, and S. Haroche, {\it Manipulating quantum
entanglement with atoms and photons in a cavity}, Rev. Mod. Phys. {\bf 73},
565-582, (2001).

\bibitem{BlHuWaGi}   A. Blais, R.-S. Huang, A. Walraff, S. M. Girvin, ad R. J. Schoelkopf,
{\it Cavity quantum electrodynamics for superconducting electrical
circuits: An architecture for quantum computation}, Phys. Rev. A {\bf 69},
062320, (2004).

\bibitem{Gau83} M. Gaudin, {\bf La Fonction d' Onde de Bethe.}  Masson, (1983).

\bibitem{YKA} E.  Yuzbashyan, V. Kuznetsov, B. Altshuler, {\it Integrable dynamics of coupled Fermi-Bose condensates.}
Phys. Rev. B {\bf 72} (2005), p. 144524.

\bibitem{JC} E. Jaynes, F. Cummings, Proc. IEEE vol. 51 (1963) p. 89.

\bibitem{BrSchMaRaHa}   M. Brune, F. Schmidt-Kaler, A. Maali, J. Dreyer, E. Hagley, J. M. Raimond, and S. Haroche,
{\it Quantum Rabi Oscillation: A Direct Test of Field Quantization in a Cavity},
Phys. Rev. Lett. {\bf 76}, 1800-1803, (1996).

\bibitem{Barankov04a}   R. A. Barankov, and L. S. Levitov,
{\it Atom-Molecule Coexistence and Collective Dynamics Near a Feshbach Resonance of Cold Fermions},
Phys. Rev. Lett. {\bf 93}, 130403, (2004)

\bibitem{Barankov04b}   R. A. Barankov, L. S. Levitov, and
B. Z. Spivak,
{\it Collective Rabi Oscillations and Solitons in a Time-Dependent BCS Pairing Problem},
Phys. Rev. Lett. {\bf 93}, 160401, (2004)

\bibitem{Eliasson90} L. H. Eliasson {\it Normal forms for Hamiltonian systems with Poisson commuting
integrals - elliptic case} Comment. Math. Helvetici 65 (1990) pp. 4-35.

\bibitem{BD11} O. Babelon, B. Dou\c{c}ot, {\it Classical Bethe Ansatz and Normal Forms in the Jaynes-Cummings Model.} 
arXiv:1106.3274

\bibitem{Yuzbashyan08} E. A. Yuzbashyan, {\it Normal and anomalous solitons in the theory of dynamical Cooper 
pairing}, Phys. Rev. B {\bf 78}, 184507, (2008).

\bibitem{DuKrNo90} B.A. Dubrovin, I.M. Krichever, S.P. Novikov, {\it Integrable Systems I.} Encyclopedia of Mathematical Sciences, 
Dynamical systems IV. Springer (1990) p.173--281.

\bibitem{BaBeTa03} O. Babelon, D. Bernard, M. Talon, {\bf Introduction to Classical Integrable systems.}
Cambridge University Press (2003).

\bibitem{Sk79} E. Sklyanin, {\it Separation of variables in the Gaudin model.} J. Soviet Math., Vol. 47,
(1979) pp. 2473-2488.

\bibitem{Audin96} M. Audin, {\it Spinning tops}, Cambridge University Press 1996.

\bibitem{Bonifacio} R. Bonifacio and R. Preparata, {\it Coherent spontaneous emission}, Phys. Rev. A {\bf 2}, 336 (1970).

\bibitem{BaDoCa09} O. Babelon, B. Dou\c{c}ot,  L. Cantini, {\it A semiclassical study of the Jaynes-Cummings model.},   J. Stat. Mech. (2009) P07011.

\bibitem{Keeling09} J. Keeling, {\it Quantum corrections to the semiclassical collective dynamics in the Tavis-Cummings model}, 
Phys. Rev. A {\bf 79}, 053825, (2009).

\bibitem{Kubo10} Y. Kubo et al., {\it Strong Coupling of a Spin Ensemble to a Superconducting Resonator}, Phys. Rev. Lett. {\bf 105}, 140502 (2010).

\bibitem{Schuster10} D. I. Schuster et al.,  {\it High-Cooperativity Coupling of Electron-Spin Ensembles to Superconducting Cavities}, Phys. Rev. Lett. {\bf 105}, 140501 (2010).

\bibitem{Amsuss11} R. Amsuss et al., {\it Cavity QED with Magnetically Coupled Collective Spin States}, Phys. Rev. Lett. {\bf 107}, 060502, (2011).

\bibitem{Bushev11} P. Bushev et al., {\it Ultralow-power spectroscopy of a rare-earth spin ensemble using a superconducting resonator}, Phys. Rev. B {\bf 84}, 060501(R) (2011)

\bibitem{Kubo11} Y. Kubo et al., {\it  Hybrid Quantum Circuit with a Superconducting Qubit Coupled to a Spin Ensemble}, Phys. Rev. Lett. {\bf 107}, 220501, (2011).

\bibitem{Wu10} Hua Wu et al., {\it  Storage of Multiple Coherent Microwave Excitations in an Electron Spin Ensemble}, Phys. Rev. Lett. {\bf 105}, 140503, (2010).


\bibitem{williamson36} John Williamson, {\it On the Algebraic Problem Concerning the Normal Forms of Linear Dynamical Systems.} American Journal of Mathematics, Vol. 58 No. 1 (1936), pp. 141-163.

\bibitem{Arnold97} V. I. Arnold, {\it Mathematical Methods of Classical Mechanics}, Springer, New-York, 1997, Appendix 6.

\bibitem{Atiyah82} M. F. Atiyah, {\it Convexity and commuting Hamiltonians.}  Bull. London Math. Soc. 14(1) (1982) pp. 1-15.

\bibitem{GuilStern82} V. Guillemin, S. Sternberg, {\it Convexity properties of the momentum mapping.} Invent. Math. 67(3) (1982) pp. 491-513.

\bibitem{San05} San V\~u Ngoc {\it Moment polytopes for symplectic manifolds with monodromy.} Adv. Math. {\bf 208} (2007), pp. 909-934

\bibitem{Duistermaat80} H. Duistermaat, {\it On Global Action-Angle variables}, Comm. Pure Appl. Math. 33 (1980) pp. 687-706.

\bibitem{San10} Alvaro Pelayo,  San V\~u Ngoc. {\it Hamiltonian dynamics and spectral theory for spin-oscillators.} 
ArXiv 1005.0439.

\bibitem{Krich83} I. Krichever, {\it ``Hessians'' of integrals of the Korteweg-De Vries Equation and Perturbations of Finite-Zone Solutions.} Soviet Math. Dokl. Vol. 27 (1983), No. 3, pp. 757-761. 

\bibitem{Audin01} M. Audin, {\it Hamiltonian Monodromy via Picard-Lefschetz theory.} Comm. Math. Phys. 229 (2002) pp. 459-489.

\bibitem{Zou92}  M. Zou, {\it Monodromy in two degrees of freedom integrable systems}, J. Geom. Phys.{\bf 10}, (1992) p. 37.

\bibitem{Zung97} N. T. Zung, {\it A note on focus-focus singularities}, Lett. Math. Phys.  {\bf 60}, (2002), pp. 87-99.

\bibitem{Zung02} N. T. Zung, {\it Another note on focus-focus singularities}, Diff. Geom. Appl. {\bf 7}, (1997), p. 123.

\bibitem{Cushman01} R. Cushman and J. J. Duistermat, {\it Non-Hamiltonian monodromy}, J. Diff. Eqs. {\bf 172}, (2001) p. 42.


\bibitem{BaTa03} O. Babelon, M. Talon, {\it Riemann surfaces, separation of variables and classical and quantum integrability.} Phys. Lett. A. 312 (2003), pp. 71-77.


\bibitem{Griffiths78} P. Griffiths and J. Harris, {\it Principles of algebraic geometry}, Wiley, New-York (1978), chapter 2.

\bibitem{BaTal07} O. Babelon, D. Talalaev, {\it On the Bethe Ansatz for the Jaynes-Cummings-Gaudin model.} 
J. Stat. Mech. (2007) P06013.

\end{thebibliography}
 \end{document}